# Martian cloud climatology and life cycle extracted from Mars Express OMEGA spectral images

**André Szantai*, Joachim Audouard+, Francois Forget*, Kevin S. Olsen+, Brigitte Gondet[x], Ehouarn Millour*, Jean-Baptiste Madeleine*, Alizée Pottier+*, Yves Langevin[x], Jean-Pierre Bibring[x]**

* *Laboratoire de Météorologie Dynamique/IPSL, (Sorbonne Université, École Normale supérieure / PSL Research University, École polytechnique, CNRS), Paris, France.*
*+ Laboratoire Atmosphères, Milieux, Observations spatiales, (CNRS/UVSQ/IPSL), Guyancourt, France*
*x Institut d'Astrophysique Spatiale, (CNRS, Université Paris-Saclay), Orsay, France.*

*Corresponding author: André Szantai: andre.szantai@lmd.ipsl.fr*

*Keywords : Mars climate, Mars atmosphere, clouds, Mars Express, OMEGA.*

## Abstract

A Martian water-ice cloud climatology has been extracted from OMEGA data covering 7 Martian years (MY 26-32) on the dayside. We derived two products, the Reversed Ice Cloud Index (ICIR) and the Percentage of Cloudy Pixels (PCP), indicating the mean cloud thickness and nebulosity over a regular grid (1° longitude × 1° latitude × 1° Ls × 1 h Local Time). The ICIR has been shown to be a proxy of the water-ice column derived from the Mars Climate Database.

The PCP confirms the existence and location of the main cloud structures mapped with the ICIR, but also gives a more accurate image of the cloud cover. We observed a more dense cloud coverage over Hellas Planitia, the Lunae Planum region and over large volcanoes (Tharsis volcanoes, Olympus and Elysium Montes) in the aphelion belt.

For the first time, thanks to the fact that Mars Express is not in Sun-synchronous orbit, we can explore the clouds diurnal cycle at a given season by combining the seven years of observations. However, because of the eccentric orbit, the temporal coverage remains limited. Identified limitations of the dataset are its small size, the difficult distinction between ice clouds and frosts, and the impact of surface albedo on data uncertainty. We could nevertheless study the diurnal cloud life cycle by averaging the data over larger regions: from specific topographic features (covering a few degrees in longitude and latitude) up to large climatic bands (covering all longitudes). We found that in the tropics (25°S – 25°N) around northern summer solstice, the diurnal thermal tide modulates the abundance of clouds, which is reduced around noon (Local Time). At northern midlatitudes (35°N – 55°N), clouds corresponding to the edge of the north polar hood are observed mainly in the morning and around noon during northern winter (Ls = 260° – 30°). Over Chryse Planitia, low lying morning fogs dissipate earlier and earlier in the afternoon during northern winter. Over Argyre, clouds are present over all daytime during two periods, around Ls = 30° and 160°.

## 1. Introduction

Understanding the water cycle of current and past Martian climate and weather remains an objective of planetary climate research. The major components of the Martian water cycle are water vapor, water-ice clouds and surface water deposits, which include the perennial polar caps, frost deposition and sublimation, and near subsurface processes (diffusion, adsorption, desorption, condensation and sublimation in the pores of the soil). This study focuses on the life cycle of water-ice clouds, and, more specifically, on their diurnal cycle. It is based on an extensive use of data from the OMEGA imaging spectrometer onboard Mars Express, from which only a small subset has been used for cloud studies previously (Langevin et al., 2007 ; Madeleine et al., 2012). Mars Express (Chicarro et al., 2004) has been in operation around Mars over the second longest period after Mars Odyssey, starting in 2004 and is still operational in 2019.

White spots near the limb of the planet have been observed with terrestrial telescopes since the 19th century and were suspected to be clouds. Based on telescopic observations of clouds between 1924 and 1971, Smith and Smith (1972) identified the main features of the seasonal cloud life cycle, in particular over major volcanoes (Olympus Mons, Elysium Mons) and the Hellas basin.

On images from the Mariner 6 and 7 spacecrafts, which made a short flyby, some features could be identified as water-ice clouds (Peale, 1973). The nature of water-ice clouds was unambiguously confirmed with the spectrometer onboard Mariner 9, the first Martian artificial satellite (Curran et al., 1973).

Mariner 9, the Viking 1 and 2 orbiters and their successors have observed cloud features of various types and at different scales. These include clouds associated to major volcanoes (in the Tharsis area, and Elysium Mons), low-level fogs and hazes (in particular in Valles Marineris), the polar hoods (clouds above the polar caps and at their edges during the corresponding fall, winter and spring seasons) and the aphelion belt (a large cloudy area that forms during the northern spring and summer, mainly in the tropics) (French et al., 1981 ; Kahn, 1984) .
Abundant observations from various instruments onboard Mars Global Surveyor (MGS), Mars Odyssey (ODY) and Mars Reconnaissance Orbiter (MRO) spacecrafts combined with the reprocessing of older satellite data from Viking orbiters and Mariner-9 led to a precise characterization of the location of clouds and their annual cycle in the 2000s (Wang and Ingersoll, 2002 ; Tamppari et al., 2003 ; Liu et al., 2003 ; Smith, 2004).
However, unlike Mars Express, put into orbit in December 2003, MGS, ODY and MRO were placed on heliosynchronous orbits. As a consequence, they always cover the (non-polar) surface of Mars at the same Martian local time, approximately 2-3 h and 13-15 h local time (LT). Therefore, they are not suited for the characterization of the diurnal lifecycle of clouds. In this paper, we present the complete climatology of clouds as observed by the OMEGA (Observatoire pour la Minéralogie, l'Eau les Glaces et l'Activité) imaging spectrometer onboard Mars Express. The major objective is the determination of the lifecycle of Martian water-ice clouds. Pixels covering ice clouds are detected by the presence of an absorption band of a water ice absorption band at 3.1 μm in their spectra. Generally, the opacity of these clouds is estimated from the comparison of two reflectances, one at the bottom of this absorption band, and the other one, with a larger wavelength, outside of it. The methodology used, in practice based on the slope of the absorption band, is detailed in section 2.1.

In Section 2 of this article we describe the OMEGA original data and the method for the derivation of spectral image-based and gridded indicators characteristic of water-ice clouds, and also some of their limitations.
The construction of a representative yearly cloud climatology, using data covering several consecutive Martian years (MY 26 to 32), is based on the assumption that the current climate does not change much from one year to another. Previous observations, mainly from satellites, have shown its validity (Smith, 2004, 2009 ; Hale et al. 2011). A major limitation to this assumption is the occurrence of global dust storms (GDSs), which take place every few Martian years. A GDS occurred during the northern winter of MY 28. It had a significant impact on surface and tropospheric temperatures, atmospheric dust content, integrated water vapor and clouds. Therefore we did not include data covering the MY 28 GDS period.

In the following section we analyze the two OMEGA-derived water-ice cloud indicators, the reversed ice cloud index and the percentage of cloudy pixels. We compare this ice cloud index to two other climatological datasets, the integrated water ice optical thickness derived from the Thermal Emission Spectrometer (TES) onboard MGS (Smith, 2004) and model predictions of the water-ice column retrieved from the Martian Climate Database (MCD) (Millour et al., 2018) derived from the Martian Global Climate Model (MGCM) of the LMD (Forget et al., 1999 ; Navarro et al., 2014).

In Section 4, we determine the daily cloud life cycle over a series of selected geographical areas. Selected daily cloud life cycles are interpreted in the context of Martian atmospheric physics and dynamics with the help of the LMD MGCM.

The resulting OMEGA-derived cloud products (4-dimensional arrays) are formatted into an accessible database.

In conclusion, a possible application is the comparison with the coming generation of high-resolution general circulation models in order to investigate small-scale processes. Another potential application is the extraction and the validation of data from Exomars / Trace Gas Orbiter (TGO) instruments, in particular from the Atmospheric Chemistry Suite (ACS) and the Nadir and Occultation for MArs Discovery (NOMAD) spectrometers.

## 2. Data and Methodology

### 2.1 Original OMEGA data

The OMEGA instrument provides spectral image cubes along an orbit at wavelengths between 0.38 and 5.1 µm. In practice, several 2D images of variable width (powers of 2 pixels, from 16 to 128 pixels) are taken along a part of an orbit ; a spectrum at 352 wavelengths in 3 channels, the visible (VIS: 0.38 - 1.05 µm ), the “C” (1.0 - 2.77 µm ) and “L” (2.65 - 5.1 µm ) near-infrared channels, is extracted for each pixel. The technical characteristics, and the performance of the instrument were described by Bibring et al. (2004) in an ESA special publication.

Detection of water ice is based on the identification of corresponding absorption lines in the spectrum. Langevin et al. (2007) first applied this principle for the detection of water ice in the south polar region by calculating the depth of a water ice absorption band near 1.5 µm. However, the water ice absorption band at 3.1 µm is better adapted to the detection of water-ice clouds, composed of small-sized ice particles (a few µm) (Langevin et al. 2007), especially in the tropics and at mid-latitudes. In a study using a small number of OMEGA spectral images, Madeleine et al. (2012) detected and identified 4 types of water-ice clouds: morning hazes, topographically controlled hazes, cumulus clouds and thick hazes.

The objective of the present study is to detect and characterize water-ice clouds from all available OMEGA spectral images, i.e. covering almost 6 Martian years (14/01/2004 – 27/04/2014: from MY 26, Ls=333.1° to MY 32, Ls=122.4°) and to build a daily and yearly cloud climatology mapped on a regular spatio-temporal grid. Since this last date, the OMEGA instrument has been activated only along a very small number of orbits, and the quality of the resulting data was insufficient to be included in this study.

### 2.2 Definition of the ice cloud index

The original ice cloud index ICI (Langevin et al. 2007, Madeleine et al. 2012) is representative of the slope on the edge of the 3.1 µm water ice absorption band. It is defined by the following relation:

$$ICI = R_{3.38} / R_{3.52} \qquad (1)$$

with $R_{3.38}$ and $R_{3.52}$: reflectances at 3.38 and 3.52 µm respectively.

This simple definition has been preferred to indexes based on a difference of radiances or on a relation involving another continuum reflectance on the other side of the water ice absorption band (close to 3.0 µm), because of other possible absorption of atmospheric or surface components. This definition was used in previous and current studies using OMEGA data (Langevin et al., 2007 ; Madeleine et al., 2012 ; Audouard et al., 2014b, Olsen et al., 2019).

In the current study we prefer to use the reversed ice cloud index ICIR:

$$ICIR = 1 - ICI \qquad (2)$$

which enables an easier identification of clouds (the more clouds / the denser / the thicker the clouds, the higher the ICIR).

In a first step, an ice cloud index value is calculated for each pixel of orbits with all types of scans (nadir, along-track, across-track and inert / limb scan). Pixels considered as inconsistent are removed from further processing when:

- their incidence and emergence value (w.r.t. the Mars surface ellipsoid) are too high (incidence >= 85° or emergence >= 50°) ;

- they are in the first and last lines of an image ;
- their signal/noise ratio is too low (SNR < 20) or infinite ;
- they belong to portions of identified orbit files of width of 128 pixels with corrupted values ;
- their ICIR is too low (ICIR < -1). This case corresponds in practice to corrupted data.

Complementary selection tests include:

- the removal of orbit files with a large spacecraft distance from the reference surface of the planet (distance > 7000 km) ;
- the removal of all orbit files during the MY 28 global dust storm (orbits 4436 to 4739 ; Ls = 261.9 - 312°), in order to build the standard cloud climatology (Wang and Richardson, 2013 ; Montabone et al., 2015).

Figure 1 compares the ICIR image containing clouds from a selected orbit file to a corresponding color image derived from the VIS channel: collocated water-ice clouds can be seen on both images.

Based on the study by Madeleine et al. (2012) who identified situations on a limited dataset of OMEGA images where water-ice clouds were present, we determined a minimal threshold value for the ice cloud index, $ICIR_{cld_min} = 0.28$ , and made some qualitative checks on data from other OMEGA orbit files. A pixel is considered as 'certainly cloudy' (i.e. covered by dense clouds, with a large or total coverage of the pixel area by clouds) if its ICIR exceeds this threshold, and will simply be qualified as 'cloudy' in the text that follows. Note that this threshold is more stringent than the one used by Madeleine et al. (2012), who used a value equivalent to an ICIR of 0.20 in tropical regions. With our threshold, pixels with ICIR value just below 0.28, which may correspond to thin clouds will not be selected. On the other hand, averaged ICIR values below our threshold may result from a mixture of cloudy (i.e. above) and non-cloudy (i.e. below the threshold) individual pixel values.

In theory, the ICIR can take values between 0 and 1. A value of zero corresponds to equal reflectances at 3.38 and 3.52 μm, and a value of one corresponds to a total absorption by water ice, due to a zero reflectance value at 3.38 μm, which is practically never observed. We found some rare cases where the ICIR of a pixel had a negative value, but close to 0. This can be the case if the reflectance at 3.52 μm is lower than at 3.38 μm, due to the absorption of another atmospheric or surface component close to the former wavelength. It may be caused in particular by the presence of $CO_2$ ice clouds (Vincendon et al., 2011).

In a recent study, Olsen et al. (2019) used our ICIR dataset, reconverted into its original Ice Cloud Index form. Based on theoretical considerations confirmed by statistical studies of the OMEGA cloudy spectra, they demonstrated that the Ice Cloud Index is a good proxy for the column mass of the water ice aerosols (see their figure 10). This results from the fact that at 3.1 μm the cloud spectral signature is primarily controlled by absorption, which itself is proportional to the mass of ice in the optical path. Olsen et al. (2019) showed that a given cloudy spectrum can reasonably be fitted by different pairs of effective radius and optical thickness (see their figure 2), but that each pair corresponds to a single value of the ice mass in the column.

The corresponding relation was used to estimate the water ice column for the entire OMEGA data set, covering Mars years 26 to 32, and these data were binned and averaged according to Ls and latitude to create a water ice climatology for Mars, (valid for relatively small particles, of effective radius between 3.5 and 7 μm), that features the aphelion cloud belt and polar hoods, in agreement with climatologies produced from other instrumental data sets (Smith, 2004, 2009, Willame et al., 2017). Here we complement their work by exploring the diurnal cycle of clouds in different regions and periods of interest.

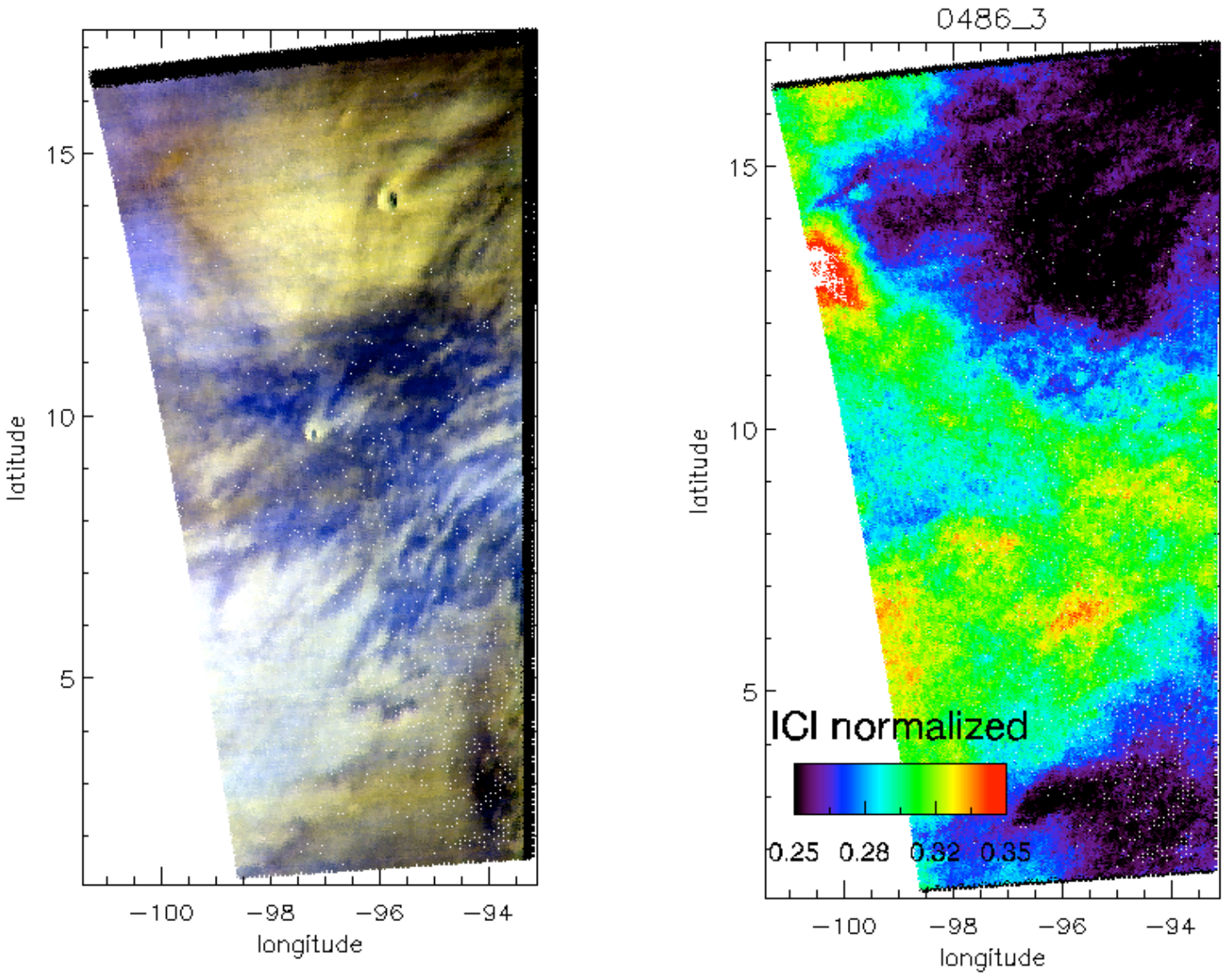

**Figure 1:** RGB OMEGA image at 3 visible wavelengths, with water-ice clouds in bluish white and light purple (left) ; corresponding IceCloudIndex image (clouds from blue: 0.28 to red: 0.35) (right). OMEGA orbit file 0486_3, location: east of Ascraeus and Pavonis Mons. The term 'ICI normalized' is a former designation of the Reversed Ice Cloud Index (ICIR).

### 2.3 Generation of the 4-dimensional ice cloud index database

In order to build a daily and yearly cloud climatology, a 4-dimensional cloud database is defined with the following grid characteristics:

Longitude ΔLON = 1° ; latitude ΔLAT = 1° ; solar longitude ΔLs = 5° ; local time ΔLT = 1 h.

These values are of the same order of magnitude as current high-resolution global climate models (e.g. Pottier et al., 2017), and allow relatively short computation times for the generation of the 4-dimensional database.

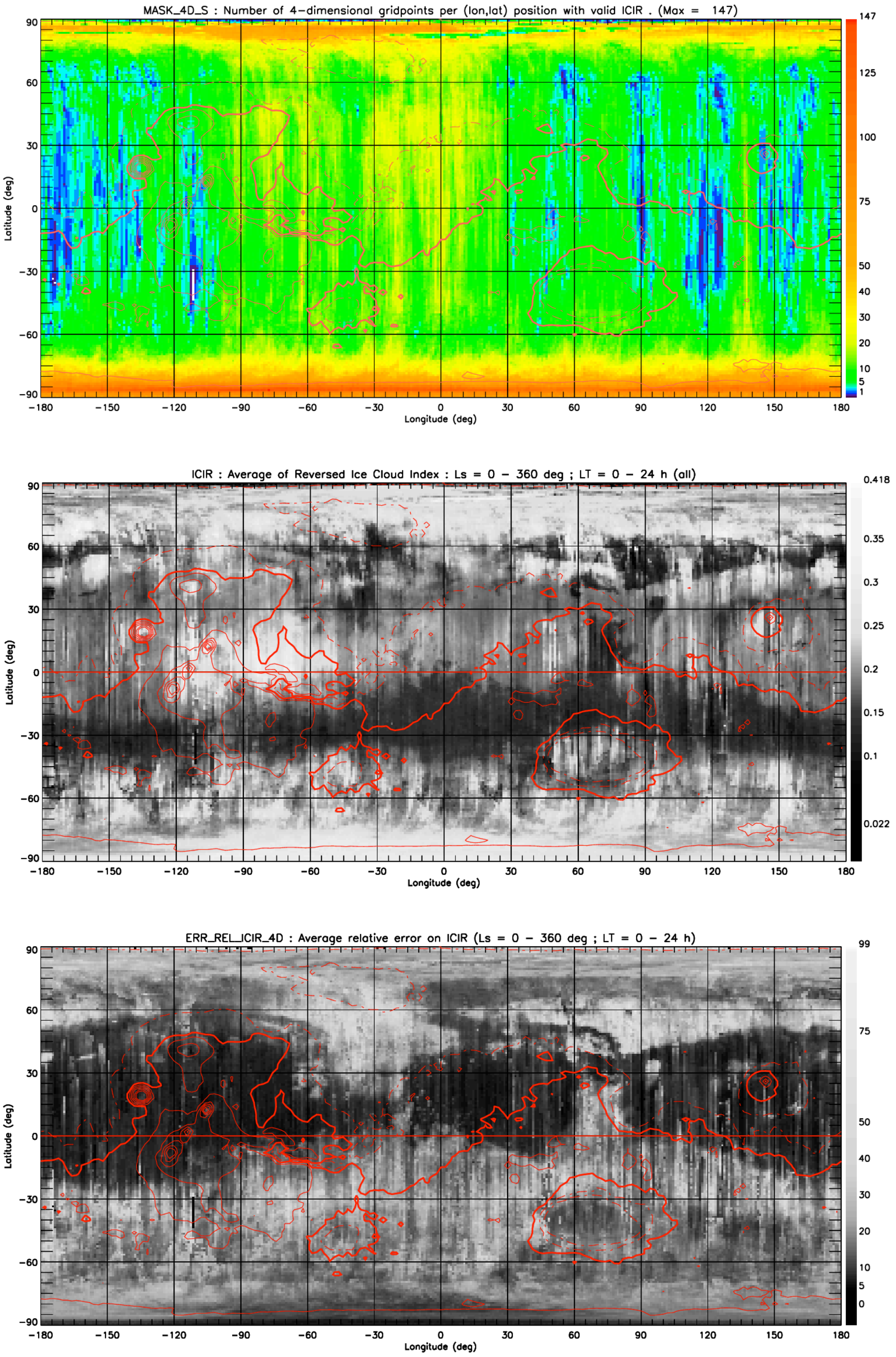

a)

b)

c)

**Figure 2:** a) Map with number of orbits summed over the entire day and solar longitudes from all Martian years. b) Averaged Reversed Ice Cloud Index (ICIR) over all solar longitudes and local times of the Martian year whenever valid data was available. c) Corresponding relative error in percentage. (The small number of relative error values of the 4-dimensional dataset that exceeded 100 % has been set to 100 % prior to the averaging.) The equator, and contour lines (in red) corresponding to surface elevation are superimposed on this figure and on following maps: thick line: datum (0 m) ; thin continuous line: elevation > 0 ; thin dashed line: elevation < 0.

Consecutive contour lines are spaced every 2500 m. The highest and the lowest values of the scale correspond to the averaged minimum and maximum of the ICIR.

All the quality-controlled pixels are then binned onto this 4-dimensional grid, i.e., they are counted and their ICIR averaged at each gridpoint. Figure 2 shows maps of the number of orbits per square degree used for the ICIR calculation, and the temporally averaged values of the ice cloud index.

At this stage, one may note the good spatial coverage over the planet. Some regions are more often covered by the satellite: the high latitudes and a 100°W – 30°E longitudinal band in the tropics (covering Chryse Planitia, and specially Valles Marineris, Mawrth Vallis and Meridiani Planum, where clays and sulfates, and also CO2 clouds (Vincendon et al., 2011) have been detected.). Everywhere data are naturally only available between sunrise and sunset, typically between 6 h LT and 18 h LT in the tropics. Further analyses (not shown) indicate that the seasonal coverage (Ls) is relatively homogeneous at the regional scale, although there are regularly-distributed gaps in the Ls-LT grid, as seen in the figures shown below in this paper. Within that context, thanks to the careful binning described above, the average values shown on Figure 2b can be considered to be a good representation of the annual-mean, daytime mean cloud distribution on Mars. One can identify large cloud structures: the aphelion belt and the polar cloud belts / polar hood edges.

### 2.4 Definition and calculation of the percentage of cloudy pixels

For each 4-dimensional gridpoint, we define the percentage of cloudy pixels PCP with the following relation:

$$PCP = 100 \times N_{cloudy} / N_{all} \qquad (3)$$

with Nall, Ncloudy: number of pixels of any type (cloudy or non-cloudy), respectively of cloudy type.

The percentage of cloudy pixels is an indicator of the cloud coverage of a gridpoint to illustrate the kilometer scale nebulosity and cloud fraction resulting from the effect of gravity waves and convection. It can be useful for comparison or use in general circulation models aiming at taking into account this subgrid scale nebulosity (Pottier et al., 2015). The map on Figure 3 shows the value of the PCP for each 1° x 1° gridpoint averaged over the day and the Martian year whenever data was available. Not surprisingly, the areas representing the cloud cover (Fig. 3) are basically the same as those covered by the ICIR (on Fig. 2).

4-dimensional gridpoints with a PCP above 0.1 %, represent 23.5 % of all gridpoints covered by valid OMEGA ICIR data.

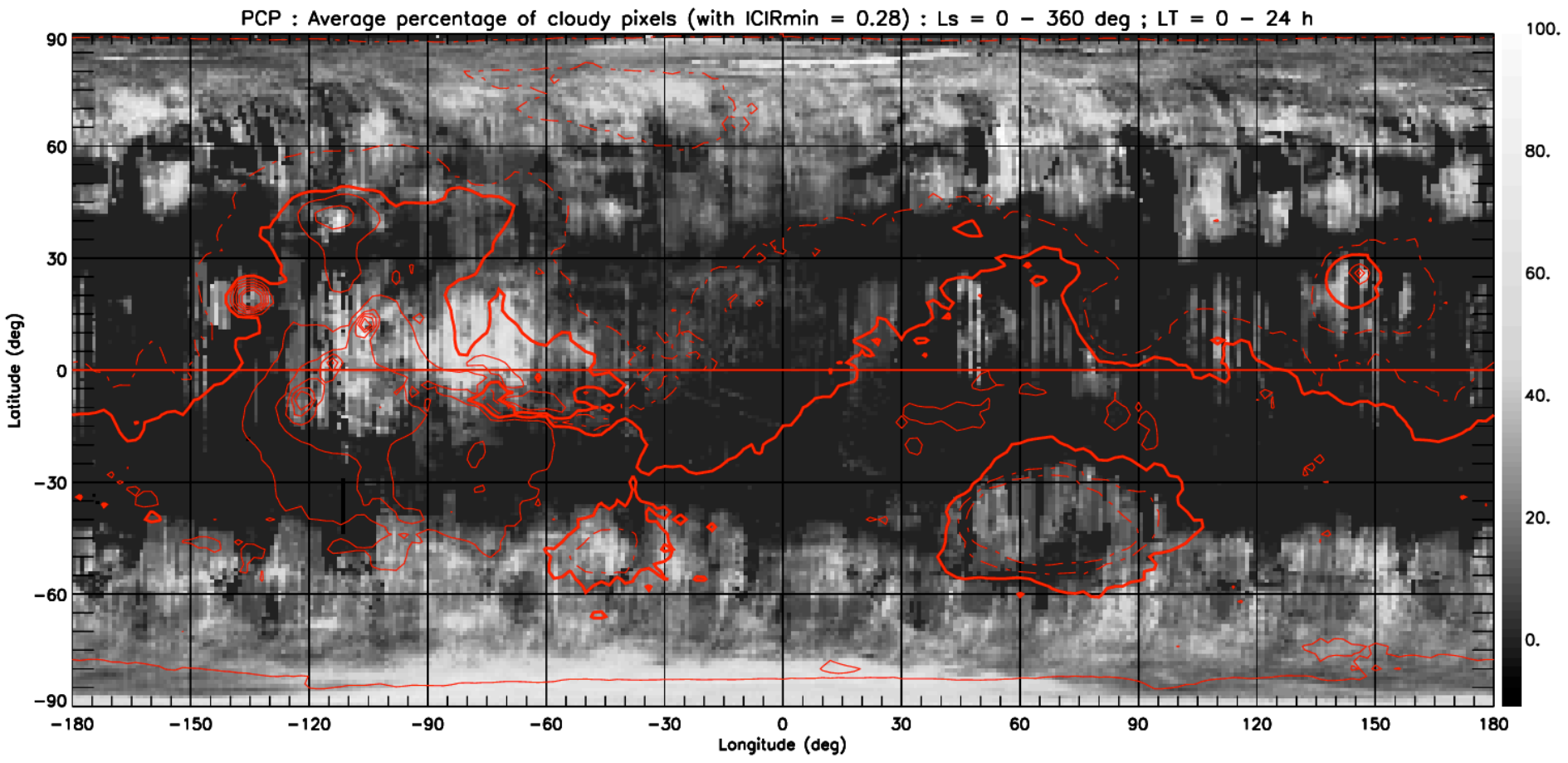

**Figure 3:** map of the average percentage of cloudy pixels (PCP) over the Martian day and year. The equator, and contour lines (in red) corresponding to surface elevation are superimposed: thick line: datum (0 m) ; thin continuous line: elevation > 0 ; thin dashed line: elevation < 0. Consecutive contour lines are spaced every 2500 m.

An important point is that some gridpoints of the 4-dimensional ICIR array may take values below the chosen threshold value of 0.28 defining original cloudy OMEGA pixels, and that these gridpoints nevertheless represent areas and periods partially covered by clouds. ICIR values of ‘cloudy’ pixels (i.e. with a value above the threshold) averaged with ICIR values of ‘cloudless’ pixels (with value below the threshold) may give an ICIR value below the threshold value of 0.28 for the gridpoint. Such situations are even more acute when gridded ICIR values are averaged again on 2 spatial or temporal dimensions in order to produce 2-dimensional figures, and canbe observed on some of the ICIR figures that follow.

### 2.5 General results about the 4-dimensional databases

The overall coverage of the 4-dimensional grid by OMEGA data is small: 1%, i.e. about 2% of the possible daytime gridoints. With such a limited coverage, on question arises : are there any (spatial) 2D gridpoints, for specific solar longitudes, where we have OMEGA-derived and -averaged data over a large portion of the day ? The Sun can illuminate regions close to the poles over a duration from 0 hour, corresponding to the polar night around the winter solstice, up to 24 hours a day, corresponding the midnight sun period around the summer solstice. In practice, the highest number of hourly-averaged observations by OMEGA over a gridpoint is 14, corresponding to 6 2D high-latitude gridpoints (at 86° to 89°N), at periods shortly after the Northern summer solstice (at Ls = 95° and 100°).

On the other hand, the Sun illuminates the tropics (25°S - 25°N) during about 12 hours a day. In this case, we found that the highest number of daily observations at a gridpoint is 4, corresponding to 16 2D gridpoints at various tropical latitudes during several periods of Northern spring and summer (at Ls = 55°, 80° and 150°).

In both climatic zones, (and also at midlatitudes, 25° - 65°, N and S, not shown), the daytime coverage of 2D gridpoints is never complete. In the most favorable cases, it represents 58,3 % of the theoretically possible longest series of instants (in the polar zones), and 33,3 % (in the tropics). And in practice, the daily coverage of single 2D gridpoints, for a fixed Ls, is always below these percentages.

This limited daily and annual data coverage for a single gridpoint led us to define larger spatial regions over which ICIR and PCP values are integrated and averaged. Based on a visual identification of the cloudy regions at various periods of the Martian year, we could define large regions of interest of different type and extension, delimited by their longitude and latitude, with a name corresponding to the main topographic (or climatic) feature. The characteristics of the selected Regions Of Interest (ROI) are listed in Table 1.

These regions of interest, where water-ice clouds have been observed, are the following:

- large climatic zones:

  * tropical (25° S - 25° N). In the tropics, the aphelion belt is the dominant large-scale cloud structure during northern spring and summer.

  * mid-latitudes (25° N – 65° N ; 25° S – 65° S). The poleward part of midlatitude areas undergo an annual cycle related to the advance and retreat of the polar hoods (Tamppari et al., 2008, Benson et al., 2010 ; Benson et al., 2011). The equatorward part is impacted by the aphelion belt around the northern summer solstice period.

  * polar regions (65° N – 90° N ; 65° S – 90° S). These regions are covered over more than half of the Martian year by a polar hood, which has its largest extension around the winter solstice. However this period also corresponds largely to nighttime at those latitudes and thus these regions are not observable except on the equatorward edge of the polar hood.

- large impact craters: Hellas, Argyre ;

- around individual volcanoes: Olympus Mons, Elysium Mons, Arsia Mons, and groups of volcanoes: Tharsis volcanoes (Ascraeus, Pavonis and Arsia Mons). Clouds are abundant over major volcanoes mainly during northern spring and summer. Clouds are also present in lesser amount over Arsia Mons around northern winter solstice (Benson et al., 2003 ; Benson et al., 2006).

- regions in the northern hemisphere where water ice or clouds have been observed or could be present: Lunae Planum, South-East Tharsis, Valles Marineris (Benson et al., 2003), Tempe Terra, Syrtis Major

- a region in the southern hemisphere where clouds have been observed: the cloud bridge zone, connecting the aphelion belt with the cloud belt around the South Pole or the edge of the south polar hood.

| | **Name of the region of interest** | **Latitude (degrees)** | **Longitude (degrees E)** | **Correlation ICIR vs. MCD WIC Ls=[60-120°]** | **Percentage of 'Certainly Cloudy' pixels Ls=[60-120°]** |
|---|---|---|---|---|---|
| 1 | **Olympus Mons** | 10 N ; 30 N | -145 ; -125 | **0,736** | 58 |
| 2 | **Alba Patera** | 30 N ; 50 N | -125 ; -105 | **0,666** | 21 |
| 3 | **Arsia Mons** | -15 S ; 0 N | -130 ; -115 | **0,814** | 78 |
| 4 | **Elysium Mons** | 15 N ; 35 N | 135 ; 155 | **0,747** | 54 |
| 5 | **Tharsis Montes** | -15 S ; 20 N | -125 ; -100 | 0,471 | 75 |
| 6 | **Tharsis South-East** | -20 S ; 10 N | -105 ; -80 | **0,707** | 75 |
| 7 | **Tempe Terra** | 30 N ; 50 N | -90 ; -55 | **0,713** | 8 |
| 8 | **Chryse Planitia** | 20 N ; 50 N | -60 ; -30 | **0,625** | 22 |
| 9 | **Lunae Planum** | 0 N ; 25 N | -70 ; -40 | 0,325 | 80 |
| 10 | **Arabia Terra West** | 10 N ; 40 N | -35 ; 10 | **0,608** | 21 |
| 11 | **Syrtis Major** | 0 N ; 35 N | 55 ; 80 | 0,355 | 50 |
| 12 | **Valles Marineris** | -15 S ; -5 S | -90 ; -45 | 0,422 | 65 |
| 13 | **SH Cloud bridge** | -35 S ; -20 S | -150 ; -60 | 0,549 | 18 |
| 14 | **Tyrrhena Terra** | -10 S ; 0 N | 70 ; 85 | 0,525 | 57 |
| 15 | **Argyre Planitia** | -55 S ; -35 S | -65 ; -25 | 0,305 | 53 |
| 16 | **Hellas Planitia** | -55 S ; -25 S | 40 ; 105 | **0,838** | 70 |
| 17 | **Tropics 25S – 25N** | -25 S ; 25 N | -180 ; 180 | **0,601** | 44 |
| 18 | **Midlatitudes 25N – 55N** | 25 N ; 55 N | -180 ; 180 | 0,489 | 5 |
| 19 | **Midlatitudes 25S – 55S** | -55 S ; -25 S | -180 ; 180 | **0,732** | 24 |
| 20 | **N Polar 65N – 89N** | 65 N ; 89 N | -180 ; 180 | -0,056 | 26 |
| 21 | **S Polar 65S – 89S** | -89 S ; -65 S | -180 ; 180 | X | X |

**Table 1:** characteristics of the regions of interest. Correlation between the OMEGA ICIR and the Water Ice Column (WaterIceCol, or WIC – see Section 4.2) from the MCD (values above 0.6 are in **bold**), and percentage of 'certainly cloudy' 4-dimensional gridpoints (i.e. that have an ICIR above the threshold value of 0.28) around the Northern summer solstice period.

These selected spatial regions of interest will be used in Section 4 to determine the diurnal cloud life cycle. The number of 4-dimensional gridpoints of a region is defined spatially by the covered angular surface (Δlongitude x Δlatitude). The number of 2-dimensional gridpoints covering a region varies between 150 square degrees (for the Tyrrhena, region, number 14) and 18000 square degrees (for the Tropical region, number 17).

### 2.6 Impact of the number of pixels per gridpoint on gridded OMEGA data

The number of original pixels from the 7436 valid orbit-segment files used to calculate the gridded and averaged ICIR values varies between one single pixel up to 98093 pixels per gridpoint. For about 14 % of the gridpoints, less than 10 pixels are averaged. Various factors have an impact on this number of pixels : the number of orbits (from one to 6), the filters that remove inconsistent pixels (see Section 2.2), the distance of the satellite to the surface (a pixel covers a larger surface of the planet when it is taken from a higher altitude) and the latitude of the pixel (a square degree covers a decreasing surface on the planet with increasing latitude). ICIR gridded values are expected to be more significant from a statistical point of view if they are computed from a sufficiently large number of original ICIR pixels.

It is difficult to identify the impact of each factor on the number of pixels per gridpoint. The main relation we identified is the impact of the latitude of the gridpoint on the number of pixels. At high latitudes (above 85°), the number of pixels is always small, below 100 pixels. The proportion of gridpoints with less than 100 pixels decreases with decreasing latitude. The number of pixels is large (above 100) at midlatitudes and in the tropics for a large majority of gridpoints. A complete description is given in Appendix A.

This relation with latitude must be kept in mind when relative error data is analyzed. It has also a more direct impact on the PCP, which depends in its definition on two numbers of pixels, the number of pixels of any kind (cloudy or non-cloudy) and the number of cloudy pixels.

## 3. Seasonal ICIR and PCP maps

### 3.1 Seasonal mapping of water-ice clouds

Geographical maps of clouds can be derived from 4-dimensional ICIR and PCP datasets by averaging data over time, i.e. over selected periods of the Martian day and year. Figure 4a and 5a, resp. 4b and 5b, show examples of PCP and ICIR maps centered on the solstices (Ls = 90 and 270°) and averaged over 90 degrees of solar longitude.

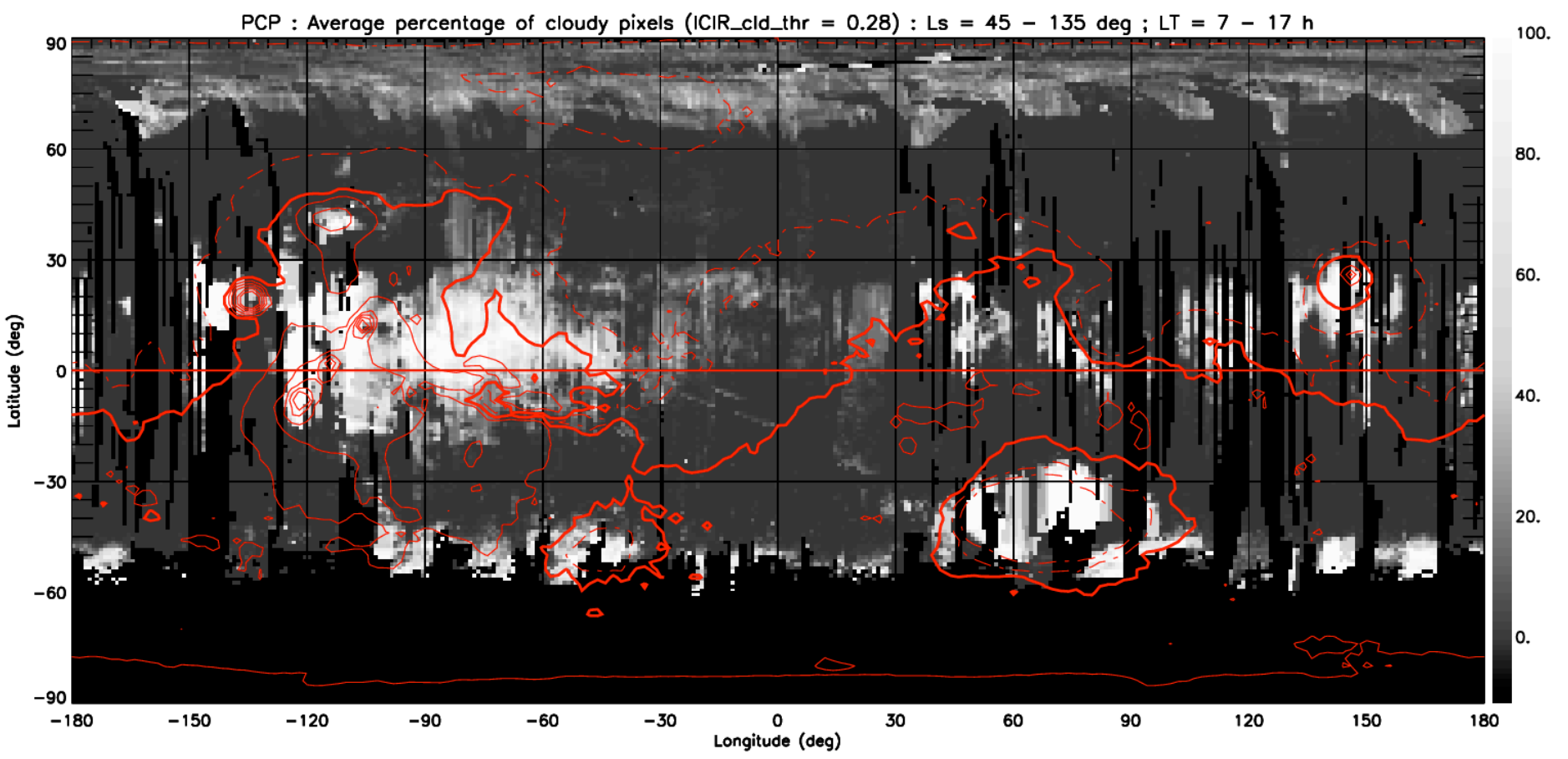

a)

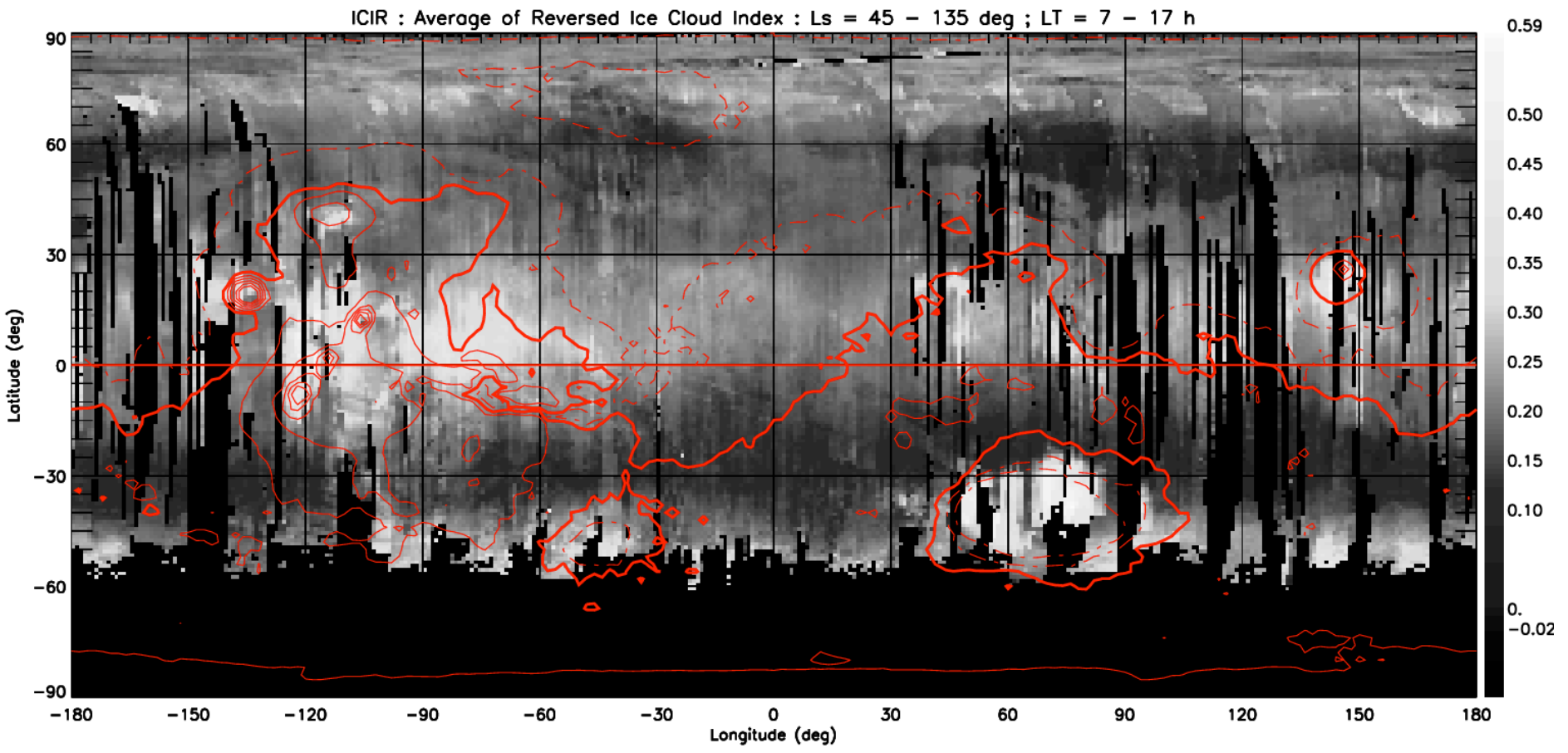

b)

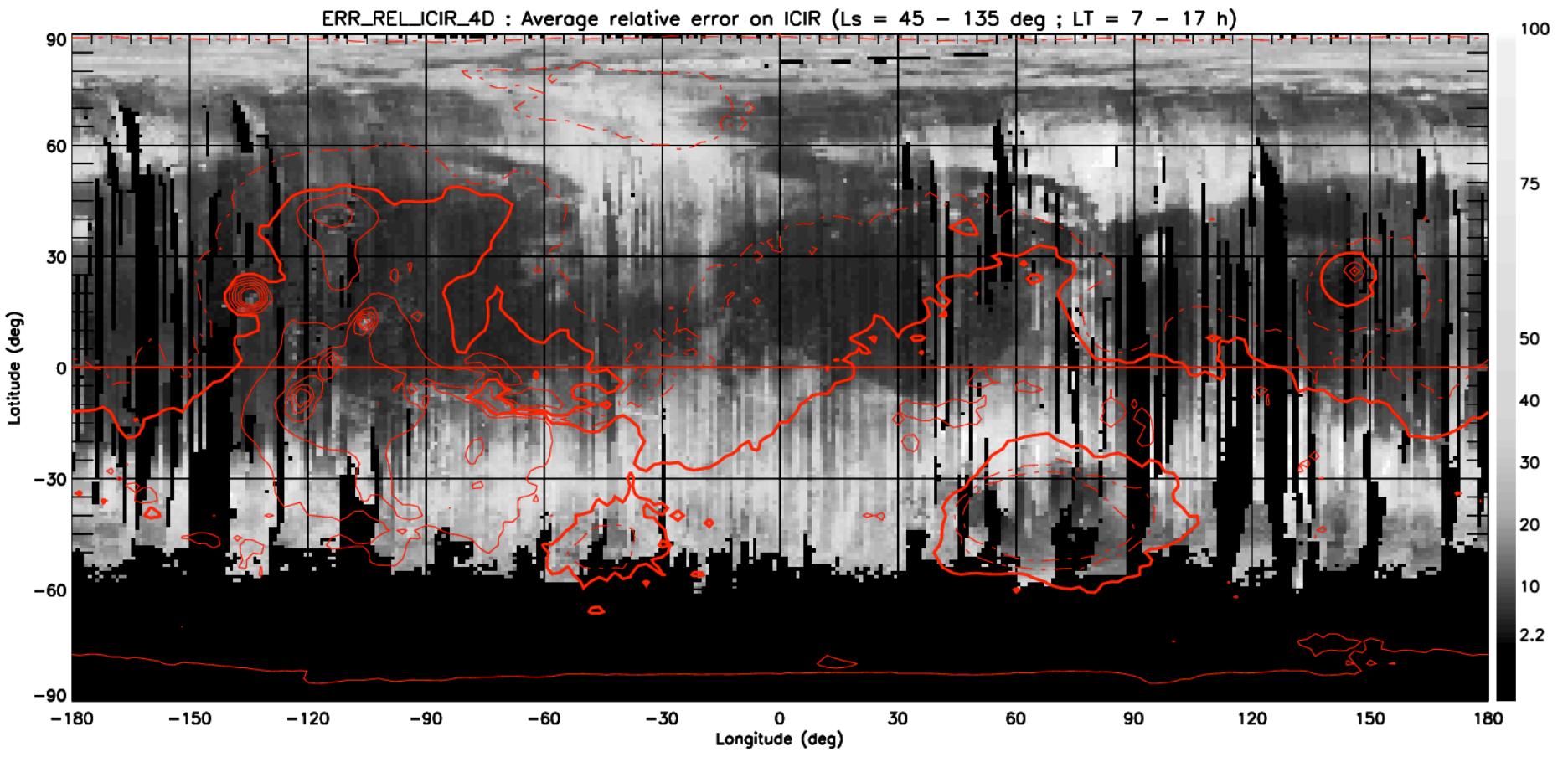

c)

*(continued on next page)*

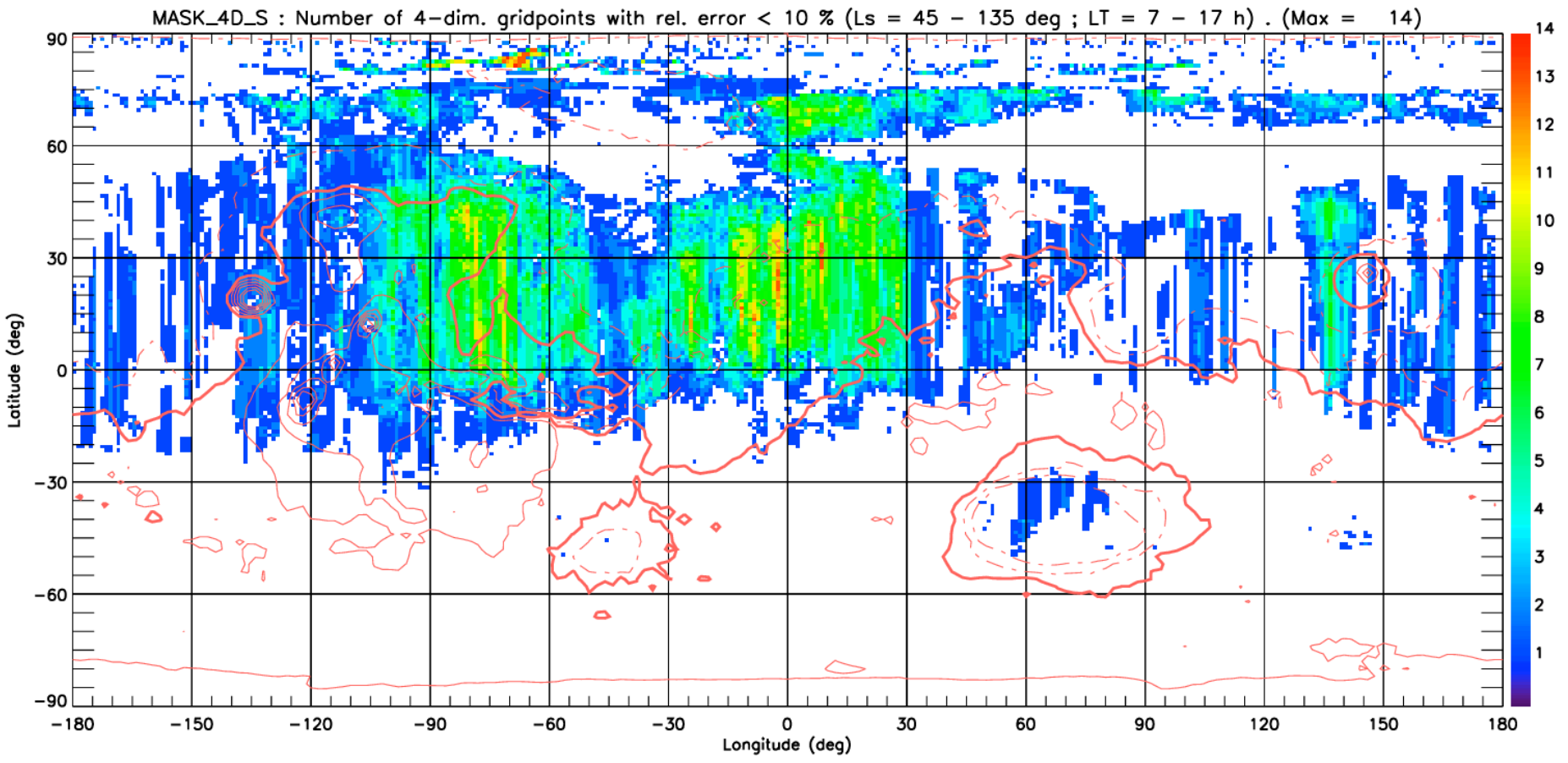

d)

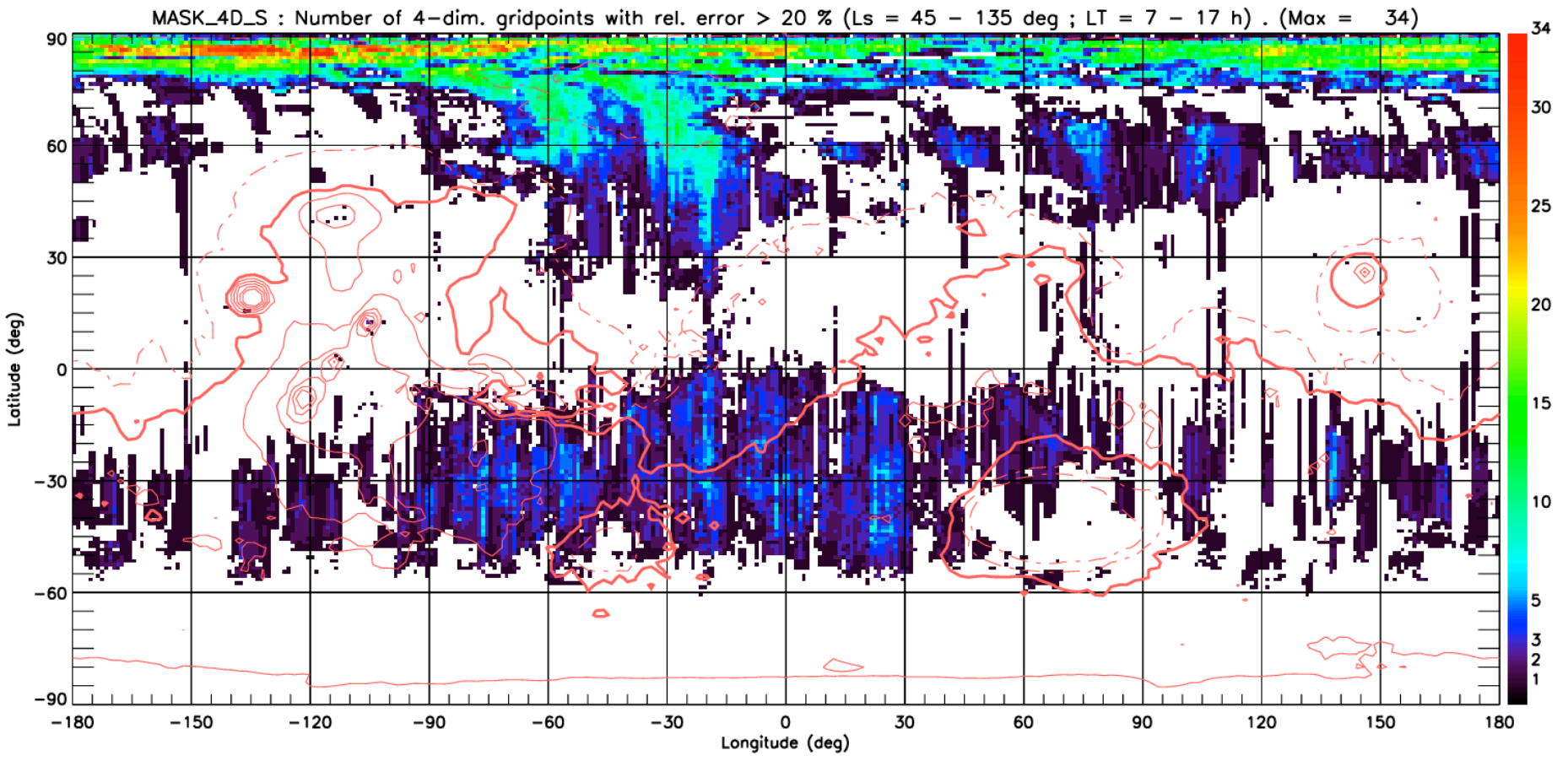

e)

**Figure 4:** average PCP (in %) (a: top), ICIR (b), relative ICIR error (in %) (c), location and number of valid gridpoints with ICIR with low relative error (< 10 %) (d), and with high relative error (> 20 %) (e), around Ls = 90°. The first and the last value of each scale correspond to the minimum and maximum value of each variable. The equator, and contour lines (in red) corresponding to surface elevation are superimposed on this figure and on following maps: thick line: datum (0 m) ; thin continuous line: elevation > 0 ; thin dashed line: elevation < 0. Consecutive contour lines are spaced every 2500 m.

The aphelion belt, including large volcanoes (Olympus and Elysium Mons, the Tharsis volcanoes and Alba Patera) and Arabia Terra, Hellas Planitia, the north polar hood edge and the southern polar cloud belt stand out as major cloudy areas around the northern summer solstice (Fig. 4b). No other outstanding cloud feature is present in between, i.e. in the tropics and mid-latitudes (below 35° N) during the period around the northern winter solstice (Fig. 5b). In comparison to the ICIR figures, the PCP (Fig. 4a and 5a) filters out smaller and less prominent cloud structures while retaining only the main ones. This filter-effect can be explained by the averaging over the two temporal dimensions (Ls and LT). For the ICIR, the average concerns 4-dimensional gridpoints with any degree of cloudiness (or absence of clouds), whereas for the PCP, the average concerns only gridpoints with dense clouds (i.e. calculated only with original OMEGA pixels above the ICIR threshold), or without dense clouds.

Note that the same major cloud features are present on the overall map of the ICIR covering the whole day and the whole year (Fig. 2b), but appear with less contrast. The contours of cloudy areas on such maps are impacted by the limited number, or absence, of OMEGA orbits covering the Martian surface. The main effect is vertical striping in the tropical and mid-latitude areas. In some cases, diagonal stripes due to unusual (non-nadir) pointing modes or peculiar orbits may also be present.

**3.2 Uncertainty estimation of the Ice Cloud Index**

The error bar on the ICIR is also calculated for each 4-dimensional gridpoint, or bin. It is the quadratic sum of the uncertainty related to the measure by the OMEGA instrument ($\Delta ICIR_i$) for N observations (pixels) used for the gridpoint and the standard deviation of the ICIR values of the N observations ($\sigma_{ICIR_bin}$). It is given by the following formula (derived from Audouard et al., 2014a):

$$\delta_{ICIR_bin} = \left[ \frac{1}{N} \left( \sum_{i=1}^{N} \Delta ICIR_i^2 \right) + \sigma_{ICIR_bin}^2 \right]^{1/2} \tag{4}$$

The first standard deviation term reflects the impact of the uncertainties on the ICIR related to the original measurements, i.e. the reflectance values at 3.38 µm and 3.52 µm used to calculate the ICIR. They include in particular the instrumental noise on OMEGA spectels. The second term reflects the variability of the ICIR pixel values used to calculate the average (reversed) ice cloud index value on the 4-dimensional grid. The maximal value for $\delta_{ICIR_bin}$ is 0.36 .

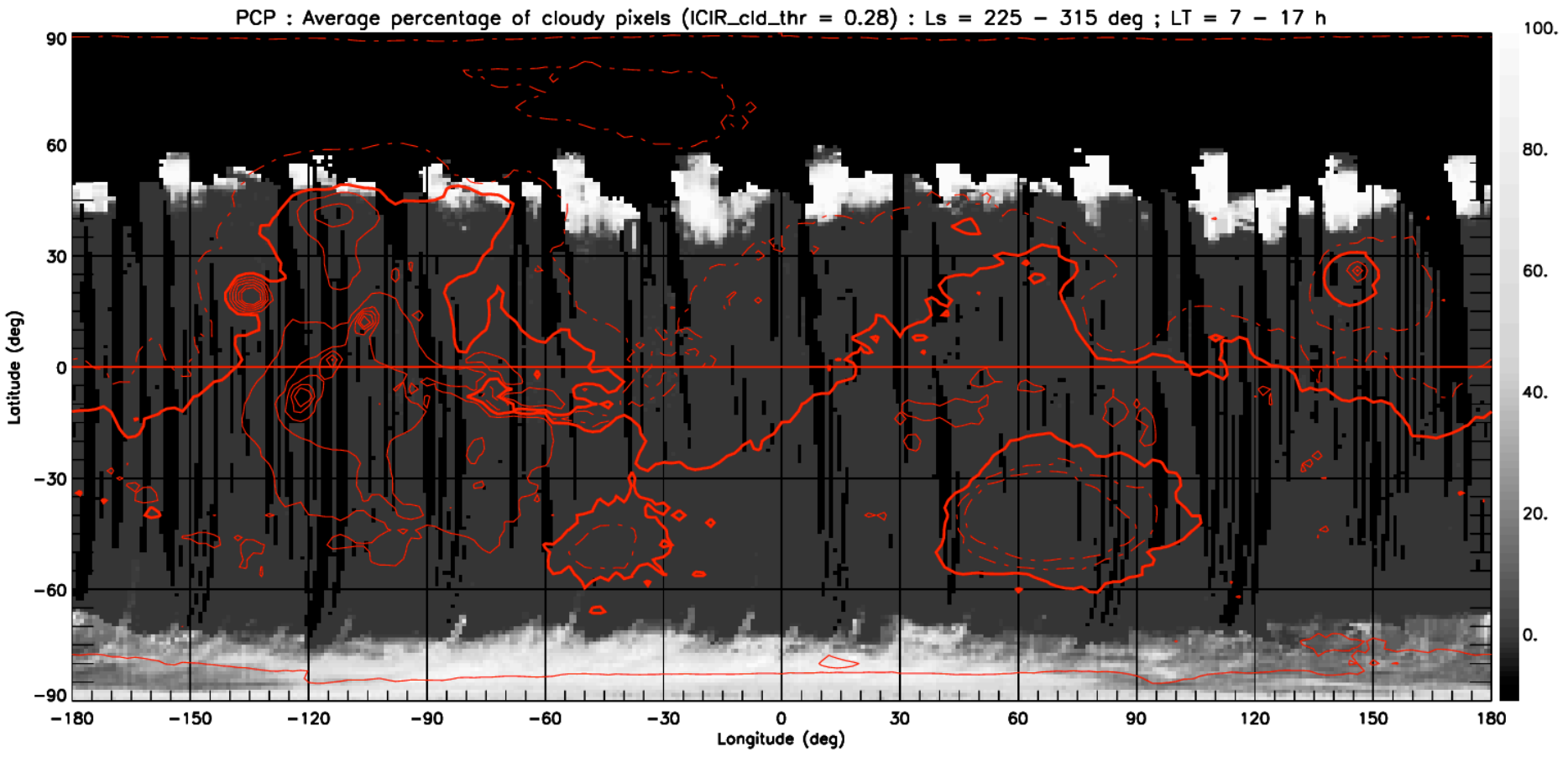

a)

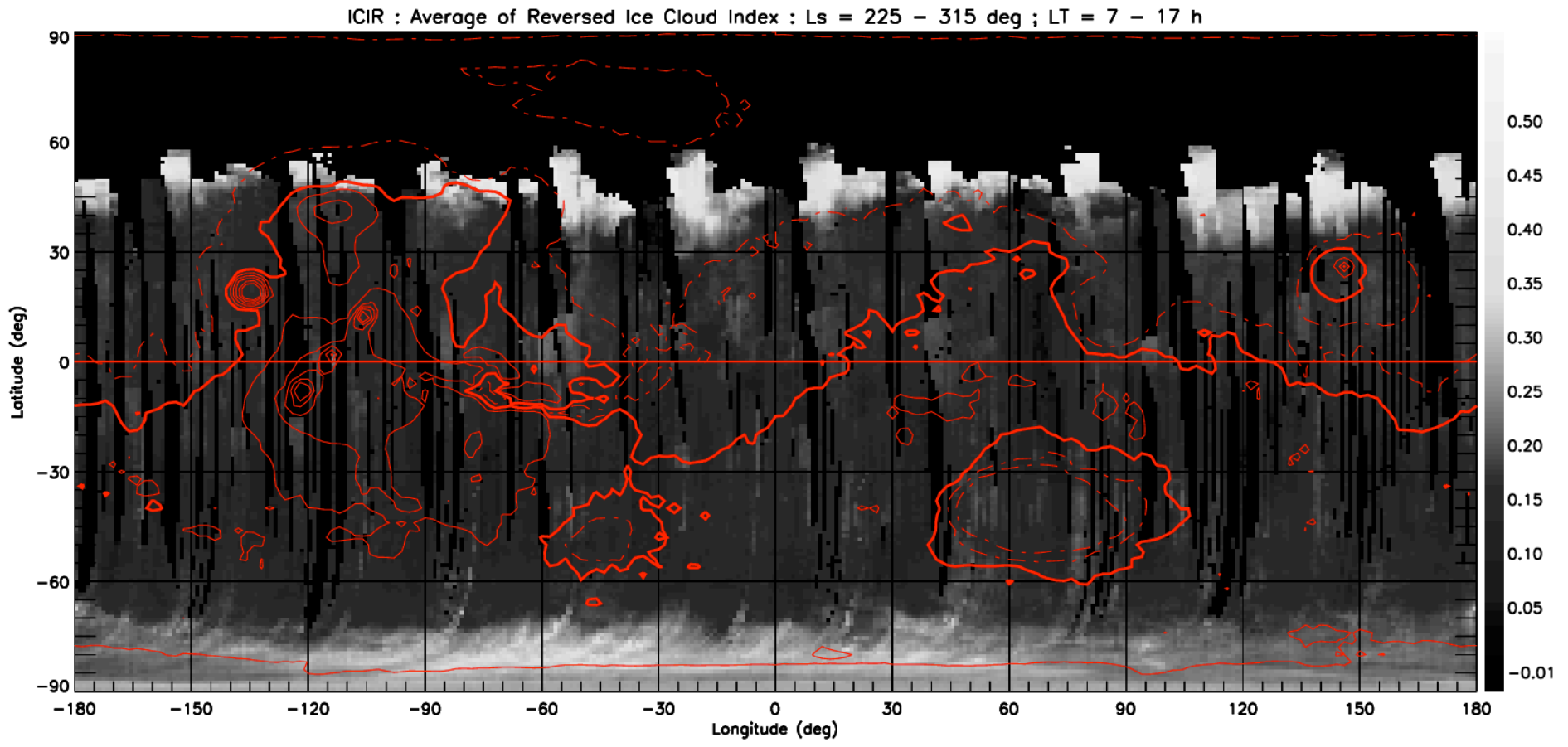

b)

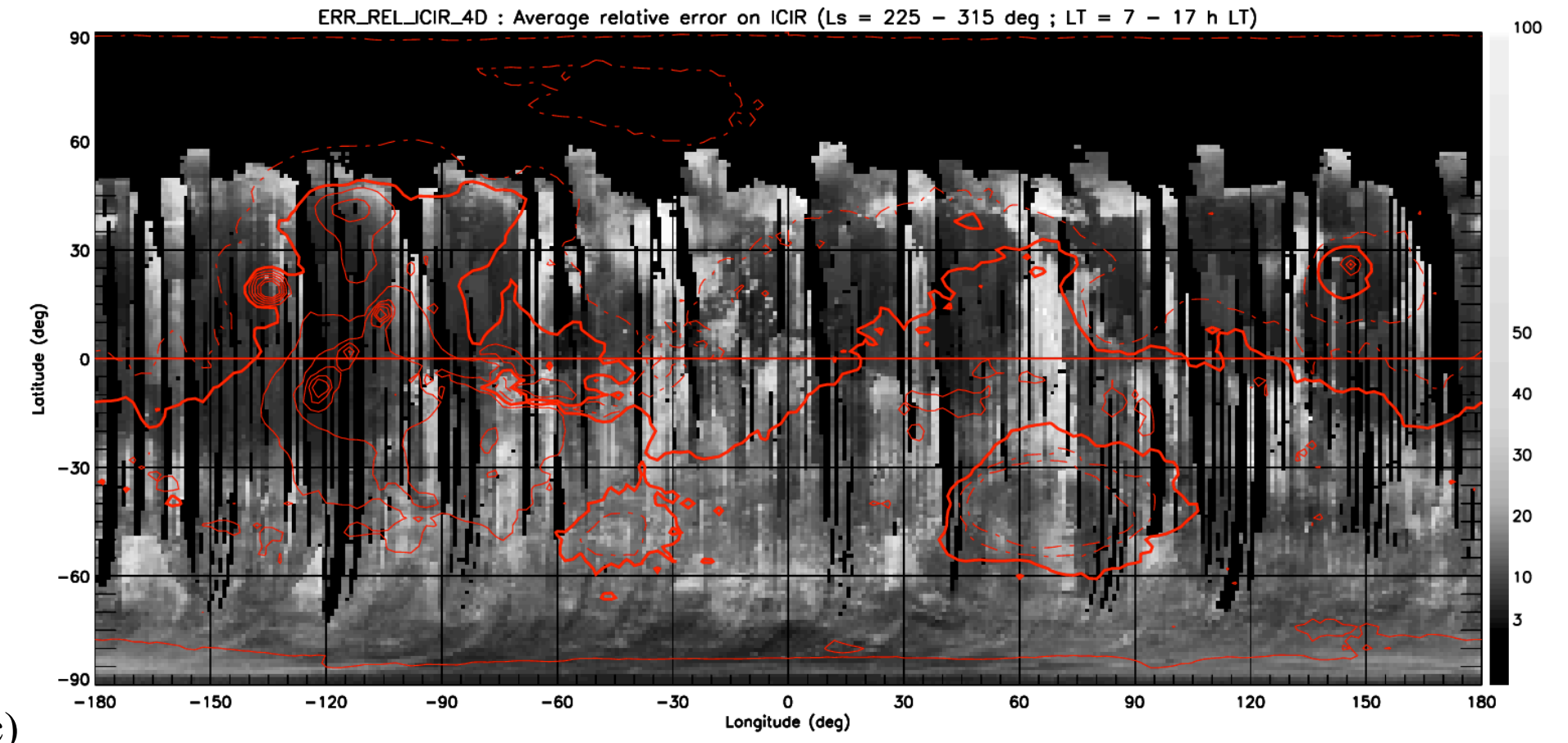

c)

**Figure 5:** average PCP (in %) (a: top), ICIR (b: middle) and relative ICIR error (in %) (c: bottom) around Ls = 270°. (Note that 4-dimensional relative ICIR error values above 100 % have been set to 100 % prior to the averaging to 2 dimensions.) The equator, and contour lines (in red) corresponding to surface elevation are superimposed on this figure and on following maps: thick line: datum (0 m) ; thin continuous line: elevation > 0 ; thin dashed line: elevation < 0. Consecutive contour lines are spaced every 2500 m. The first and the last value of each scale correspond to the minimum and maximum value of each variable.

The dominant element in the uncertainty is the instrumental error of the OMEGA instrument at the wavelengths around 3.4 μm used to calculate the ICIR. This is observed for ~84 % of valid gridpoints (Table 2). It indicates that the standard deviation of the original OMEGA pixels, used to calculate the gridded (and averaged) ICIR, is in general a less important component of the error bar.

Figure 2c represents the relative uncertainty of the ICIR variable, ($\delta_{ICIR}$ / ICIR , in %), averaged over all available solar longitudes and local times (fig. 2a). Figure 4c and 5c represent seasonal relative error values averaged around the solstice periods (northern summer: Ls = 45 – 135°, labeled NSS ; southern summer: Ls = 225 – 315°, labeled SSS) and during daytime (LT = 7 – 17 h). These two figures can be compared visually to the corresponding ICIR fields (Fig. 4b and 5b).

From the visual comparison with these figure pairs, we can note the following relations:

- The higher the ICIR values, the lower their relative error. This is clearly visible specifically for the aphelion belt around the NSS.
- Conversely, the lower the ICIR values, the higher their relative error. This can be observed in particular at southern latitudes (around 30° S), and in the northern plains where areas around 60° N have low reflectances (around the NSS).
- Exceptions to these rules are the cloudy edges of the polar hood around the corresponding winter: high ICIRs can be associated to relatively high error values. This can be related to the original low reflectance values used to calculate the ICIR. During the winter season, these low values result from high incidence angles. Thus the resulting reflectances have a higher instrumental error, which is forwarded into the ICIR error.

We also used the 4-dimensional relative error on the ICIR in order to :

- determine the location and impact of high/low ICIR values,
- detect extreme cases,
- determine the impact of instrumental noise on data.

Table 2 shows the number and percentage of valid 4-dimensional gridpoints with specific values of the ICIR, the relative error (ERR_REL_ICIR), the instrumental error of the OMEGA data used to calculate the gridded ICIR, and the standard deviation resulting from averaging the ICI values from original OMEGA pixels. Values and percentages are given for the whole dataset (all solar longitudes and all local times), and for the northern summer solstice period (Ls = 45 – 135° and LT = 7 – 17 h).

From the 4-dimensional ICIR data and its relative error we also made the following observations, summarized in Table 2. Remember that the data coverage by OMEGA is not homogeneous, and increased between 100°W and 15°E, and close to the poles, as shown also on Figure 2a :

- Gridpoints with a negative ICIR are exceptional. They represent 0.018 % of all valid gridpoints. They occur mainly in the afternoon and evening (14 – 22 h LT), at midlatitudes, and mainly in areas of low albedo.

- ICIR gridpoints with small relative error (< 10 %) :

They represent about 30 % of valid gridpoints. On Figure 4d, they are mainly located in the tropics and midlatitudes. Around the southern summer solstice, some gridpoints are also present close to the South Pole (not shown). Corresponding regions can easily be matched with regions of high albedo on maps, such as e.g. the albedo map of the planet derived from OMEGA data in the Visible and near-IR (C-channel) spectral domains (Vincendon et al., 2015).

- ICIR gridpoints with large relative error (> 20 %) :

They represent about a third of valid gridpoints. Concerning the location of gridpoints with large relative error, Figure 4e is nearly the complementary image of Figure 4d (over regions where valid ICIR data is present). Regions with low albedo (Chryse and Acidalia Planitia, Syrtis Major, Utopia Planitia and the south hemisphere mainly midlatitude region) have the highest relative error.

- Gridpoints with an intermediate error (between 10 and 20 %) have a spatial coverage close to the coverage of the gridpoints with a large error (> 20 %), but also have a small area in common with the gridpoints with a small error (< 10 %) (not shown). Note that the number of counted gridpoints for a specific range of values (e.g. Figure 4d and 4e) also depends on the coverage by the OMEGA instrument, as shown on Figure 2a.

- Gridpoints with an extremely high relative error, above 100 %, form a small fraction (0.5 %) of the valid gridpoints and belong to the same low albedo regions. The highest relative error is found when the absolute error $\delta_{ICIR}$ exceeds the ICIR (i.e. the relative error is above 100 %). This is the case when the ICIR is very small, close to zero ; this situation is more likely to occur over dark (low albedo) and cloudless regions.

- Table 2 also shows that the statistics of the relative error and the components of the error bars of the complete dataset (all Lss and LTs) and those of the smaller NSS dataset are in the same order of magnitude.

| **Data description** | **Number of valid gridpoints (all Ls and LTs)** | **Percentage of valid gridpoints (all Ls and LTs)** | **Number of valid gridpoints (Ls = 45 - 135° ; LT = 7–17h)** | **Percentage of valid gridpoints (Ls = 45 - 135° ; LT = 7–17h)** |
|---|---|---|---|---|
| **All valid ICIR** | 1113807 | 100 % | 229339 | 100 % |
| **ICIR < 0** | 204 | 0.018 % | 36 | 0.016 % |
| **ERR_REL_ICIR<10%** | 338231 | 30.4 % | 61009 | 26.6 % |
| **10%<ERR_REL_ICIR<20%** | 402663 | 36.1 % | 75668 | 33 % |
| **ERR_REL_ICIR > 20%** | 37293 | 33.5 % | 92662 | 40.4 % |
| **ERR_REL_ICIR > 100 %** | 5191 | 0.5 % | 1173 | 0.51 % |
| **ΔICI > STDDEV(ICIR)** | 933302 | 83.8 % | 182118 | 79.4 % |
| **ΔICI < STDDEV(ICIR)** | 180505 | 16.2 % | 47221 | 20.6 % |

**Table 2:** number of 4-dimensional gridpoints with specific ICIR, relative error (ERR_REL_ICIR), instrumental error (ΔICI), and standard deviation of the average ICIR calculation (STDDEV(ICIR)) values. Values are counted and calculated for the complete dataset (all Lss and LTs) and the Northern Summer Solstice period (NSS : Ls = 45 -135° ; LT = 7 –17 h).

### 3.3 Location of partially cloud-covered areas

Are there some specific regions partially covered by clouds and where are they located ? In order to answer these two questions we have constructed 2D maps with the frequency of occurrence of various degrees of cloud cover, characterized by a PCP interval defined by an upper and a lower threshold (PCPmin and PCPmax).

We have applied the following 3-step procedure:

1) For each 5° Ls bin, we define:

* 1< PCP =< 5%: very small cloud cover

* 5 < PCP =< 50%: (not too small) minority of clouds

* 50 < PCP =< 95%: (not too large) majority of clouds

* 95 < PCP =< 100%: (almost) total cloud cover.

* 1 < PCP =< 100%: presence of clouds in any amount (extremely low values of PCP, below 1%, are not taken into account).

2) For each 2D (spatial) location, we count the number of (temporal) occurrence of the corresponding PCP values between the upper and lower bounds, i.e. the number of gridpoints: NGP($PCP_{min}$, $PCP_{max}$).

3) The frequency of occurrence is the ratio between the number of occurrences observed over a specific PCP interval and the total number of cloudy occurrences. It can be expressed as a percentage by the following formula:

$$f_{occurrence}(PCP_{min}, PCP_{max}) = 100 \,.\, NGP(PCP_{min}, PCP_{max}) \;/\, NGP(1\%, 100\%) \qquad (5)$$

The maps of Fig. 6a to 6d represent 4 levels of cloud coverage expressed by the frequency of occurrence, observed around the NSS and during daytime (LT = 7 – 17 h).

The highest cloudiness, with a PCP close to 100%, can be observed over a limited number of regions (Fig. 6d), namely:

- the aphelion belt over the Tharsis rise (150°W – 40°W), including the Tharsis volcanoes, Olympus Mons and Alba Patera,
- Elysium Mons,
- Arabia Terra and Syrtis Major,
- the Hellas basin.

Along two latitudinal bands, around 55°N and 30°S (with Hellas excluded), only reduced partial cloud coverage (PCP < 50%) can be observed, or no cloud at all.

In the regions covered by the North polar cloud cover, by the edge of the South polar hood, and by a large part of the aphelion belt, all densities of cloud coverage can be observed (from 1% to 100%). Only two regions have a peculiar configuration:

- the Hellas basin, completely covered by clouds (Fig. 6d), is surrounded by a thin ring of partial cloud cover (Fig. 6a-c),
- an area on the eastern part of the Tharsis rise and the western part of Lunae Planum (around 10°N, 75°W) where partial cloud cover is absent and which is visible as a dark hole on figures 6a-c. In this area, we found only a complete cloud cover (PCP > 95%).

A similar comparison (not shown) has been conducted around the southern summer solstice period (Ls = 225 – 315°). Only two large, cloudy areas can be observed, the reduced southern polar hood (70°S – 90°S) and the southern edge of the northern polar hood (35°N – 60°N). Cloud coverage frequencies of all types, with a PCP between 5% and 100% are located over the same parts of these cloudy areas. Cases of PCP between 1% and 5% are also present over the same areas but are less frequent.

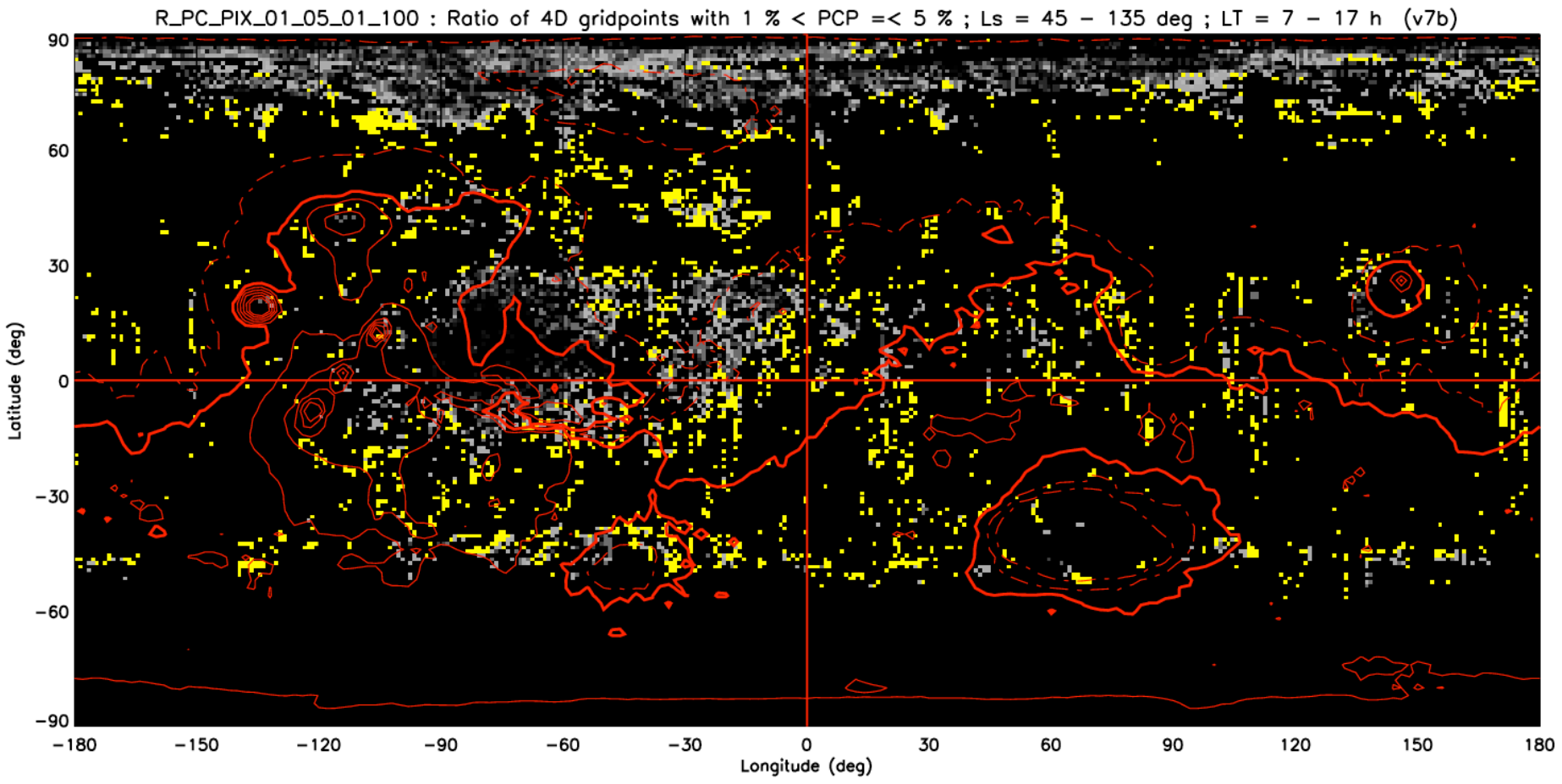

a)

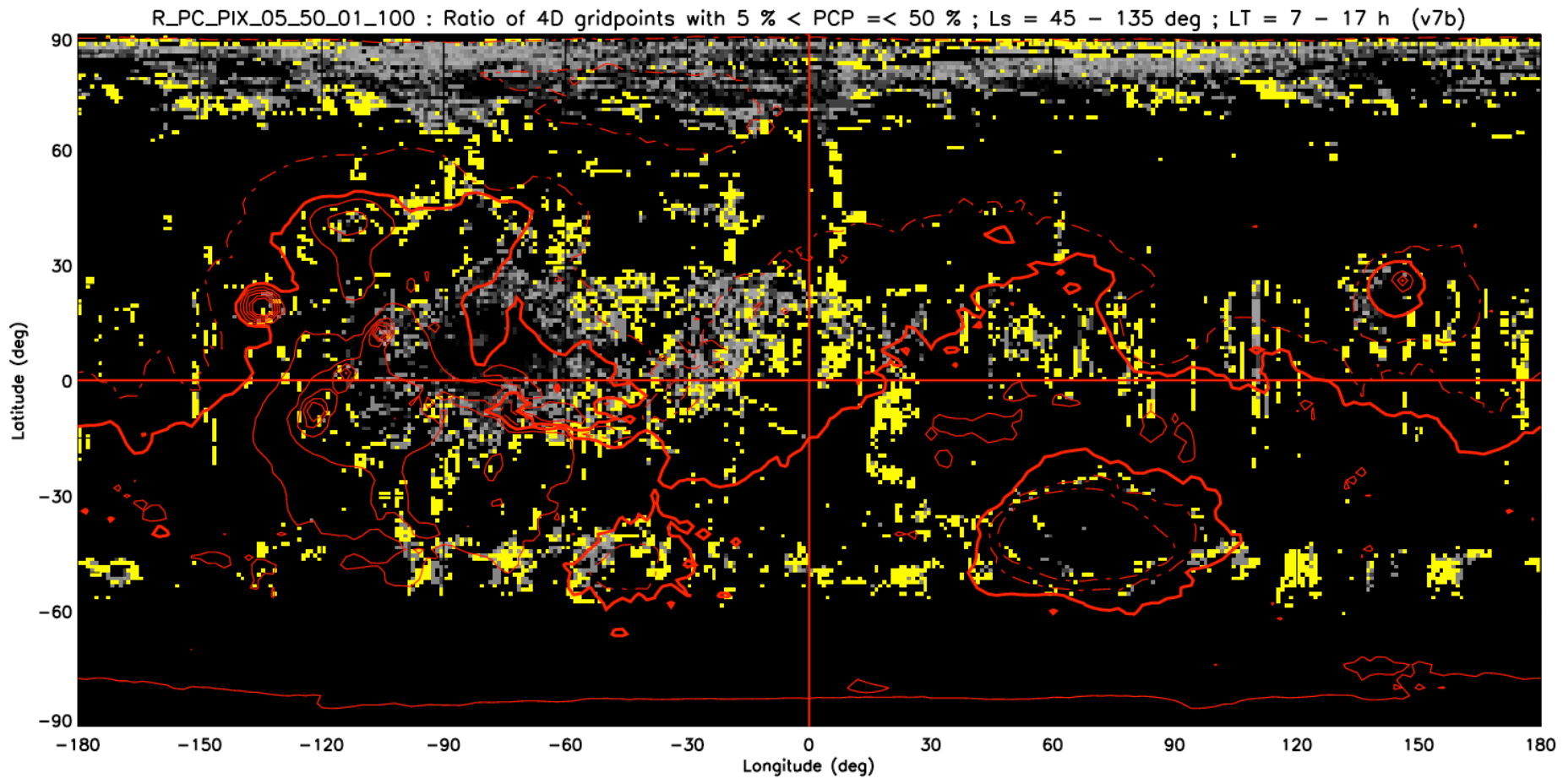

b)

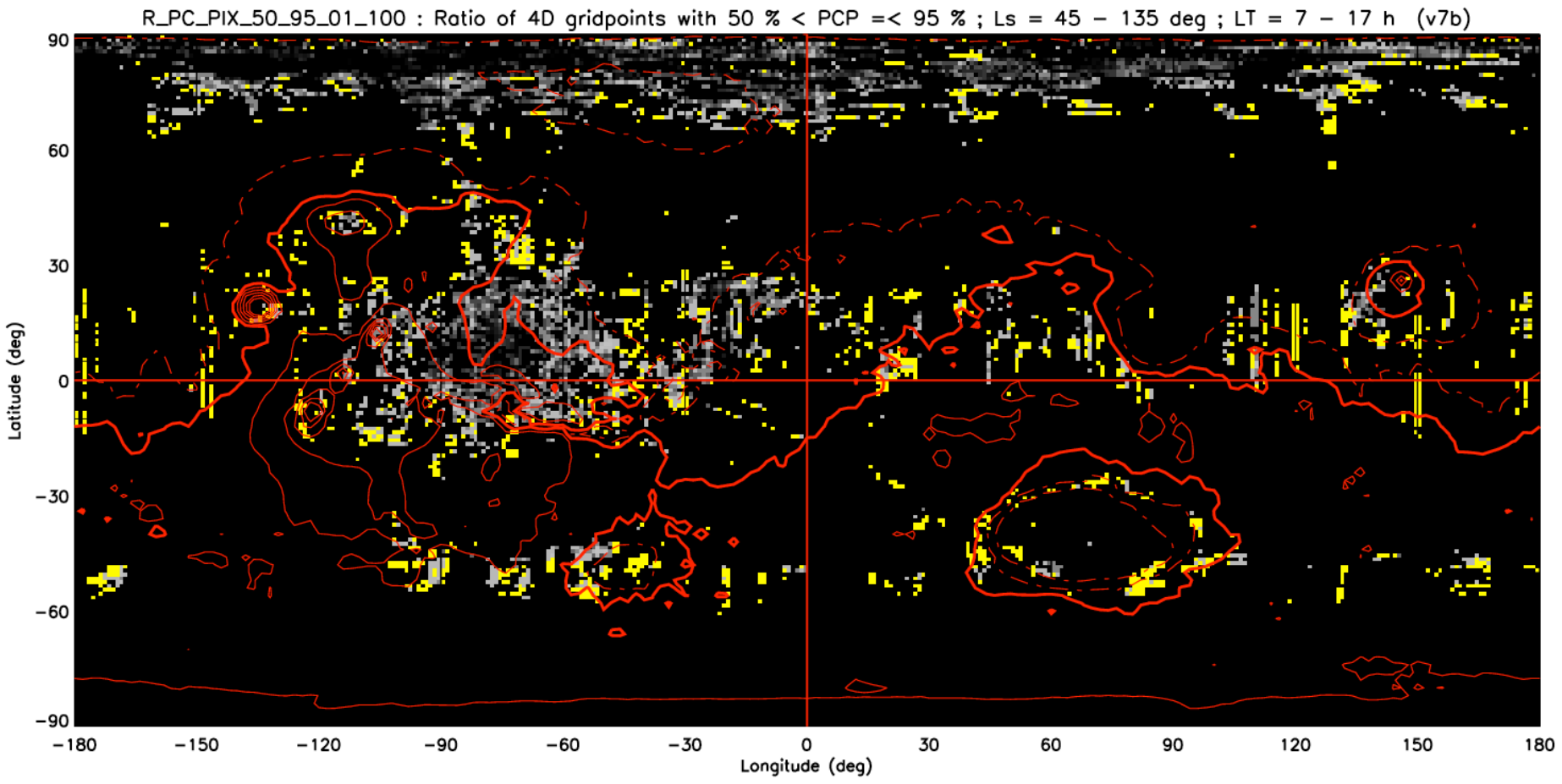

c)

*(continued on next page)*

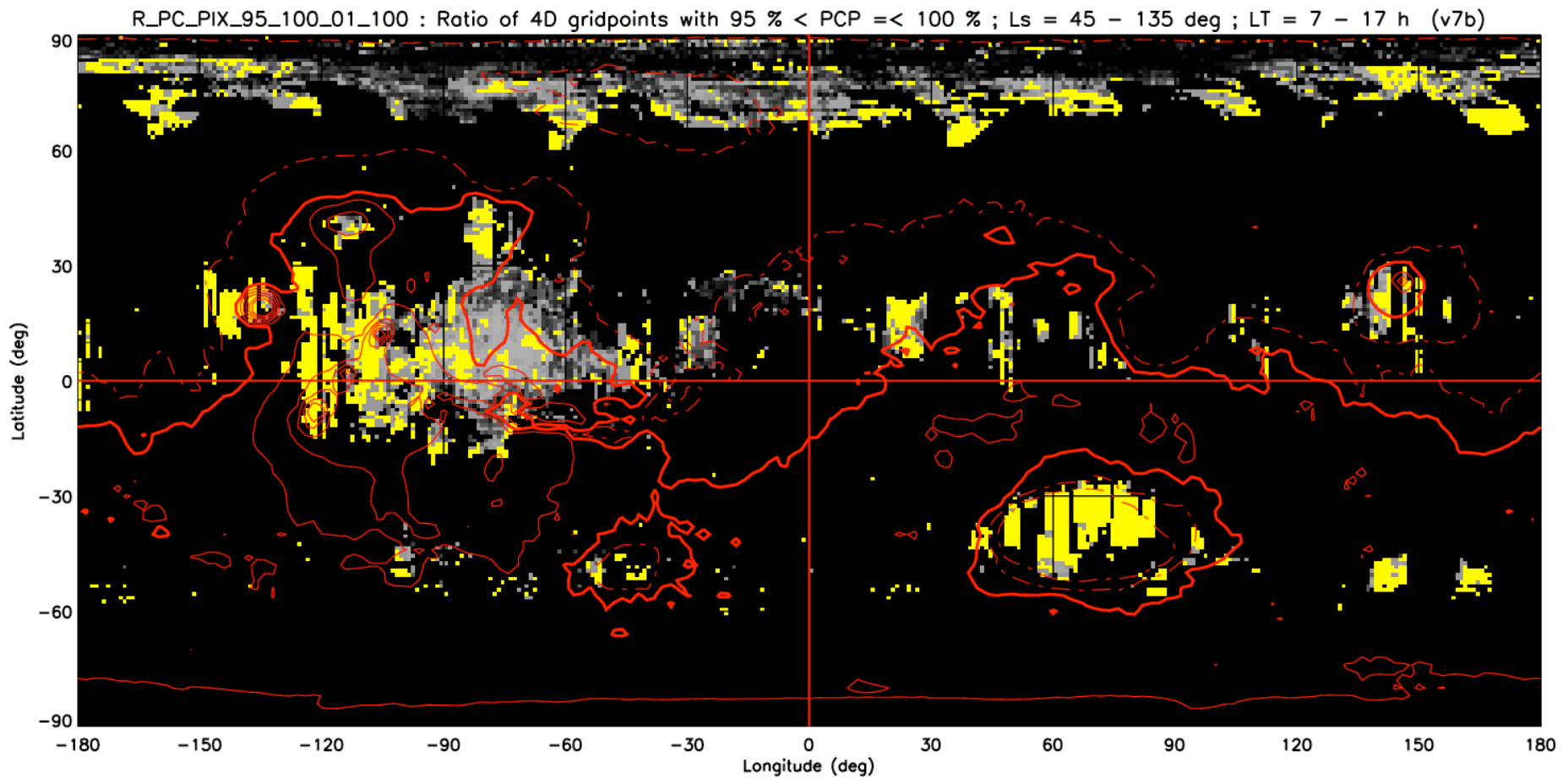

d)

**Figure 6 :** location of 4 cloud cover types during Ls = 45° - 135° and daytime (LT = 7 – 17h). (a) Very small (1% < PCP =< 5%) ; (b) minority of clouds (5% < PCP =< 50%) ; (c) majority of clouds (50% < PCP =< 95%) ; (d) total cloud cover (PCP > 95%). The color range corresponds to the frequency of occurrence (count) of the cloud cover type, ranging from black (no occurrence or absence of data) up to light gray, and yellow for the maximal number of occurrences on the corresponding figures. The equator, and contour lines (in red) corresponding to surface elevation are superimposed on this figure and on following maps: thick line: datum (0 m) ; thin continuous line: elevation > 0 ; thin dashed line: elevation < 0. Consecutive contour lines are spaced every 2500 m.

## 4. Comparison of ICIR data with TES observations and MCD model predictions

### 4.1 Comparison of ICIR data with TES optical thickness mapped climatology

In order to validate the ICIR as a climatological product, we compared it to the water ice optical depth derived from TES data.

The water ice optical depth ( $\tau_{TES_wice}$) is one of the products derived from thermal infrared spectra (6 – 50 µm) provided by the TES instrument. It has been calculated for a wavenumber of 825 $cm^{-1}$ (wavelength : 12.1 µm), corresponding to a peak absorption of water ice aerosols (Smith, 2004). In this study we used the water ice optical depth from the TES mapped climatology derived from nadir measurements. TES-derived data have been remapped for each Martian year (from MY 24 to 27) onto a regular grid: longitude ΔLON = 7.5° ; latitude ΔLAT = 3° and solar longitude ΔLs = 5°. The main source of uncertainty in the retrieval of optical depth comes from errors in the retrieval algorithm, whereas random instrumental noise and calibration errors have a minor impact. Scattering by water ice (and dust) particles is not taken into account, thus is is the absorption optical depth that is retrieved (Smith, 2004). Resulting from the gridding and averaging process, the uncertainty of the water ice optical depth has been reduced from the original values down to +/-0.03 or 10 %, whichever is greater. Thus the resulting relative error becomes large (above 20 %) for small values of optical depth (below 0.15). For the comparison with the averaged ICIR data, we selected data from one Martian year, the hybrid year MY 24, around 14h LT, and at latitudes below 80°. The beginning of MY 24 was missing and is replaced by the beginning of MY 25, also because a major dust storm - not taken into account - occurred during the last third of MY 25.

For this comparison, the original ICIR data was reprocessed and adapted to the TES mapped climatology grid, with a specific selection of 4-dimensional gridpoints at 14 h LT corresponding approximately to the observation time of TES, and a limitation below 75° latitude. Fig. 7 shows that the high, averaged ICIR values are collocated with high $\tau_{TES_wice}$ around the northern summer solstice period. Some differences can be observed at high northern latitudes and on the northern edge of the Hellas Basin, where ICIR values are high, whereas $\tau_{TES_wice}$ values are relatively low. The ICIR 2D map has a noisier aspect than the $\tau_{TES_wice}$ map. A possible explanation of the reduced cloudiness at higher latitudes could be the biased estimation of the optical depth obtained from very low TES-derived temperature values. A mapped image of the relative error on $\tau_{TES_wice}$ is approximately the negative of the original $\tau_{TES_wice}$ image : a small relative error (~10 %) can be observed in the aphelion belt, and large relative errors (up to 150 % over surfaces of low albedo) over the region derived from low TES-derived temperature north of 50°N. The relation between ICIR and its relative error (low, resp. high ICIR corresponds to high, resp. low relative ICIR error, with exceptions at high latitudes and a link to surface albedo in cloudless regions) has already been observed at normal resolution during the same season (summer solstice) and described in subsection 3.1 and shown on Fig 4. Note that, although the spatial resolution and the daily period covered are not the same, between 7 and 17 h LT instead of the shorter 13 – 15 h LT period compared here, the location of relative error on the ICIR is qualitatively similar.

2D histograms of ICIR and $\tau_{TES_wice}$ (Fig.8) show a linear trend and possibly indicate a physical relation between both variables, with an intermediate correlation value (0.53) for the complete TES dataset (limited to 78° in latitude), but much higher values for known cloudy areas and periods (0.83 for the aphelion belt) (Fig 8b). The high density of counts for $\tau_{TES_wice}$ below 0.04 can be related to the absence of clouds over large non-polar regions (latitude < 78°) during northern autumn and winter on the full-year histogram (Fig 8a). In comparison, histogram counts are less scattered, probably due to the more important presence of clouds (in particular in the aphelion belt) around the northern summer solstice (Fig. 8b).

Around the summer solstice, water ice is detected in both datasets and should mainly correspond to clouds located in the aphelion belt. The full-year histogram (Fig. 8a), which includes the summer solstice histogram, has also a non-negligible supplementary population of extratropical counts with significant ICIR (above 0.10) and relatively low TES optical depth values (below 0.03). The weak thermal signature of water ice on $\tau_{TES_wice}$, which can be observed when clouds and the surface below have a close temperature (fogs), or due to frost contamination of the surface, could explain the characteristics of this population.

A figure of $\tau_{TES_wice}$ (not shown) with a similar aspect to Fig. 7b, and close values of correlation with ICIR data have been obtained with averaged TES data from MY 26.

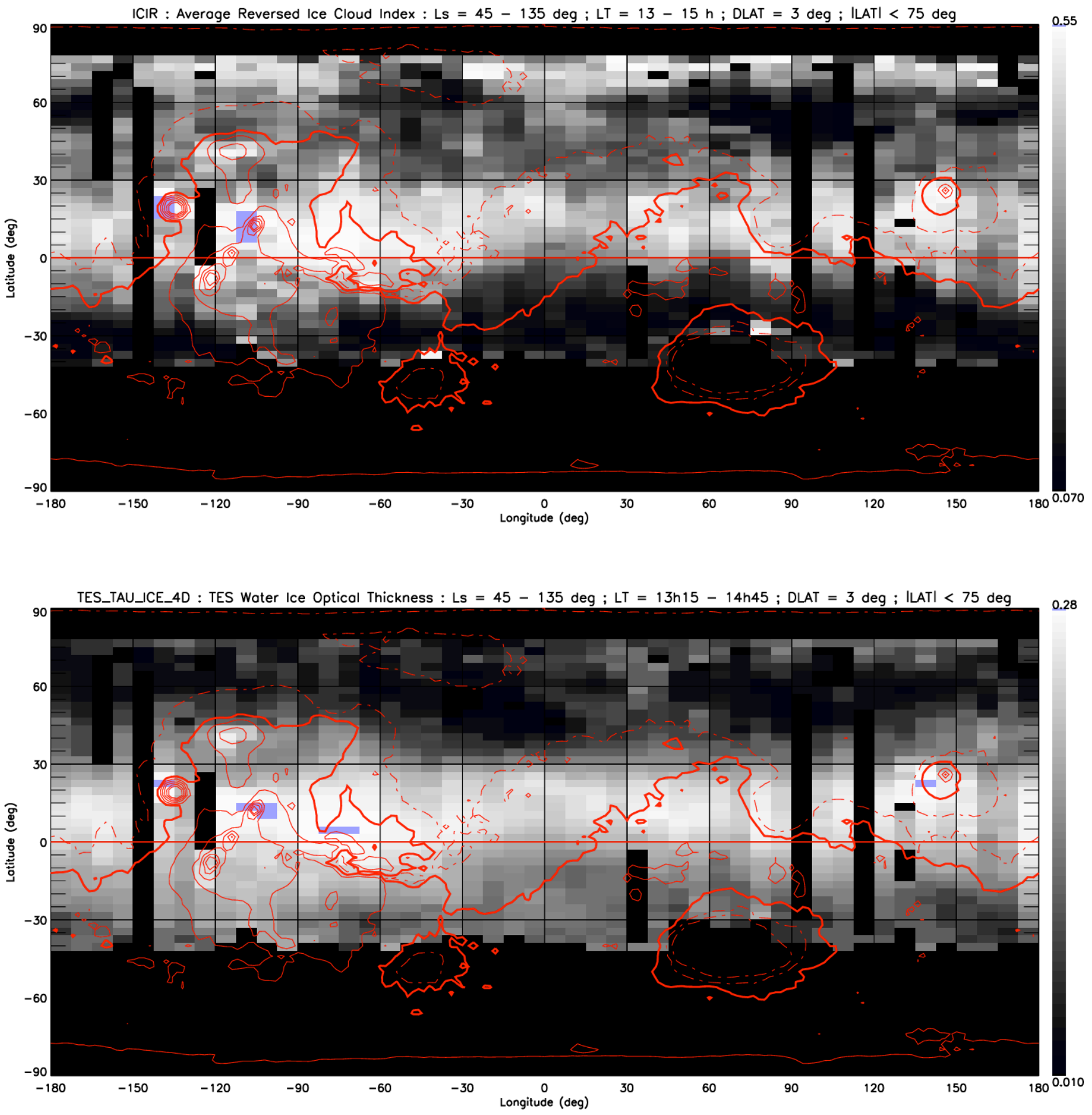

a)

b)

**Figure 7:** maps of averaged ICIR at LT=14 h (a) and optical thickness $\tau_{TES_wice}$ (b) from MY 24 hybrid during daytime (~14 h LT), Ls = 45 – 135°, limited to latitudes < 75°. The color range is from dark grey up to white, with maximal values in light blue. Minimum and maximum values are indicated at each end of the scale. The equator, and contour lines (in red) corresponding to surface elevation are superimposed on this figure and on following maps: thick line: datum (0 m) ; thin continuous line: elevation > 0 ; thin dashed line: elevation < 0. Consecutive contour lines are spaced every 2500 m.

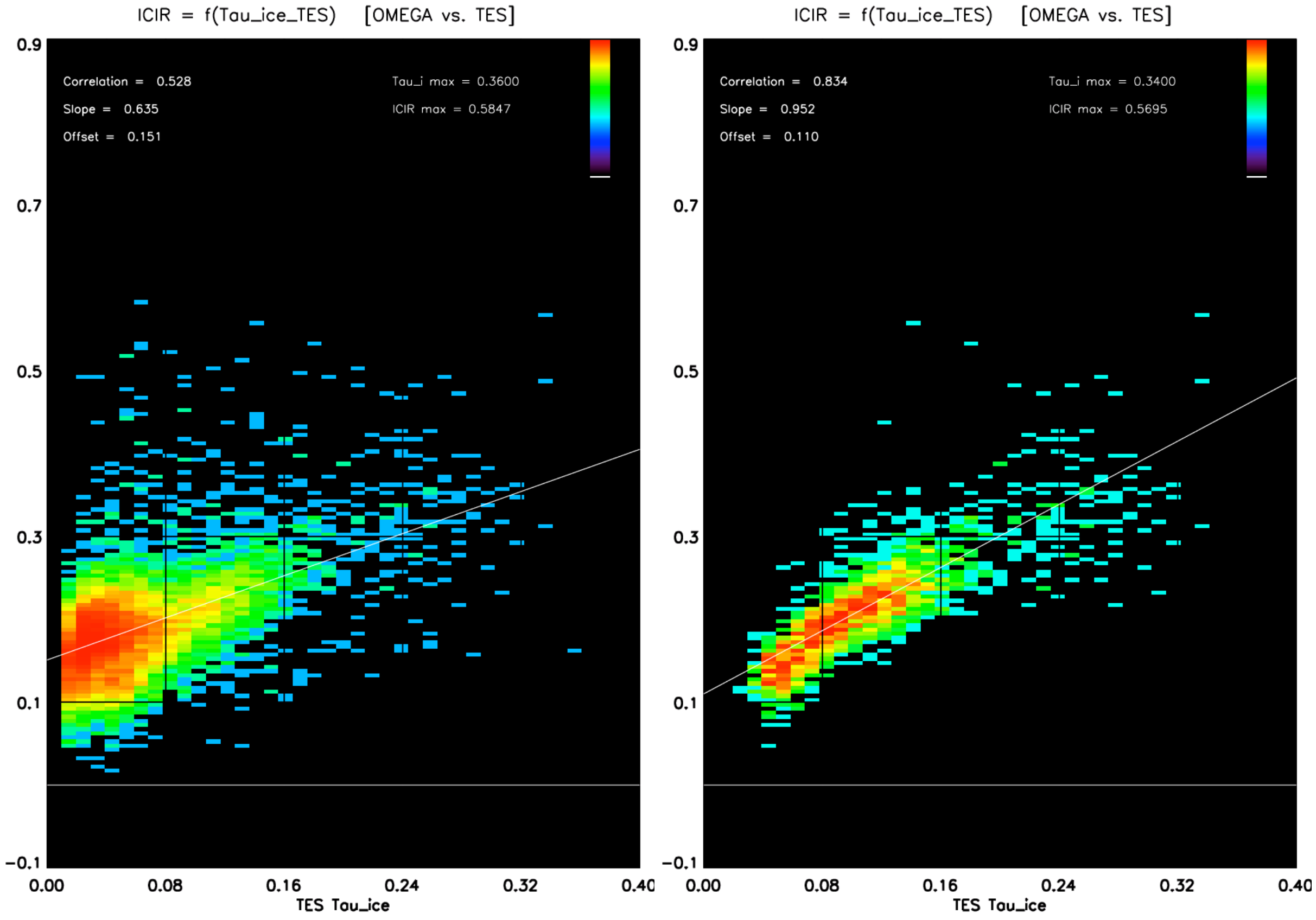

**Figure 8:** 2D histograms of ICIR vs. $\tau_{TES_wice}$ for all latitudes below 78° and all seasons (left), and in the tropics (-27° < lat < 27°) during Northern summer (Ls = 45 – 135°) (right) at the TES climatological resolution (ΔLON = 7.5° ; ΔLAT = 3°)

### 4.2 Comparison with water-ice column and integrated optical thickness from the Mars Climate Database

The Mars Climate Database version 5.3 (Millour et al., 2018) is a database of meteorological fields, computed using runs of the LMD Global Climate Model (GCM) of the Martian atmosphere (see Forget et al. (1999), for general settings, Navarro et al. (2014) for details on modeling the water cycle, and Pottier et al. (2017) and references therein for a detailed description of the latest version).

A key aspect of the GCM is that it uses a prescribed dust scenario (i.e. imposed columnar dust opacity at all locations and time). In practice, such forcings are implemented after analysis of a variety of dust observations from MY 24 to MY 31 (Montabone et al., 2015), or using a climatological dust scenario derived from these observations (Millour et al. 2018). The MCD thus provides fields calculated using various dust scenarios. For the comparisons presented in this paper, we used the MCD (version 5.3) climatological scenario, corresponding to our best guess of a typical Mars year, i.e., a Mars year without any global planet encircling dust storm. We first chose the water-ice column variable for the comparison with the OMEGA Ice Cloud Index. The GCM, which produces the water-ice column and most other variables of the MCD, calculates these original variables on a regular grid of 3.75° latitude and 5.625° longitude. An average day is calculated over Ls intervals of 30°, with a daily time-step of 2 hours.

The water-ice column (WaterIceCol) from the MCD is extracted on the same 4-dimensional grid (1° longitude x 1° latitude x 5° Ls x 1 h LT) as the ICIR and PCP. WaterIceCol values are calculated for each spatial (2D) gridpoint by bilinear interpolation, in the high-resolution mode (taking into account topography and correcting surface pressure and atmospheric temperature). Similarly, the WaterIceCol variable undergoes bilinear interpolation in order to cover the same temporal (Ls,LT) domain as the OMEGA gridded variables.

In a first stage, we averaged the water-ice column over the same daily and yearly time period in order to compare 2-dimensional daytime and seasonal maps. Fig. 9 shows the water-ice column around northern summer solstice (Ls = 45° – 135°) and around winter solstice (Ls = 225 – 315°), which can be compared to the ICIR and PCP maps of Figures 4 and 5.

Around northern summer solstice, major cloudy areas, namely the aphelion cloud belt, the edge of the south polar cloud belt, the remaining clouds of the north polar hood, and the Hellas basin clouds, are present on both datasets.

Regional differences can nevertheless be observed: the areas between Olympus Mons and the three major Tharsis volcanoes, as well as the Lunae Planum region, appear to be less cloudy in the WaterIceCol variable (on Fig. 9) than on the corresponding OMEGA-derived ICIR (on Fig. 4b). The south polar cloud belt has a more northerly edge (at ~30 – 35° S) in the GCM-derived data than in the OMEGA-derived data (at ~40 – 50° S). A similar comparison can be made for the position of the north polar hood on the northern winter solstice maps (Ls = 225 – 315°). The southern edge of this prominent feature is located further towards the pole on OMEGA-derived data (~40 – 50° N) than on GCM-derived data (~30 – 40° N). These model-observation discrepancies are the same as previously reported and analysed in Navarro et al. (2014).

The major difference between OMEGA-derived and GCM-derived datasets at any season, visible between Figs. 4b and 5b, and Figs. 9a and 9b, remains the smoother aspect of the cloudy structures in the latter dataset, which is related to the interpolation method and to the coarser spatio-temporal resolution of the Martian GCM (MGCM) used to build the MCD (ΔLon = 5.625° ; ΔLat = 3.75° ; ΔLs=30° ; ΔLT = 2 h), even though the MCD data used for the comparison has been interpolated to the same, higher resolution used for the ICIR and PCP data. This smoother aspect is still present when the OMEGA data is gridded and averaged on a grid almost identical to the original GCM-grid (with a solar longitude step of ΔLs = 15° instead of 30°). The overall similarities of cloudiness observed inside each dataset (MCD and OMEGA) at different scales has also been previously observed on GCM-simulations at standard (current MCD 5.3) and high horizontal (1° x 1°) resolution (Pottier et al., 2017).

Another way to compare both datasets consists in calculating the correlations between the ICIR or PCP and the MCD WaterIceCol. Table 1 shows that correlation values can be quite different around the summer solstice (Ls=60°–120°), depending on the observed region. A high correlation (above 0.6) is observed over a majority of cloudy areas of different sizes (from major volcano surroundings to entire latitudinal bands), but not for all of them. Around the winter solstice (Ls = 240° – 300°), high correlation values are only observed in the northern hemisphere, mainly in areas covered by the edge of the north polar hood. In areas with less than 10% cloud cover (i.e. in the tropics and midlatitudes of the southern hemisphere), correlation values are very low (below 0.4).

The ICIR and WaterIceCol are physically related to the presence of water ice particles, therefore high correlation values are expected and generally observed when clouds are present. Low correlation values are observed when no, or very few clouds are present (especially during northern winter), and the resulting (normally low) ICIR may not be representative of the absence or quantity of clouds. In this case, unlike the WaterIceCol, which is directly related to the presence of water ice and will take a value close or equal to zero, the ICIR will be more dependent on the two reflectances (at 3.38 and 2.52 μm) and thus from Martian surface properties. It can also indicate that the WaterIceCol variable does not reproduce observed cloudiness correctly, possibly in relation with lower spatio-temporal resolution of the underlying GCM used to build the MCD. This could be in particular the case over the Lunae Planum, Syrtis Major and Argyre areas around the northern summer solstice.

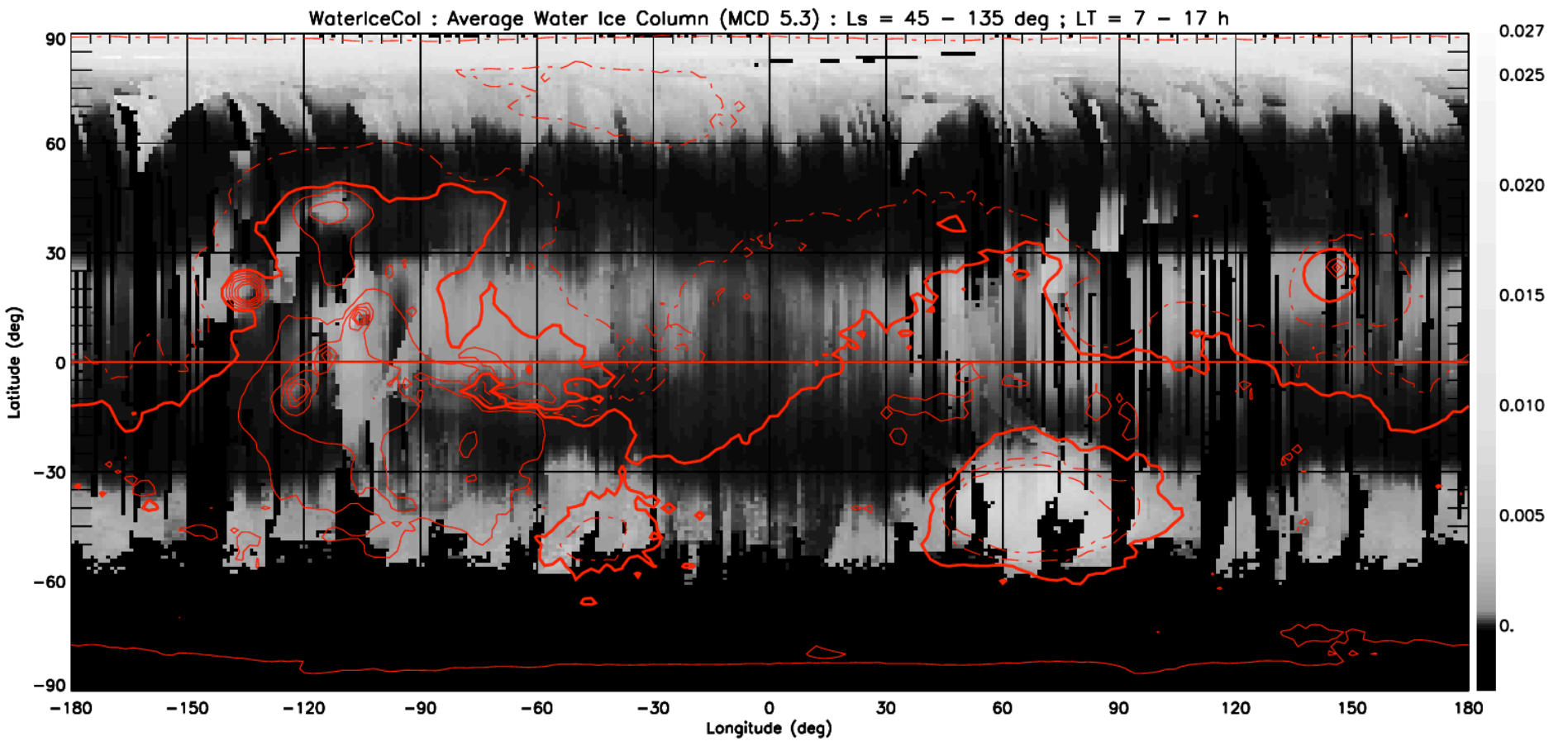

a)

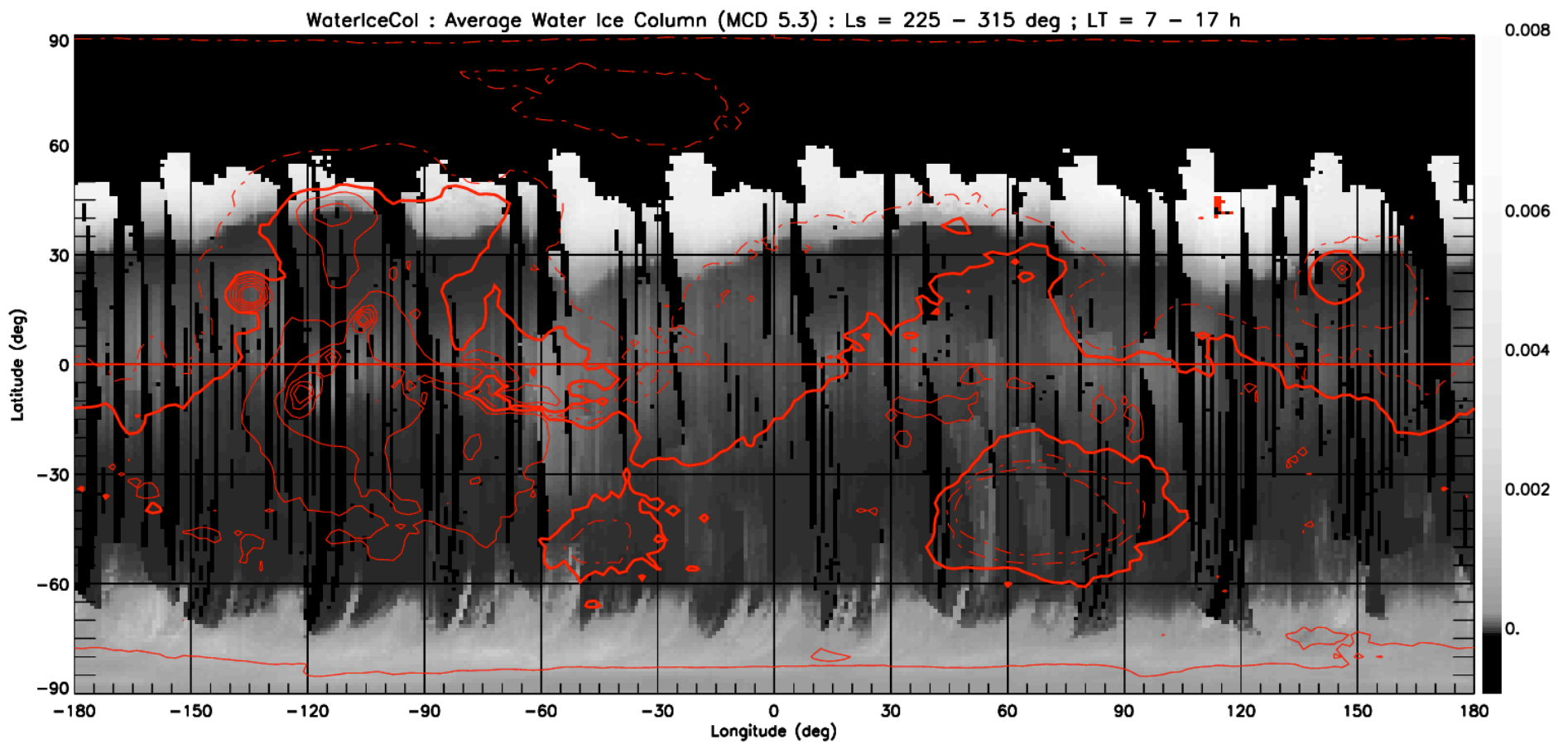

b)

**Figure 9:** maps of average MCD water-ice column (WaterIceCol, in $kg/m^2$) between Ls = 45° and 135° (a), and between Ls = 225° and 315° (b), both during daytime (7 – 17 h LT). The equator, and contour lines (in red) corresponding to surface elevation are superimposed on this figure and on following maps: thick line: datum (0 m) ; thin continuous line: elevation > 0 ; thin dashed line: elevation < 0. Consecutive contour lines are spaced every 2500 m. The first and the last value of each scale correspond to the minimum and maximum value of each variable. Values below zero (darkest black) on the scale correspond to gridpoints where OMEGA ICIR data is absent. Note that the magnitude of scale between both figures is different.

## 5. The cloud Seasonal and Diurnal cycle at global scale and over selected regions

The temporal evolution of several geographical regions listed in Table 1 is described in this section. After a description of the global seasonal cycle, we focus on the regions where there is sufficient data to describe the cloud life cycle at least during a part of the Martian year.

### 5.1 The seasonal cycle of morning, noon and afternoon clouds

The cloud seasonal cycle along a Martian year is traditionally represented by a 2-dimensional diagram as a function of solar longitude (Ls) and latitude, after averaging over all longitudes and time of a Martian day, whenever data is available. Such diagrams were extracted from TES data ($\tau_{TES_wice}$) (Smith, 2004), from SPICAM data ($\tau_{wice}$) (Willame et al., 2017) and from MGCM data ($\tau_{wice}$ and WaterIceCol) (Montmessin et al., 2004 ; Navarro et al. 2014, Pottier et al., 2017). These diagrams show the main cloud structures at planetary scales, e.g. the aphelion belt during northern spring and summer, the north polar hood during extended autumn and winter, and the south polar hood during reduced southern autumn and winter. We extracted the same type of figures covering a limited period of daytime (4 Martian hours). Figure 10 shows the ICIR in the morning (6 – 10 h LT), around local noon (10 – 14 h LT) and in the afternoon (14 – 18 h LT). The three main cloud structures are also present here. But although OMEGA data does not cover the southern hemisphere in the morning during northern summer (Ls = 80 – 170°), a major difference is outstanding between morning and afternoon hours with high ICIR values on the one hand, and the period around noon on the other hand, when the ICIR takes smaller values, indicating a reduced cloud coverage in the aphelion belt region. This is described and interpreted in further detail in the next section.

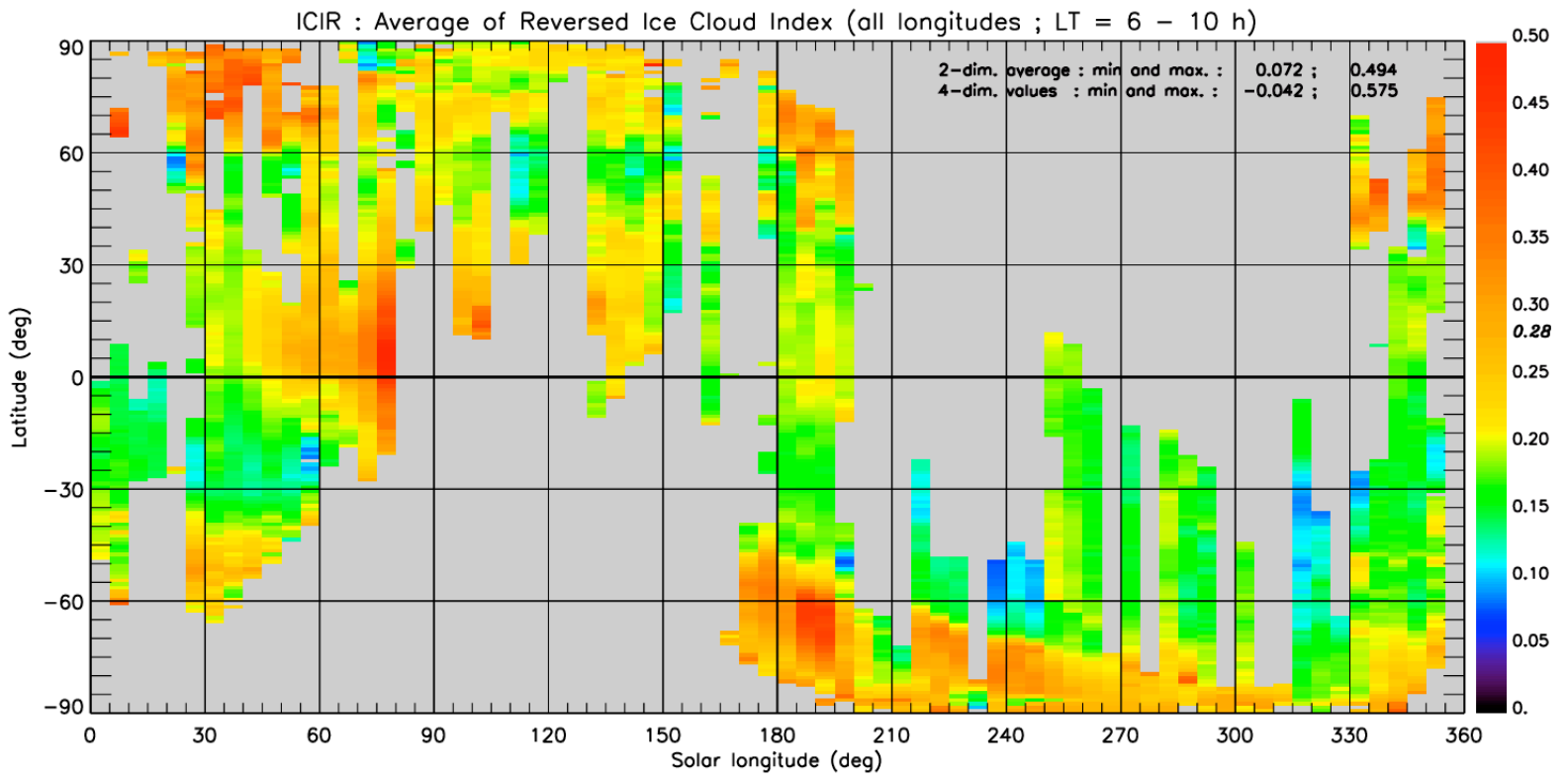

a)

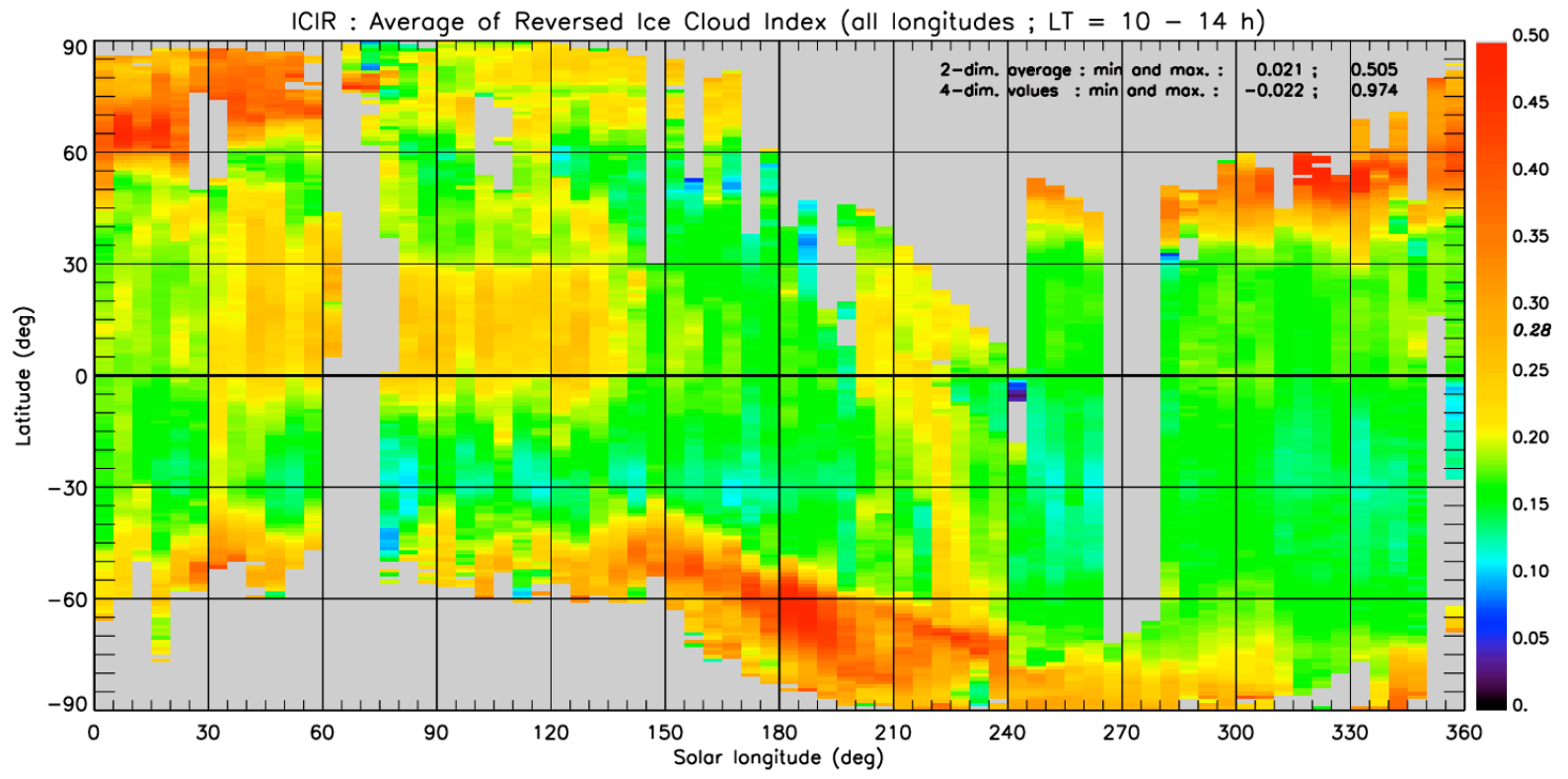

b)

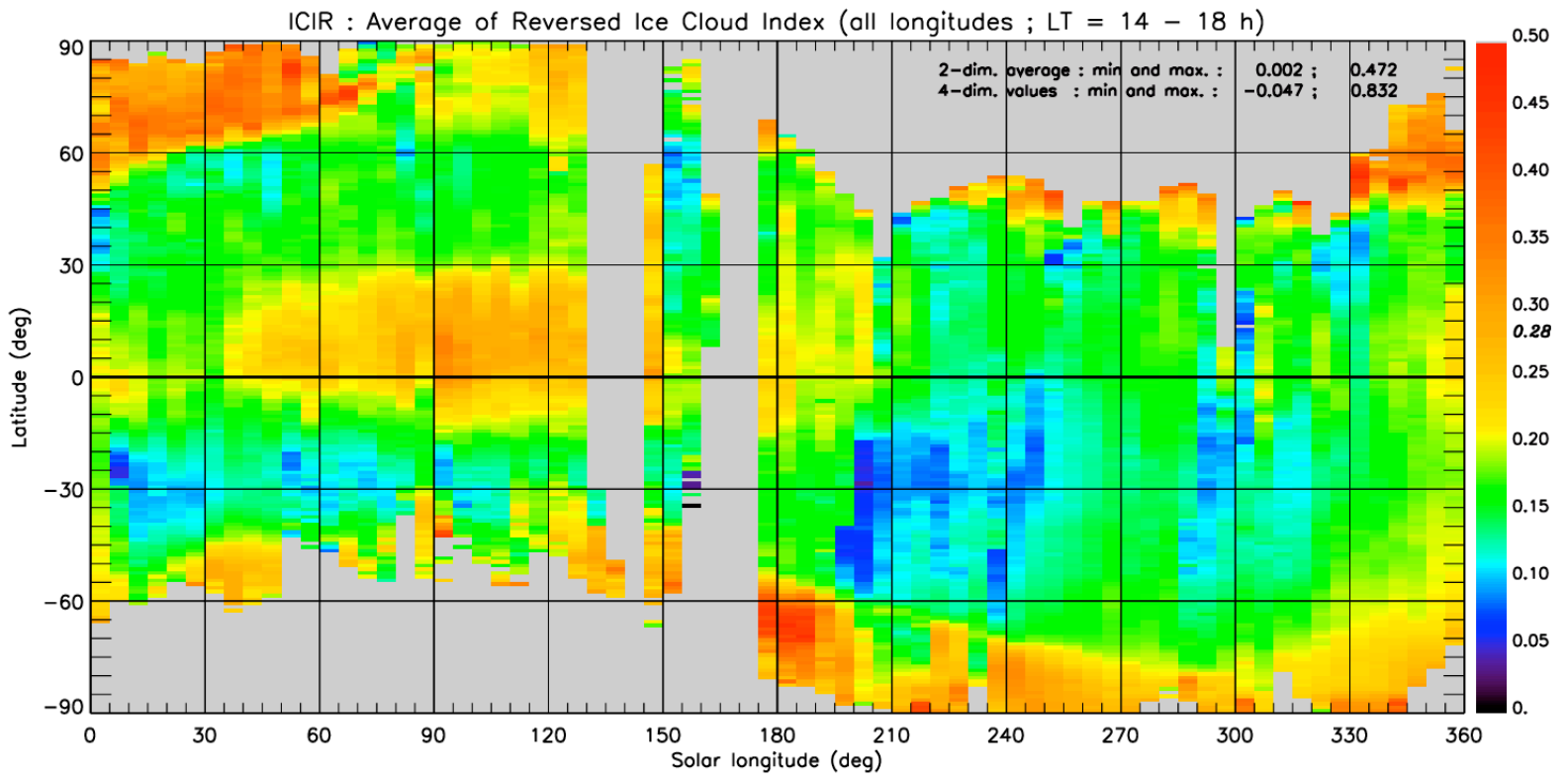

c)

**Figure 10:** reversed ice cloud index (ICIR) averaged over all longitudes and over 4-hour periods: 6 – 10 h LT (a), 10 – 14 h LT (b), 14 – 18 h LT (c). Values in the upper right part of each figure indicate the minimum (respectively the maximum) value of the averaged ICIR on the (2-dimensional) figure, and minimum (respectively the maximum) value of the 4-dimensional ICIR dataset.

### 5.2 Clouds over the tropical plains.

Here we limit the region of interest to latitudes between 25°S and 25°N, and longitude between 90°W and 120°E, i.e. located between the major volcanoes of the Tharsis bulge and Elysium Mons. Figures 11a shows the observed ICIR variable, Figure 11b the MCD-predicted WaterIceCol shown over the same gridpoints as the ICIR, and Figure 11c the MCD-predicted WaterIceCol over the complete region and at all instants. One can see on Figure 11a that:

- No clouds are present in the morning before Ls ~ 20°.

- Around summer solstice (Ls = 80°-120°), the WaterIceCol (Fig. 11b) indicates reduced cloudiness at 12 h, in comparison with morning and afternoon periods. This trend is also present but less obvious on the ICIR data (Fig. 11a), for Ls between 95 and 120°.

Figure 11c shows the percentage of cloud cover in the same coordinates. It is extensive during the largest part of the northern spring and summer (Ls = 30 – 150°), according to the MCD WaterIceCol data (Fig. 11b). This corresponds mainly to the aphelion cloud belt. Maximal PCP values of 100% are observed during this season in the morning, for Ls between 70° and 80°, corresponding to a complete dense cloud coverage of the limited area where ICIR data was available.

Note also that no clouds are observed after 16 h LT (ICIR < 0.20, on Fig. 11a) during a large part of the Martian year (between Ls = 185° and 40°). During this period, the ICIR error bar is also the most important, with a relative error above 30 % (Fig. 11d).

The ICIR (Fig. 11a), the sampled and the complete WaterIceCol (Fig. 11b and c) averaged data show basically the same evolution of cloudiness during the Martian year, with one exception : some clouds are detected on ICIR data (with ICIR > 0.20) around 13 h LT after the Northern autumn equinox (Ls = 200 – 220°) . These reliable observations, with a low relative ICIR error (below 10 %), correspond to very low values in the WaterIceCol dataset.

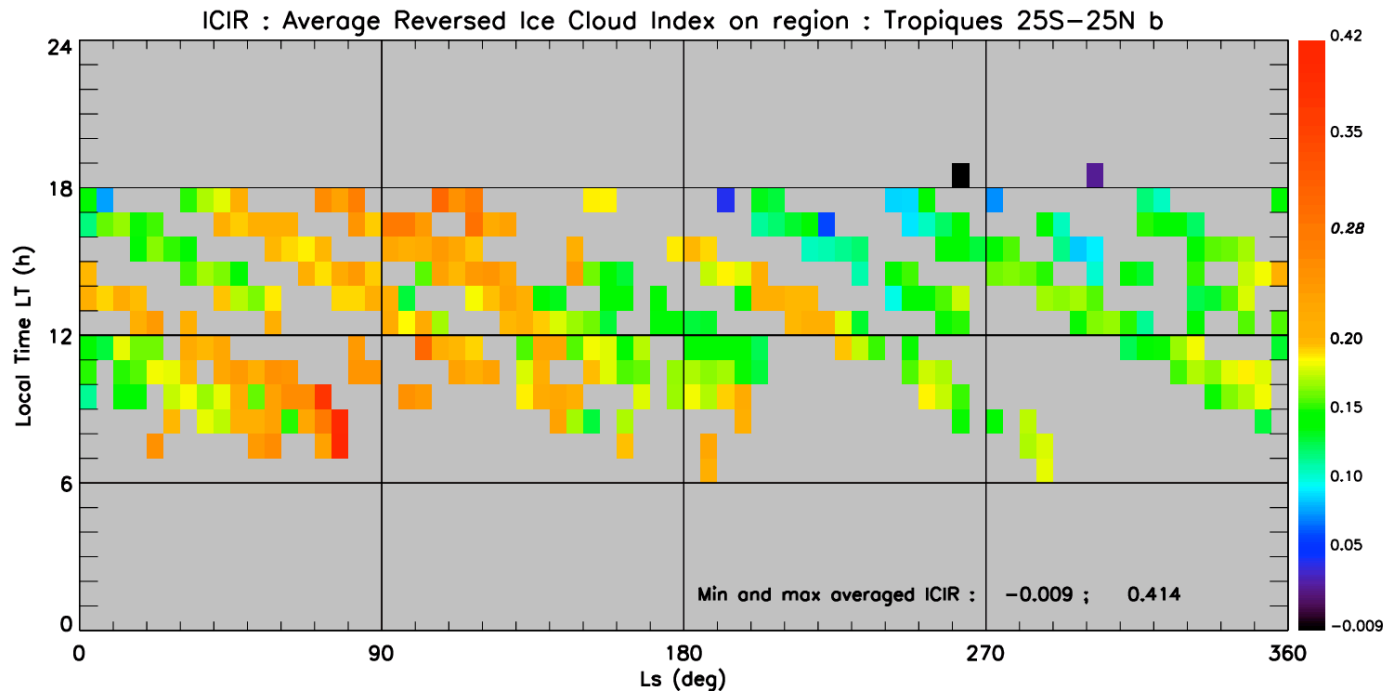

a)

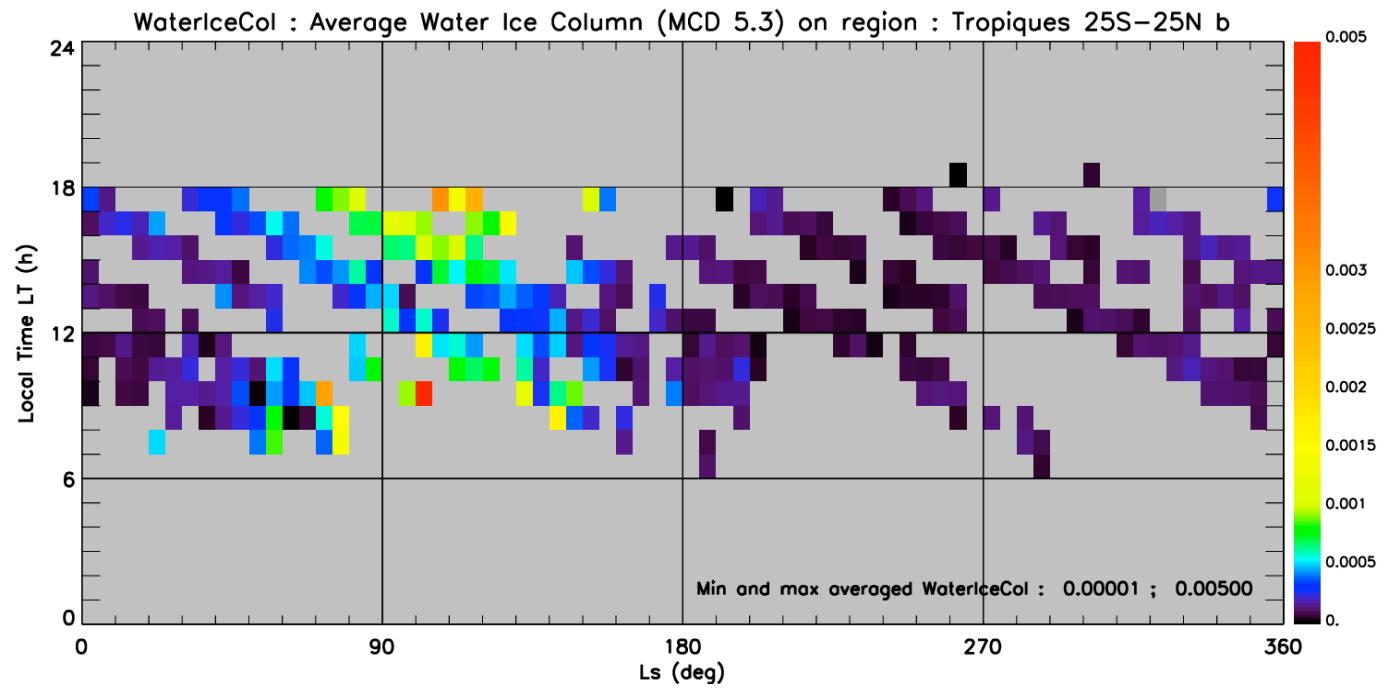

b)

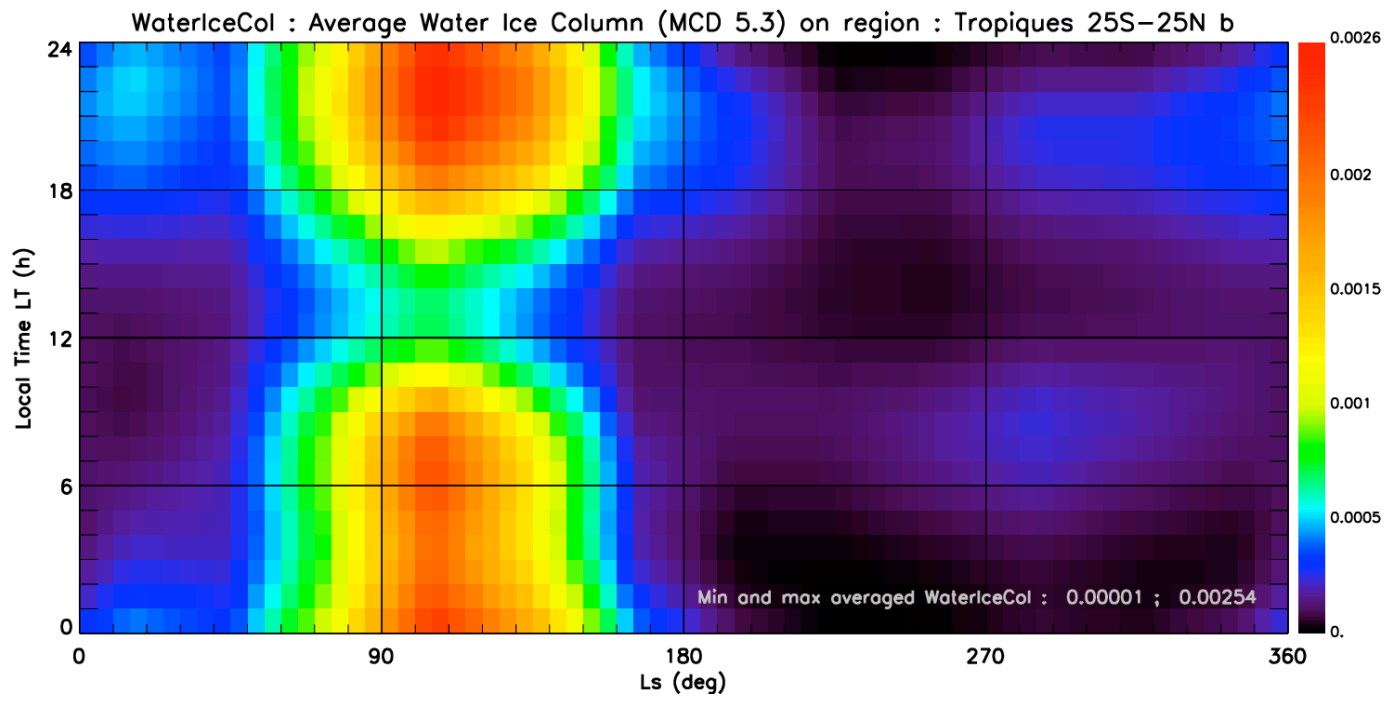

c)

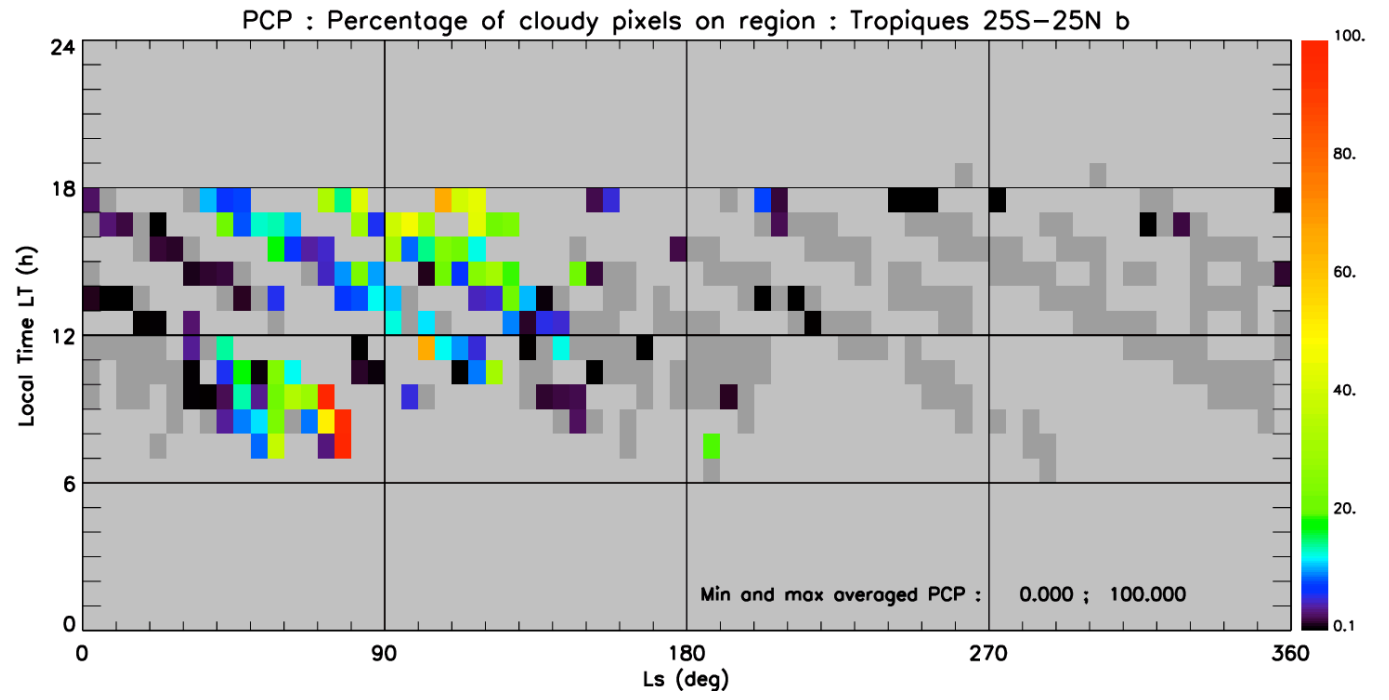

d)

*(continued on next page)*

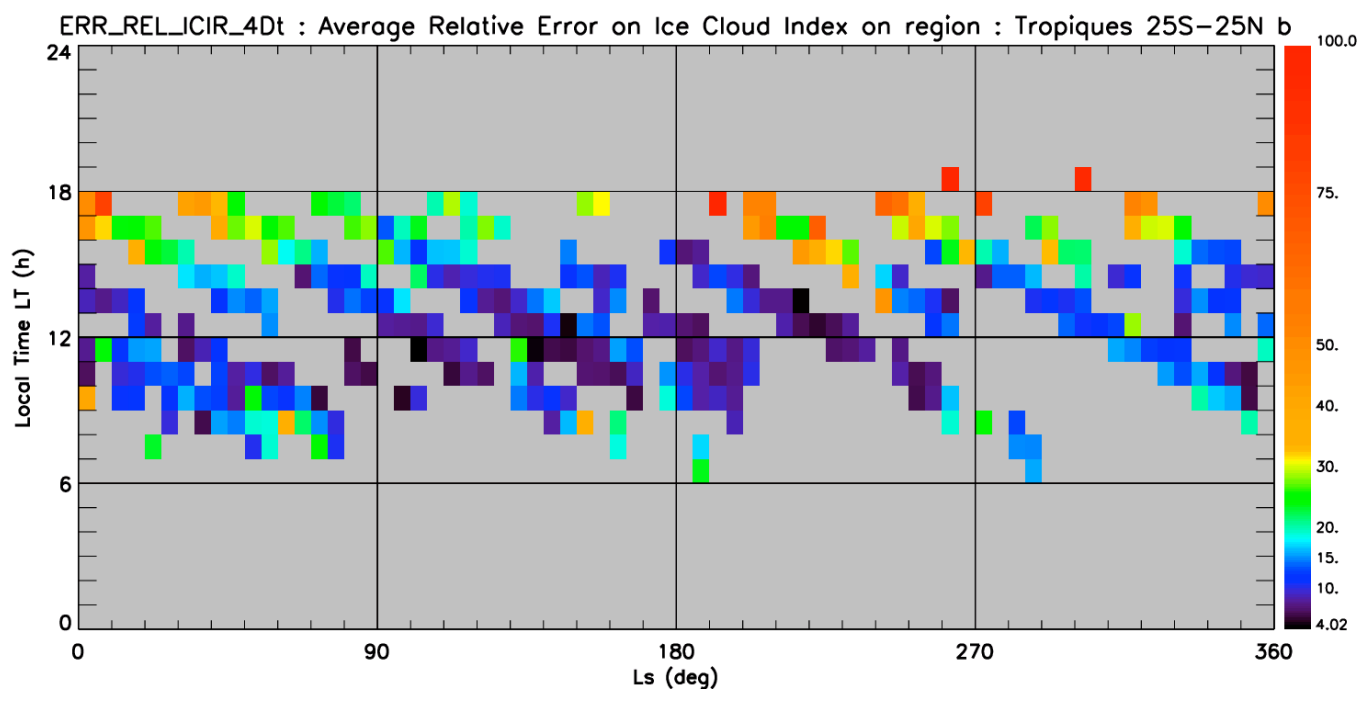

e)

**Figure 11:** diurnal cycle of the Ice Cloud Index (a) ; MCD Water-Ice Column in kg/m$^2$ (annual climatology scenario, average solar EUV and average dust conditions) with the same spatio-temporal coverage as the ICIR (b) ; MCD Water-ice column at all instants and locations of the region (c) ; Percentage of Cloudy Pixels (Dark gray corresponds to a PCP equal to 0 %) (d), and relative error of the ICIR, in % (e), over the tropical zone (25°S – 25°N ; 90° W – 120° E) during one Martian year. Light gray corresponds to missing values. The scale and the maximal value of the two water-ice column figures are different.

With the help of complementary data from the MCD and the LMD MGCM, the evolution of the clouds and the minimal cloudiness shown on Figure 11b and c can be explained by the propagation of the diurnal thermal tide which induces a temperature anomaly that controls the condensation/sublimation of the cloud ice (Fig. 12).

An almost similar figure to Figure 11 and similar observations can be made when all the longitudes, also covering the Tharsis area and the large volcanoes, are taken into account in the average calculation of the ICIR, the WaterIceCol and the PCP, instead of the limited longitudinal band between the 90°W and 120°E. The impact of the added ICIR data covering the major volcanoes, which are more often covered or surrounded by clouds than other regions at the same latitude (Benson et al., 2006), is limited. We can explain this with the small number of supplementary 4-dimensional gridpoints covering these volcanoes. On the other hand, the added ICIR data coverage from the larger set of 4-dimensional 'non-volcanic' gridpoints (between 120°E and 90°W) is more important. These gridpoints must have similar characteristics on average to those of the region limited in longitude (between 90°W and 120°E).

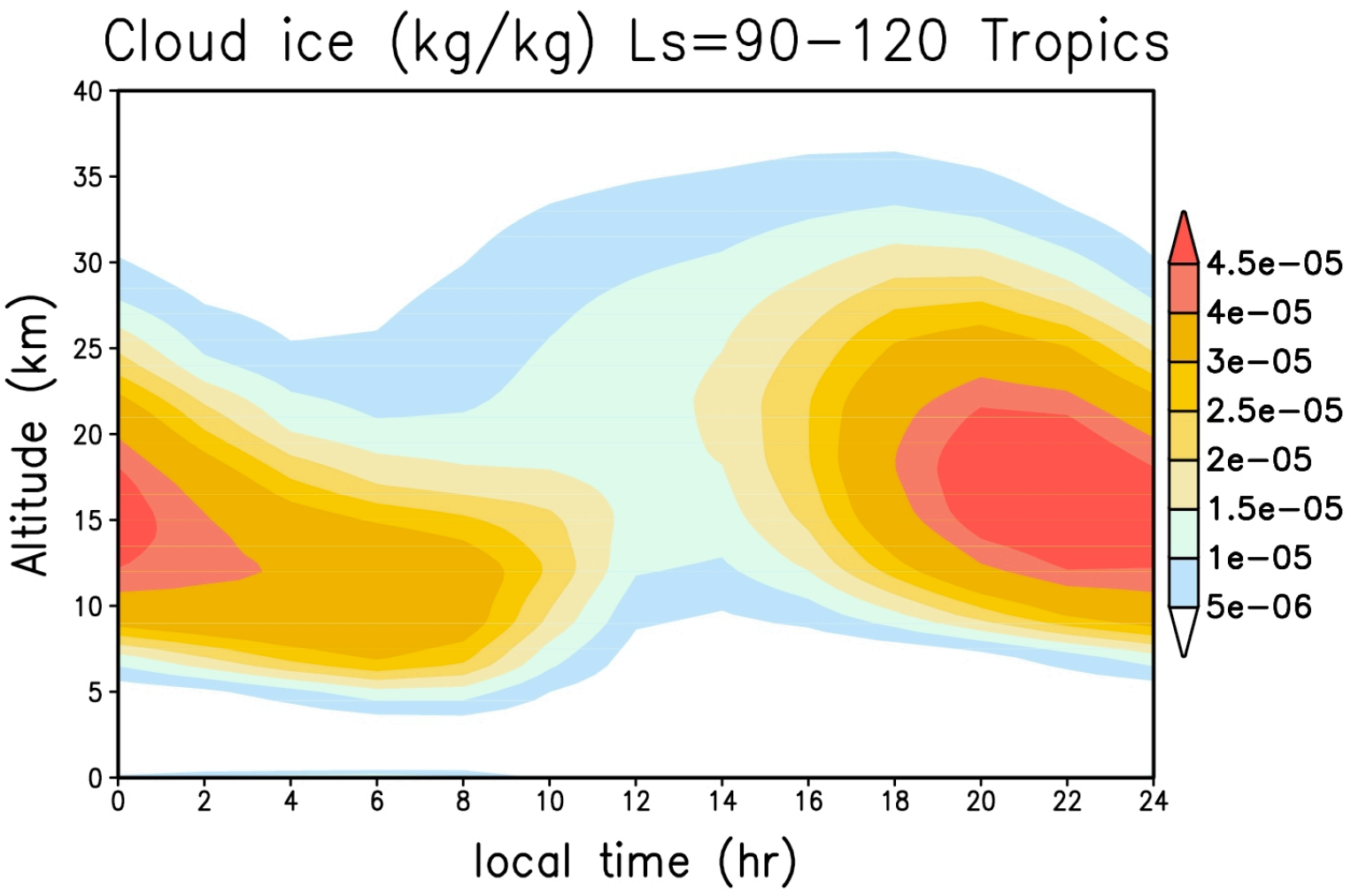

**Figure 12:** MGCM diurnal cloud ice density in the tropics (25°S - 25°N) at the beginning of summer (Ls = 90° - 120°), for longitudes between 90°W and 120°E.

5.3 **Northern midlatitudes (35° N – 55° N ; all longitudes):**

In this midlatitude band, clouds are present at the end of autumn and during winter (Ls = 250 – 360°) (Fig. 13a ; see also Fig. 5b). Observed clouds (after Ls = 310°) are mainly present in the morning and around noon. The cloudiness decreases at the end of the afternoon. Lower ICIR values are found at the beginning of northern spring (Ls = 0 – 40°), mainly in the morning. The observed clouds, also detected in the MCD water-ice column data (fig. 13b and c), must correspond to the north polar hood present during winter, with remnants retreating towards the pole at the beginning of spring (Benson et al., 2011).

During spring and summer (Ls = 40 – 150°), the ICIR (with values between 0.18 and 0.30) suggests the presence of some clouds in the morning and around noon, probably corresponding to the northern part of the aphelion belt, including Alba Patera. The WaterIceCol figures (Fig. 13b and c) with their low values (below 0.012 kg/m2) rather suggest the absence of clouds during this period, although one can observe a slight decrease of WaterIceCol values between the morning and afternoon hours – comparable to the decrease observed on ICIR data.

Over nearly the whole year, clouds are absent (with ICIR < 0.18) during the last hours of the daylight period (i.e. in the late afternoon, and in the early evening around the summer solstice period), whenever the data coverage was sufficient. The corresponding error bars on the ICIR are generally large, with a relative error above 30 %.

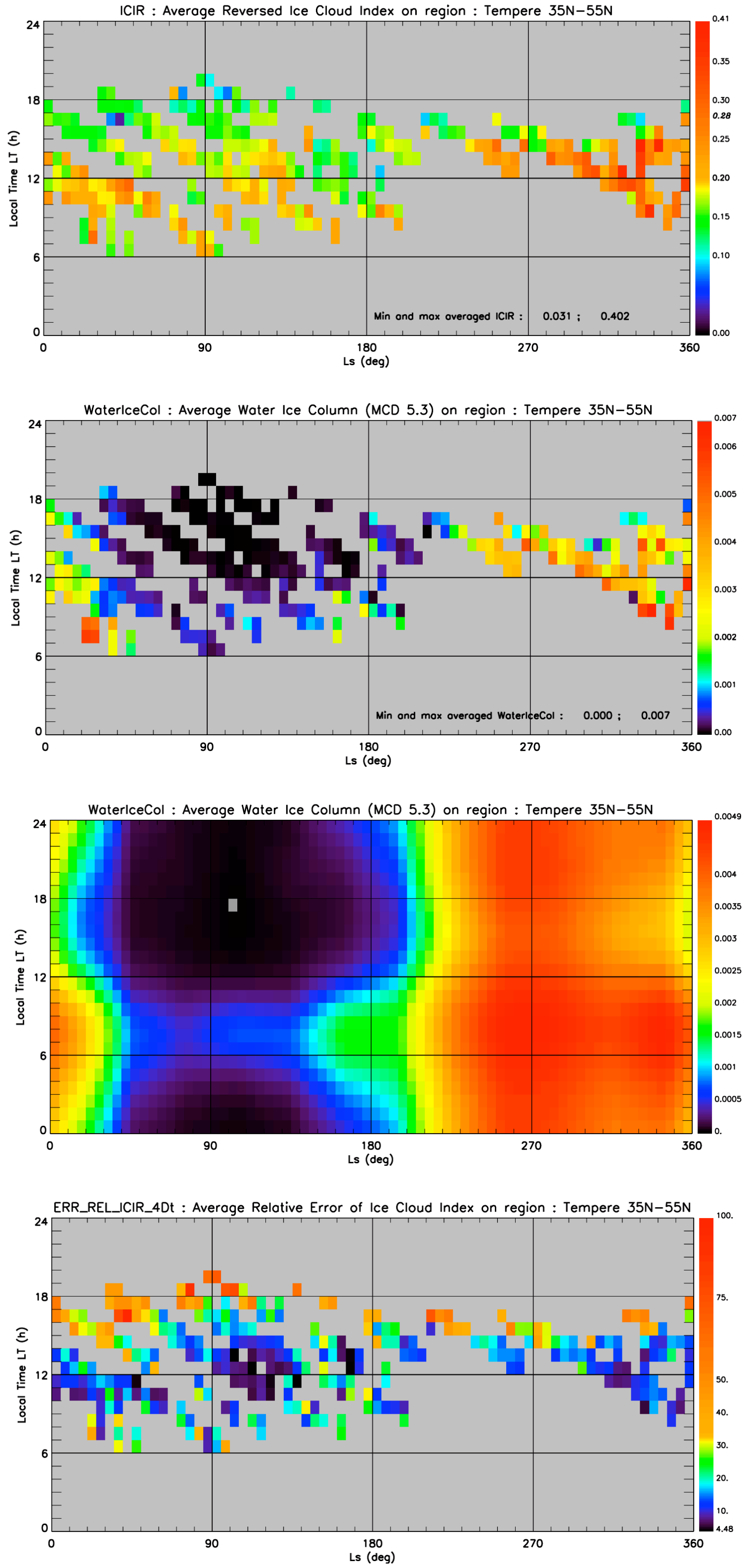

a) b) c) d)

**Figure 13:** Diurnal cycle of the average Reversed Ice Cloud Index (a: top), and of the modeled (MCD) collocated Water-ice column in kg/m$^2$ (annual climatology scenario, average solar EUV and average dust conditions) (b) ; Water-Ice Column at all instants and locations of the region (c), and relative error of the ICIR,

in % (d: bottom) over the midlatitude region (35°N – 55°N ; all longitudes). The gray gridpoint on figure b locates the minimal WaterIceCol value.

**5.4 Chryse Planitia (20° N – 50° N ; -60° E – -30° E):**

The two regions presented above gave indications of the presence of clouds over broad latitudinal bands; it is also interesting to examine results covering smaller geographical regions in order to study the presence -and possibly the formation- of clouds at local scale, with the risk of having a smaller ICIR data coverage. We present results obtained in Chryse Planitia, which partially overlaps the Northern Midlatitude area, and also the Tropical area.

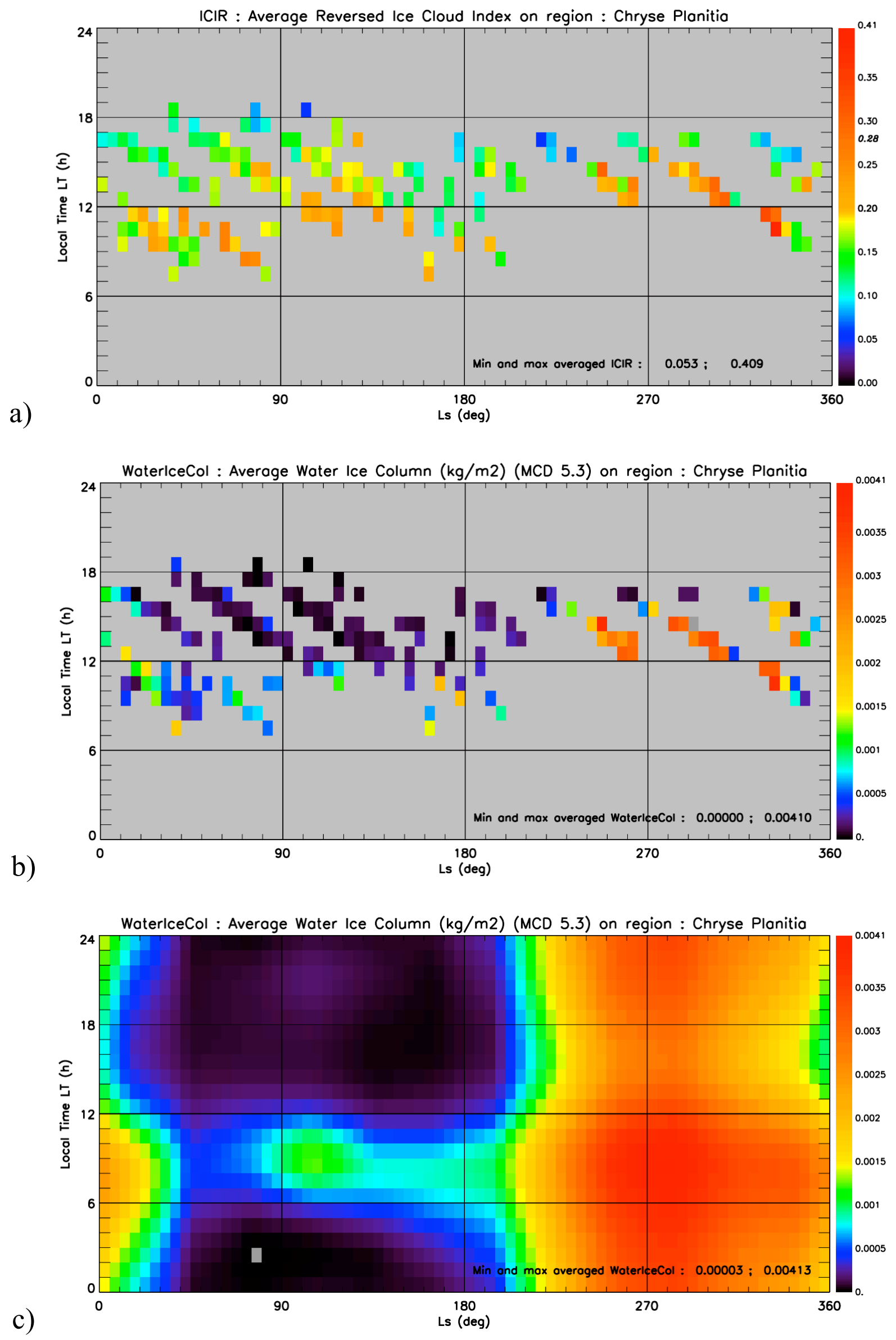

**Figure 14:** diurnal cycle of the Ice Cloud Index over the Chryse Planitia region (20°N – 50°N ; -60°E – -30°E ) during one Martian year. ICIR (a) ; MCD Water-Ice Column in kg/m$^2$ (annual climatology scenario, average solar EUV and average dust conditions) with the same spatio-temporal coverage as the ICIR (b) ; MCD Water-ice column at all instants and locations of the region. The gray gridpoint locates the minimal WaterIceCol value (c). Light gray corresponds to missing values. The scale of the two water-ice column figures is the same.

Figure 14a indicates the presence of two main periods when clouds are present, during northern spring and early summer, and around the winter solstice (Ls = 245 – 335°). Similar cloud coverage (but with less dense clouds in summer than in winter) can be observed on collocated MCD water-ice column charts (Fig 14b and c). In the model as in the mapped observations, the spring-summer clouds correspond to the tropical aphelion belt, whereas the winter clouds belong to the winter polar hood which can extend down to 20°N in Chryse Planitia.

These winter clouds are present around noon and dissipate in the middle of the afternoon. This time the MGCM suggests that the clouds are low-lying fogs (Fig. 15a). As on Earth they form near the surface during the night as the surface cools and are maximal in the morning. They only dissipate a little in the afternoon when the surface heated by the sun is warm enough to heat the atmosphere. This is why the fog thickness is minimum in the afternoon on both Figure 14 and 15a. Above 30 km, the MGCM predicts the formation of a cloud that is very thin and should be invisible to OMEGA. Nevertheless it provides an example of high cloud controlled by thermal tides but with a maximum of condensation during the day and dissipation during the night (Fig. 15b).

a)

b)

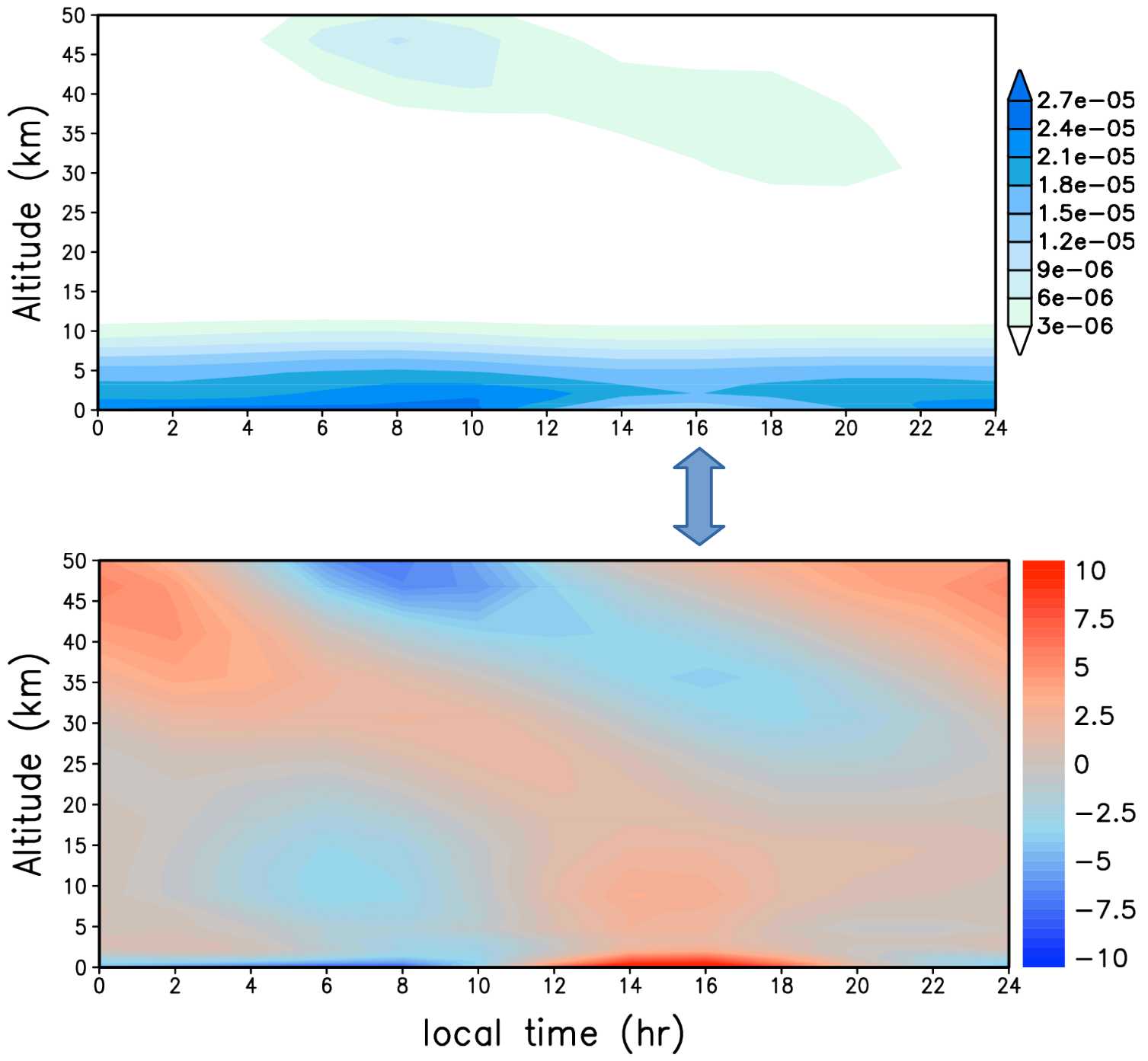

**Figure 15:** MCD Global Climate Model predictions - winter fog in Chryse region (Ls = 270° – 300°): mean water ice content (kg/kg) (a: top). Temperature diurnal anomaly (K) (b: bottom). The time at which fog dissipates (~16 h LT) is indicated by the arrow.

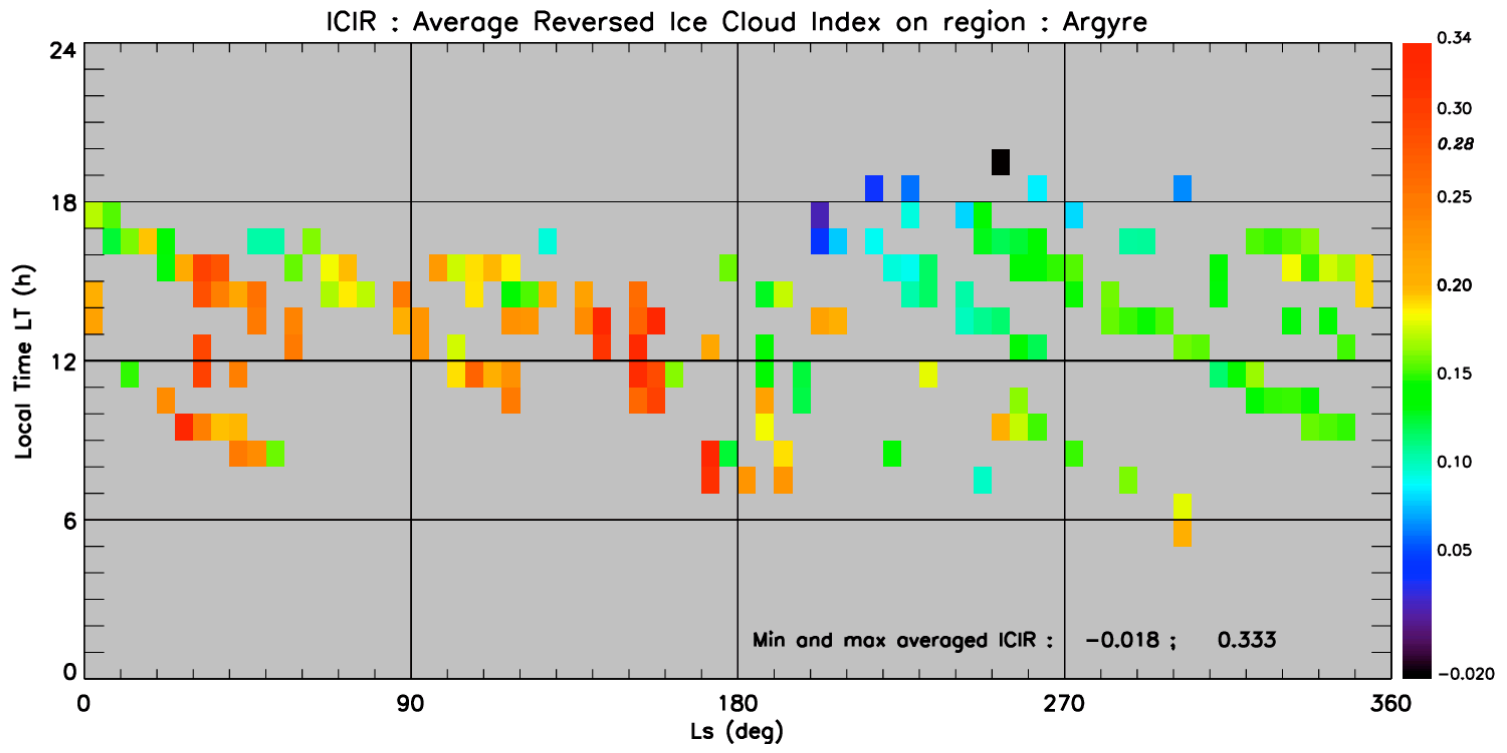

a)

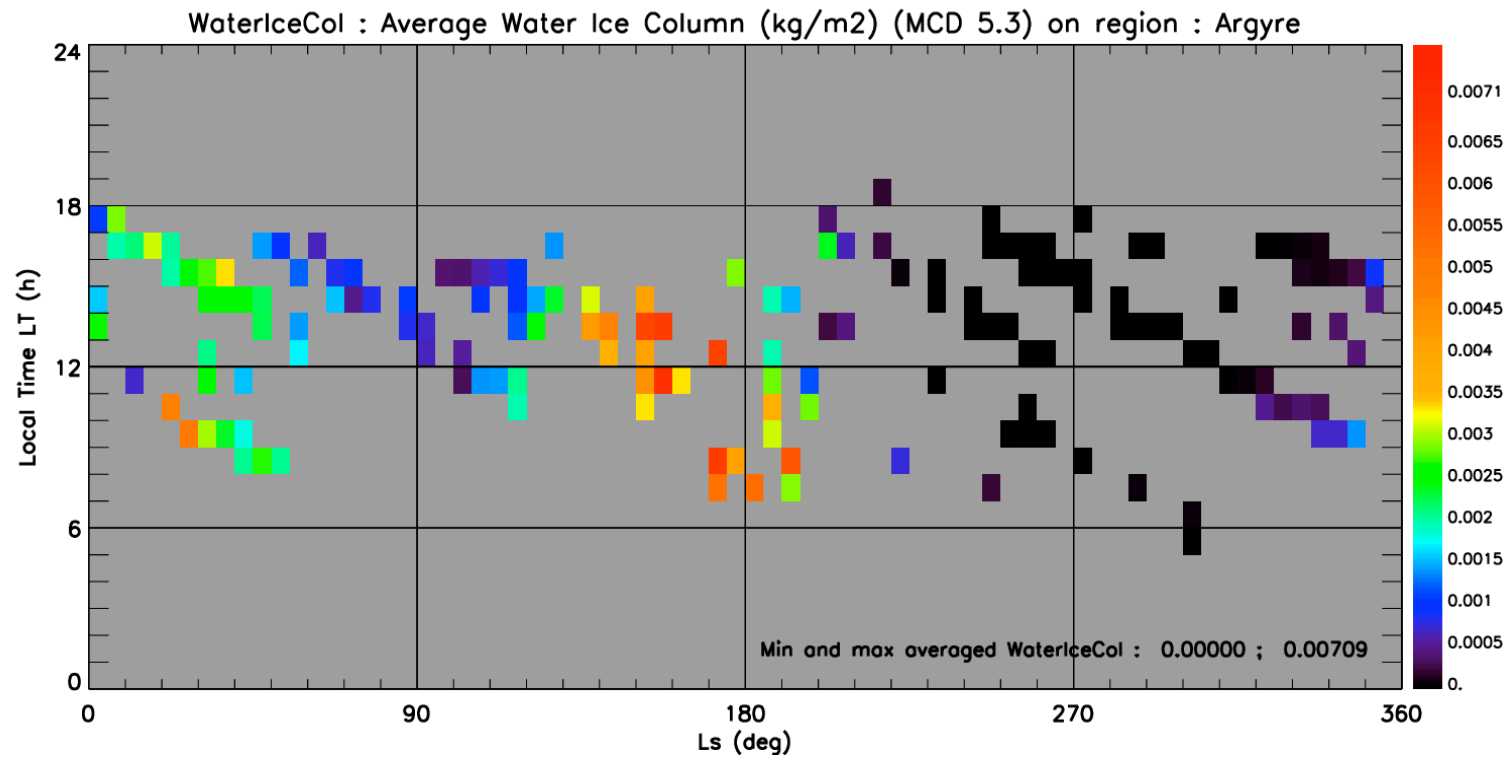

b)

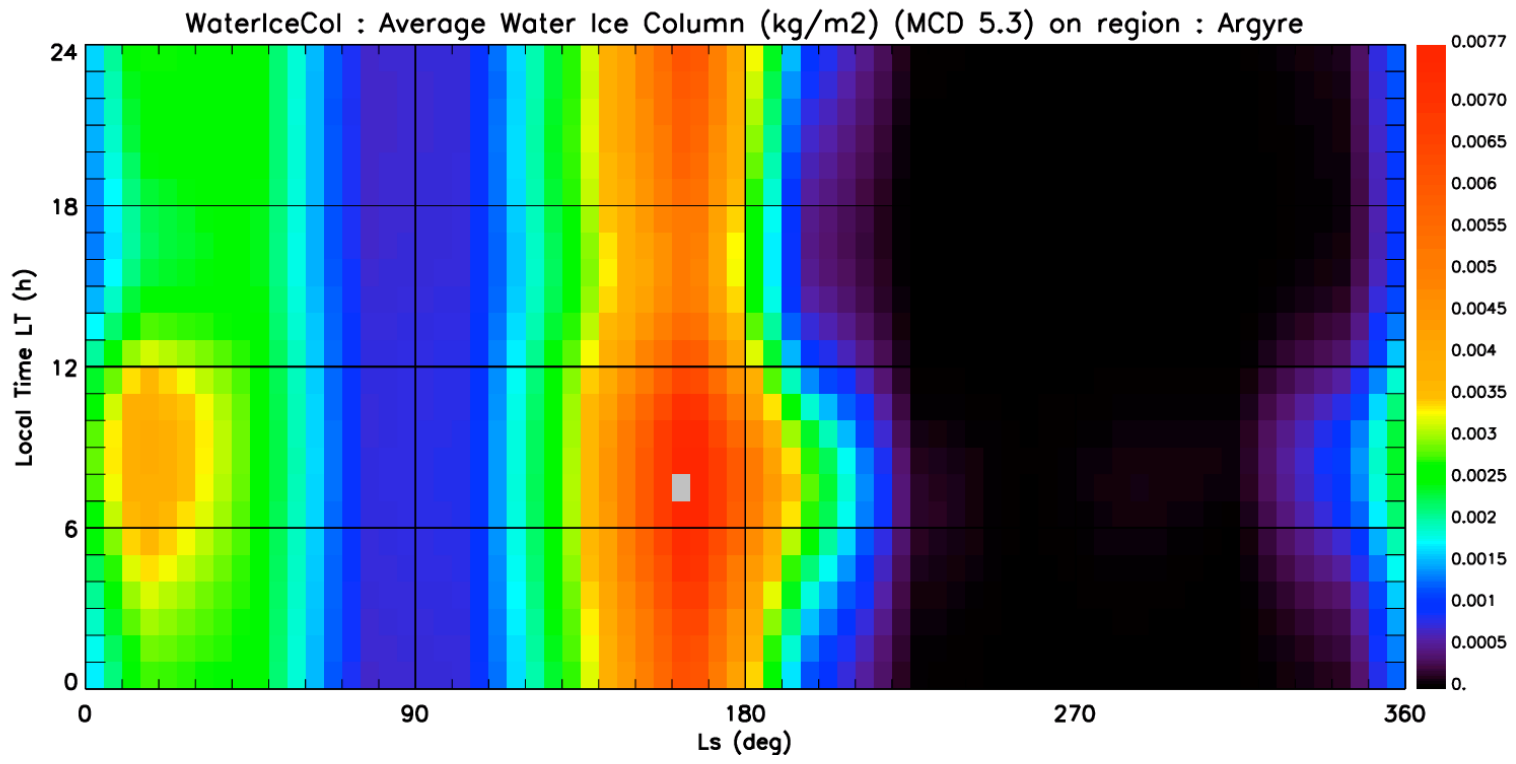

c)

**Figure 16:** diurnal cycle of the Ice Cloud Index over the Argyre Planitia region (55°S – 35°S ; -65°E – 25°E ) during one Martian year. ICIR (a) ; MCD Water-Ice Column in kg/m$^2$ (annual climatology scenario, average solar EUV and average dust conditions) with the same spatio-temporal coverage as the ICIR (b) ; MCD Water-ice column at all instants and locations of the region. The gray gridpoint locates the maximal WaterIceCol value (c). Light gray corresponds to missing values. The scale of the two water-ice column figures is the same.

### 5.5 Argyre (55° S – 35° S ; -65° E – -25° E):

The ICIR data show the existence of two major periods of significant cloudiness during the southern winter, around Ls = 45° and 135° (Fig. 16a), and confirmed by Water Ice Column data from the MCD (Fig. 16b and c). Between these periods, around winter solstice (Ls = 70 - 105°), cloudiness is slightly reduced (at least around noon and in the afternoon, when ICIR data was available). The water vapor inflow from the north polar region may be reduced by the important cloud formation in the aphelion belt, which limits the southward transport of water vapor. During both cloud peaks, clouds can be present during a large part of daytime. The MGCM also confirms the presence of clouds at 12 h LT during these peaks (Fig. 17). In southern spring and summer, clouds are absent, or strongly reduced, during all of the day.

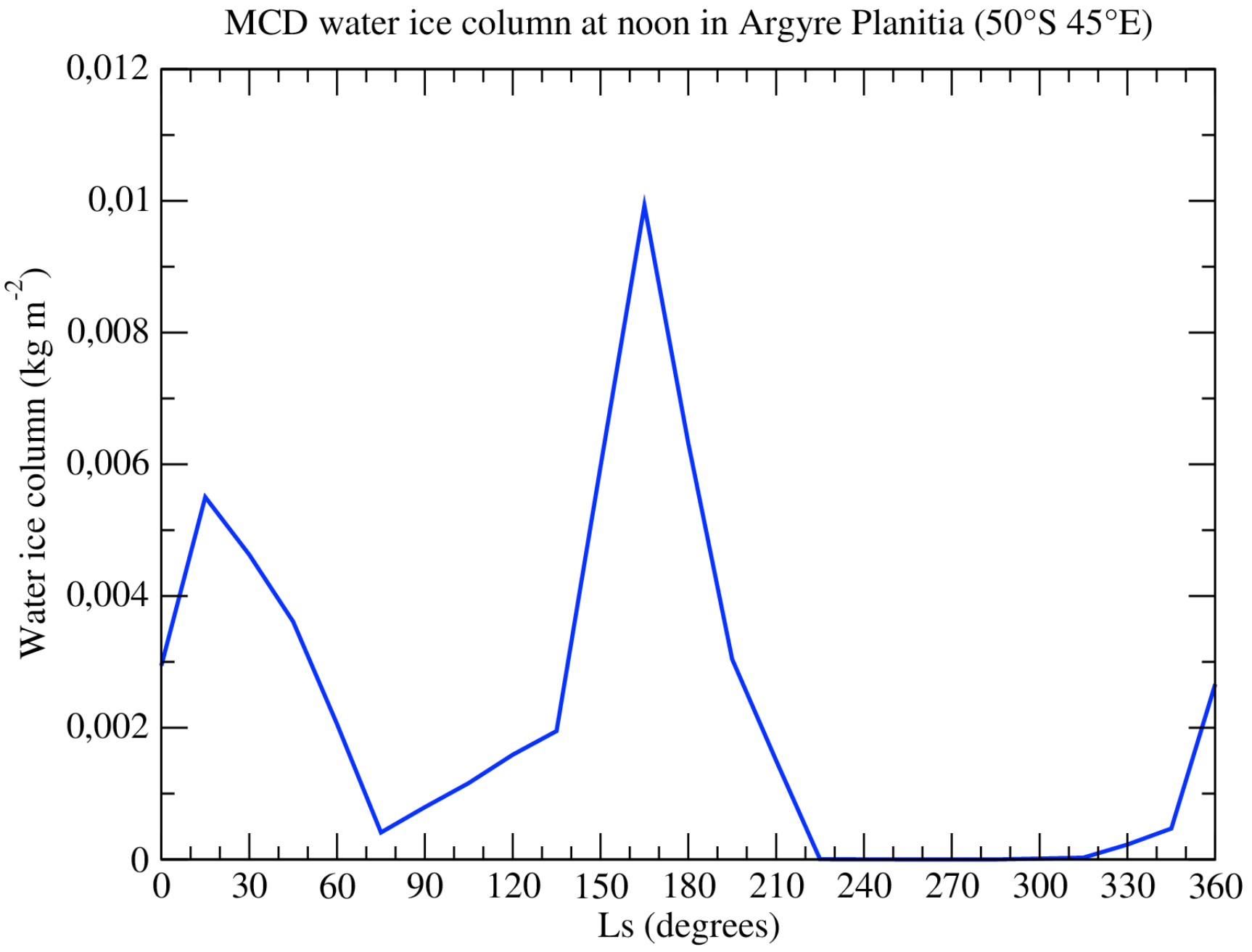

**Figure 17**: MCD prediction of water-ice column (kg/m$^2$) at (50°S. ; -45°E), the central longitude of the Argyre region, at 12h LT over the Martian year.

**5.6 The Southern Hemisphere cloud bridge (35° S – 20° S ; -150° E – -60° E):**

This region has been defined and selected because it is located between two major areas of cloudiness during northern spring and summer, the aphelion belt and the southern polar hood. Clouds and hazes have been reported in this region, in relation with occasional southward transport of water vapor (Benson et al., 2010 ; Mateshvili et al., 2007, Smith, 2004 ; Liu et al., 2003). On Figure 18a, clouds are present during three major periods:

- Spring (Ls = 0 – 60°), clouds are present in variable amounts during the morning until the middle of the afternoon at the beginning of this period. They tend to disappear earlier and earlier in the afternoon as time evolves.
- Summer (Ls = 80 – 150°), clouds are still present around noon and reduced cloudiness can be observed during the beginning and middle of the afternoon. (Few data were available in the morning.)

- Autumn (Ls = 180 – 250°), variable amounts of clouds are still observed during parts of the day, but tend to dissipate earlier and earlier as the season evolves.

In this region, the maximal observed 4-dimensional ICIR value is 0.27, reflecting the reduced cloudiness in the data sample, much lower than the values obtained for the data sample of the previously described Tropical region (0.41).

The MCD Water-ice column chart (Fig. 18b) is consistent with the Ice Cloud Index chart, except that during spring and summer in the afternoon WaterIceCol is relatively more important, and during autumn and winter it is less important.

The percentage of cloudy pixel chart (Fig. 18c) reflects the partial cloud coverage by clouds in this geographical area. The maximal value of the PCP is 25% in this 4-dimensional data sample, and during a large part of the 3 periods detected as cloudy on the ICIR figure, the PCP is close to 0%. This situation may also indicate the presence of thin clouds on some original pixels, which may be just below the threshold for the selection as cloudy pixels and therefore not increasing the PCP, which will therefore remain at low values. Low values of the PCP also can result from the averaging over a larger region, where a few gridpoints may have individual values of 100 % but are surrounded by numerous other gridpoints with a PCP of 0 %.

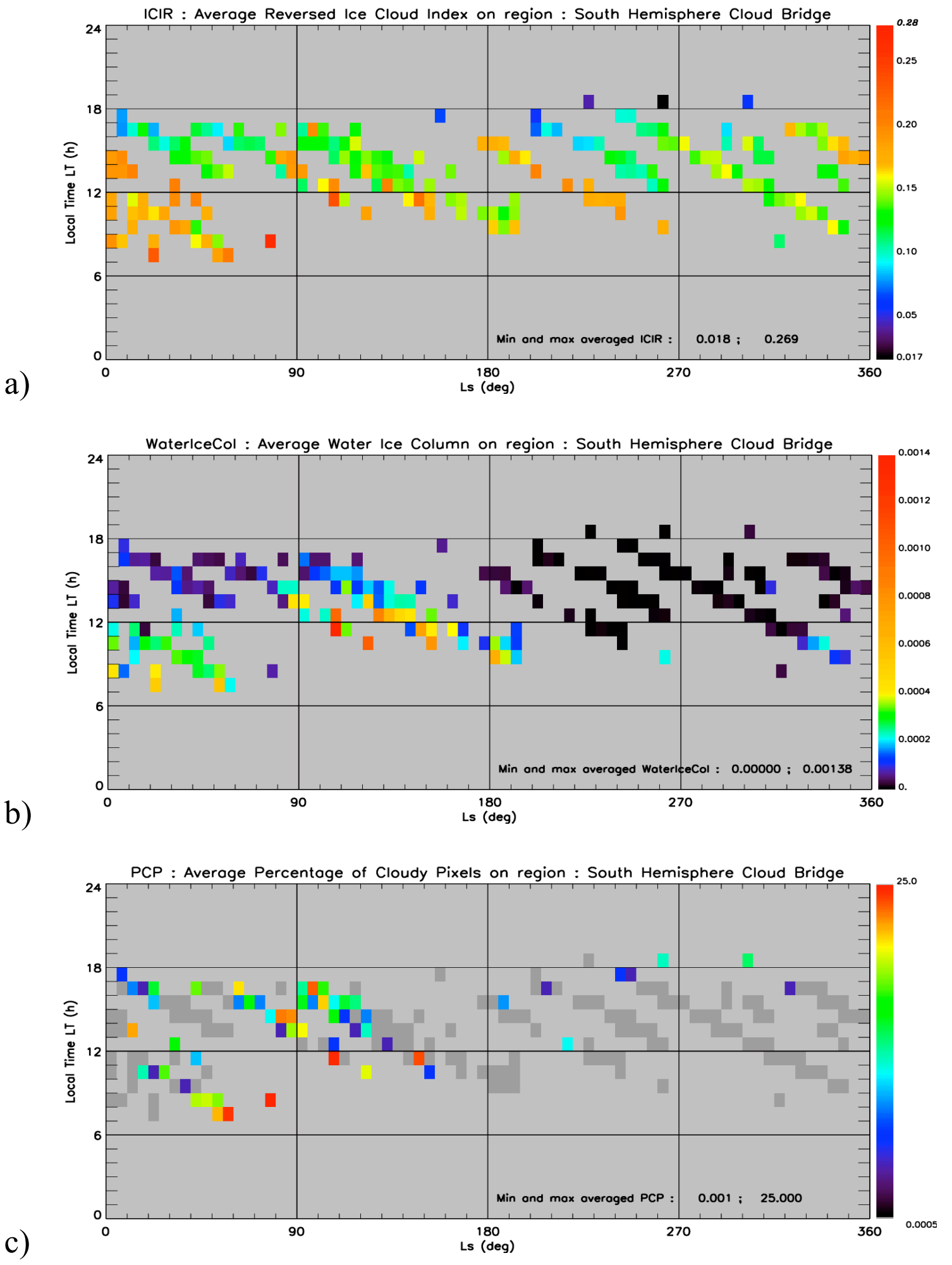

**Figure 18:** diurnal cycle of the Ice Cloud Index (a: top), MCD Water-ice column (b: middle) and Percentage of Cloudy Pixels (c: bottom) over the Southern Hemisphere cloud bridge region (35°S – 20°S ; -150°E – -60°E) during one Martian year. The color scale is strongly non-linear for c).

**5.7 Other regions:**

Among the 19 non-polar regions of interest, those for which sufficient data are available to characterize a fraction of the diurnal cycle at least during some parts of the Martian year are the three ones which have the largest spatial extension and cover all longitudes, namely the Tropics and Midlatitudes N and S (respectively number 17 to 19 in Table 1). Smaller regions, namely Tharsis South-East, Tempe Terra, Chryse Planitia, Lunae Planum, Valles Marineris, Southern Hemisphere cloud bridge, Tyrrhenea Terra, Argyre and Hellas Planitia are included (or overlap largely) in the three regions with the largest extension. Small regions give less information about the diurnal (and annual) cloud life cycle, but this information is consistent with that of the large regions in most cases. For example, the coinciding ICIR gridpoints on the large Tropical region (25°S – 25°N ; 90°W – 120°E ; Fig. 11) and on the smaller, included Valles Marineris region (15°S – 5°S ; 90°W – 45°W ; number 12 in Table 2) generally take similar or close values (Fig 19).

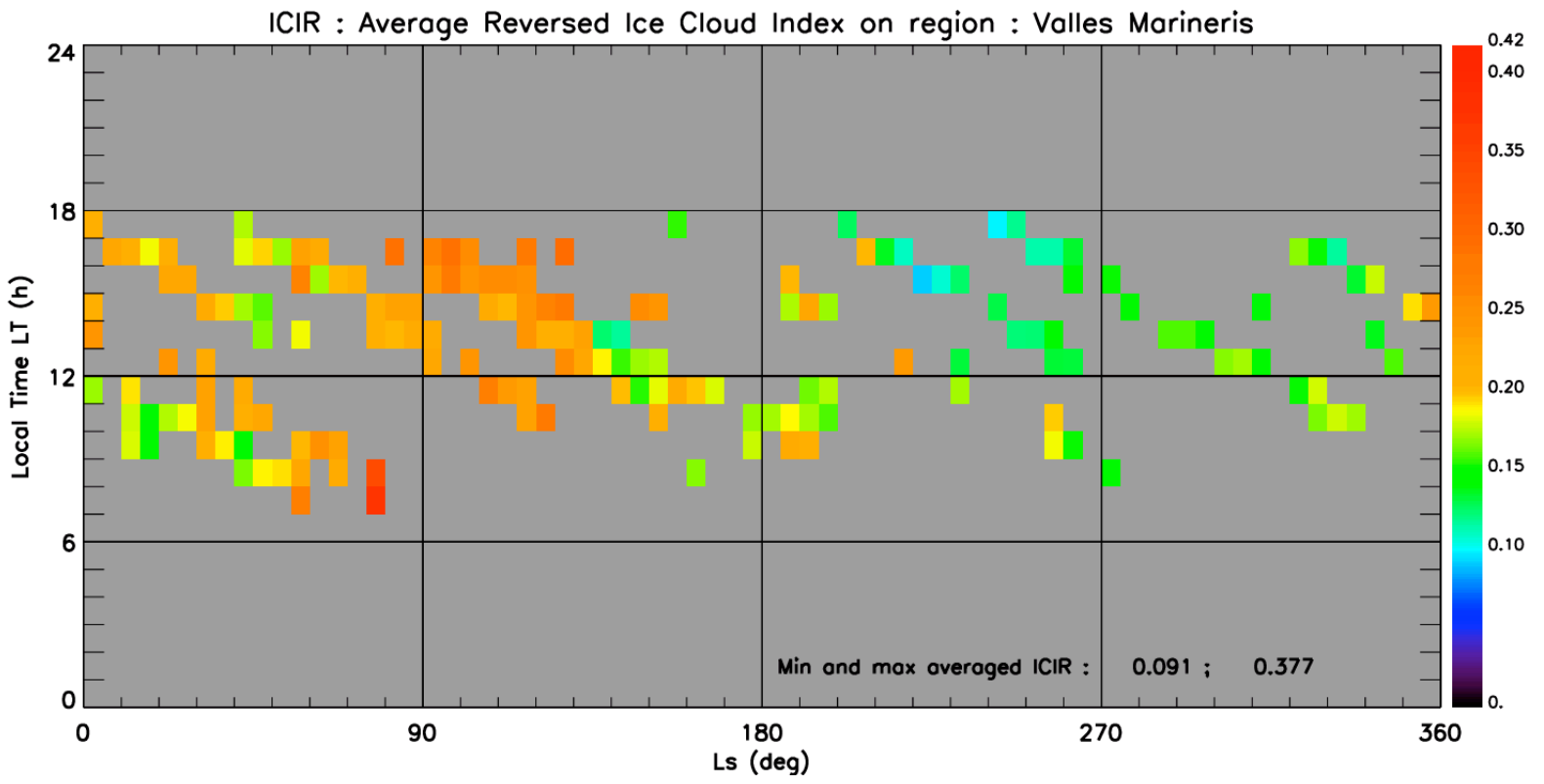

**Figure 19 :** diurnal cycle of the Ice Cloud Index over the Valles Marineris region (15°S - 5°N ; 90°W - 45°W) during the Martian year. The color scale is the same as the scale from the Tropical region scale on Figure 11.

Volcanoes (Olympus, Arsia, Elysium Mons and Alba Patera) are known to be covered or surrounded by clouds. Corresponding regions even cover a smaller surface. Their ICIR values indicate the presence of clouds at some periods of the day and the year but are not in sufficient number to show a diurnal cycle. These ICIR values are compatible with those of the three largest regions, with the exception of the Alba Patera region where fewer clouds are observed than in the largely overlapping Midlatitude N (25° – 55° N) region during winter.

## 6. Conclusion

Our high-resolution 4-dimensional gridded data products, the Reversed Ice Cloud Index and the Percentage of Cloudy Pixels, are valuable indicators for detecting and characterizing Martian water-ice clouds, although the dataset is sparsely populated (~2% of all daytime 4-dimensional grid-points). The ICIR and the PCP are complementary products representing the cloudiness (presence of clouds and abundance) in two different ways:

- The ICIR is a general indicator of cloudiness. It is a better extractor of thin clouds at pixel scale, especially from pixels with values just above the threshold used for the calculation of the PCP. Note that the gridded ICIR can take values below this threshold because it results from an average of pixels that can be cloudy or not. It has been shown to be a very good proxy for the mass of the water-ice column.

- The PCP gives subgrid scale information that may be useful for GCMs.: It can be the indicator of partial cloud coverage of a grid cell (in our case of 1° x 1°). It is less adapted for the detection of individual thin clouds at pixel scale than the ICIR, because such clouds are eliminated by the threshold-based selection prior to the PCP calculation.

The estimation of the uncertainty on the ICIR calculation depends on the error estimation on the reflectance values used to calculate the ICIR at pixel scale and on the variability of the ICIR originating from different pixels (and possibly different orbits) at the level of the grid cell. The highest uncertainty values are obtained :

- in regions under low solar illumination (i.e. at high incidence angles, with resulting low reflectance values), mainly located at high latitudes, and over all regions at dusk (end of the afternoon).
- in regions of low albedo of the surface, whenever clouds are present or not, such as Syrtis Major and the dark latitudinal band around 60°N covering Acidalia and Utopia Planitia.

About two-thirds of the 4-dimensional gridpoints have a small or intermediate relative ICIR error, i.e. below 20 %. The major contributor to the ICIR uncertainty is the instrumental error, for about 80 %, whereas the variability of meteorological situation (observed locally between data from the same orbit, or between data from several overlapping orbits) has a minor impact. We did not investigate the uncertainty estimation of the PCP, which is more complex to evaluate and depends on the threshold used to select clouds. These complementary studies were beyond the scope of this article.

In principle gridded ICIR values should be of higher quality, from a statistical point of view, if they were derived from a large number of original pixels. Gridded ICIR values at high latitudes (above 75°), and especially those close to the poles (above 85°), are always derived from a small number of pixels, and thus should be considered of lesser quality even if the corresponding ICIR standard deviation is small. Further studies are necessary to quantify the impact of the number of pixels on the representativeness of gridded ICIR values.

When integrated over a large time period of the day and the year, 2D maps can show the spatial cloud coverage of the planet over (part of) a season. The major cloud structures are similar to those observed in TES optical thickness data and those derived from the water-ice column of the Mars Climate Database.

When integrated over a specific period and larger area (covering more than one grid cell), the ICIR can also show the daily evolution of cloudiness. As an example, we found that the diurnal cycle in the tropics is characterized by important cloudiness in the morning, decreasing until noon, and increasing again in the afternoon ; this can be explained with model simulation outputs as resulting from a thermal tide.

The current 4-dimensional database covers only a small fraction of the Martian cloud climatology. A consequence of the sampling imposed by the satellite orbit and the instrument availability is the visible difference of aspect of the temporal charts of the water-ice column when all instants are integrated (Figs. 11c, 13c, 14c and 16c are smooth) and when only a temporal sample is available and integrated (Figs. 11b, 13b 14b, 16b and 18b, with the sampling of corresponding ICIR data, are more rugged). We do not now yet if this sample is a good representation of the complete diurnal and annual cloud life cycle. An argument in favor of its representativeness is that the Martian climate observed by instruments from other satellite or Earth-based telescopes indicate that at least at the annual and interannual scale (with the exception of global dust storms) the Martian climate shows little variability, less than the terrestrial climate.

Another open question is how well do the ICIR and PCP represent the water-ice clouds. The ICIR corresponds to the absorption of small (micrometer to millimeter-sized) water ice particles, which compose Martian (water ice) clouds but can also be produced by thin frost. At this stage we did not attempt to discriminate clouds and frost and in this text we considered that the ICIR is only an indicator of clouds. This assumption is likely justified in a large majority of cases, although water ice frost has occasionally been observed in some sites and during some periods of the day and the year on the surface (Langevin et al., 2007 ; Carrozzo et al., 2009).

The ICIR and PCP appear to be representative of the cloud cover when compared to other cloud-related datasets. Qualitative agreement has been observed in cloudy areas with optical thickness from TES ($\tau_{TES\ _wice}$) and with the water-ice column from the MCD. Acceptable correlations at global scale (for TES data) and regional scales (for MCD data) suggested quite logically that a relation could exist in cloudy areas between the ICIR on one side, and the $\tau_{TES\ _wice}$ and the WaterIceCol on the other side. Olsen et al. (2019) have confirmed and described this relation between the related (original) ice cloud index (ICI) and the water ice column.

The four main 4-dimensional variables derived from OMEGA, the ICIR, the PCP, the uncertainty on the ICIR and the number of Mars Express OMEGA orbit files used at each gridpoint were uploaded to the ESA Planetary Science Archive at open.esa.int/esa-planetary-science-archive/, conforming to the Planetary Data System (version 4) requirements. Non-heliosynchronous satellites, namely MAVEN (Mars Atmosphere and Volatile Evolution mission), ExoMars / Trace Gas Orbiter (Korablev et al., 2019 ; Vandaele et al., 2019), and the upcoming Emirates Mars Mission (Amiri et al., 2018), may potentially provide complementary information and data on water-ice clouds at various local times of the Martian day in the near future.

## Aknowledgement

This work has received funding from the European Union's Horizon 2020 Programme (H2020-Compet-08-2014) under grant agreement UPWARDS-633127. Michael Smith has provided the TES mapped climatology data, part of the validation.

## Appendix A : Impact of the number of pixels per gridpoint on the ICIR

The ICIR value at a 4-dimensional gridpoint is the average of a variable number of ICIR values from valid individual pixels. Could this number of pixels impact the quality and the representativeness of gridded ICIR values ?

We identified several factors that define this number of pixels :

- The number of orbits providing valid pixels : from one to 6 orbits.
- The filters applied on pixels, described in section 2.2, which remove corrupted or inconsistent pixels and thus reduce the number of valid pixels.
- The removal of pixels during the global dust storm period of MY 28, in order to build a standard cloud climatology, also reduces the number of pixels during the dust storm season.
- Due to Mars Express's elliptical orbit, the area on the surface covered by a single pixel varies with the satellite's altitude. Derived from an instantaneous field of view (IFOV) value of 1.4 mrad x 1.1 mrad (Bonello et al., 2005) and for a scan at nadir, one pixel covers ~0.17 km2 from an altitude of 330 km, and ~75 km2 from an altitude of 7000 km. Larger surfaces can be covered with other scan modes (across-track, along-track…). On the other hand, one square degree covers a surface on the planet that decreases with increasing latitude : one square degree covers 3468 km2 at the equator, 341 km2 at latitude 84°, and only 31 km2 at the pole. Therefore, a single pixel observed from an altitude of 7000 km2 may cover more than 2 square degrees at the pole (and only 1/46 square degree at the equator).

We found that the number of averaged pixels per gridpoint is highly variable, from one single pixel up to 98093 pixels. From a statistical point of view, gridded ICIR values should be more significant and better represent clouds when they result from the average of a large number of pixels. But at this stage, no simple relation can be established between the number of pixels per gridpoint and the averaged ICIR.

We then tried to locate areas where gridded ICIR values are derived from a small, respectively from a large number of pixels. For this purpose, we constructed maps (defined by their longitude and latitude) with the frequency of occurrence of a number of pixels in a delimited range, with reference to the total number of (Ls,LT) occurrences at the gridpoint.

Figure A1 shows that gridpoints derived from one or 2 pixels exist, but are overall rare, observed only in a few percent of the cases. They are located mainly at high latitudes, above 75°. The highest frequencies can be observed near the North Pole, possibly resulting from the coverage of small degrees on the surface and / or from a position of Mars Express at a high altitude.

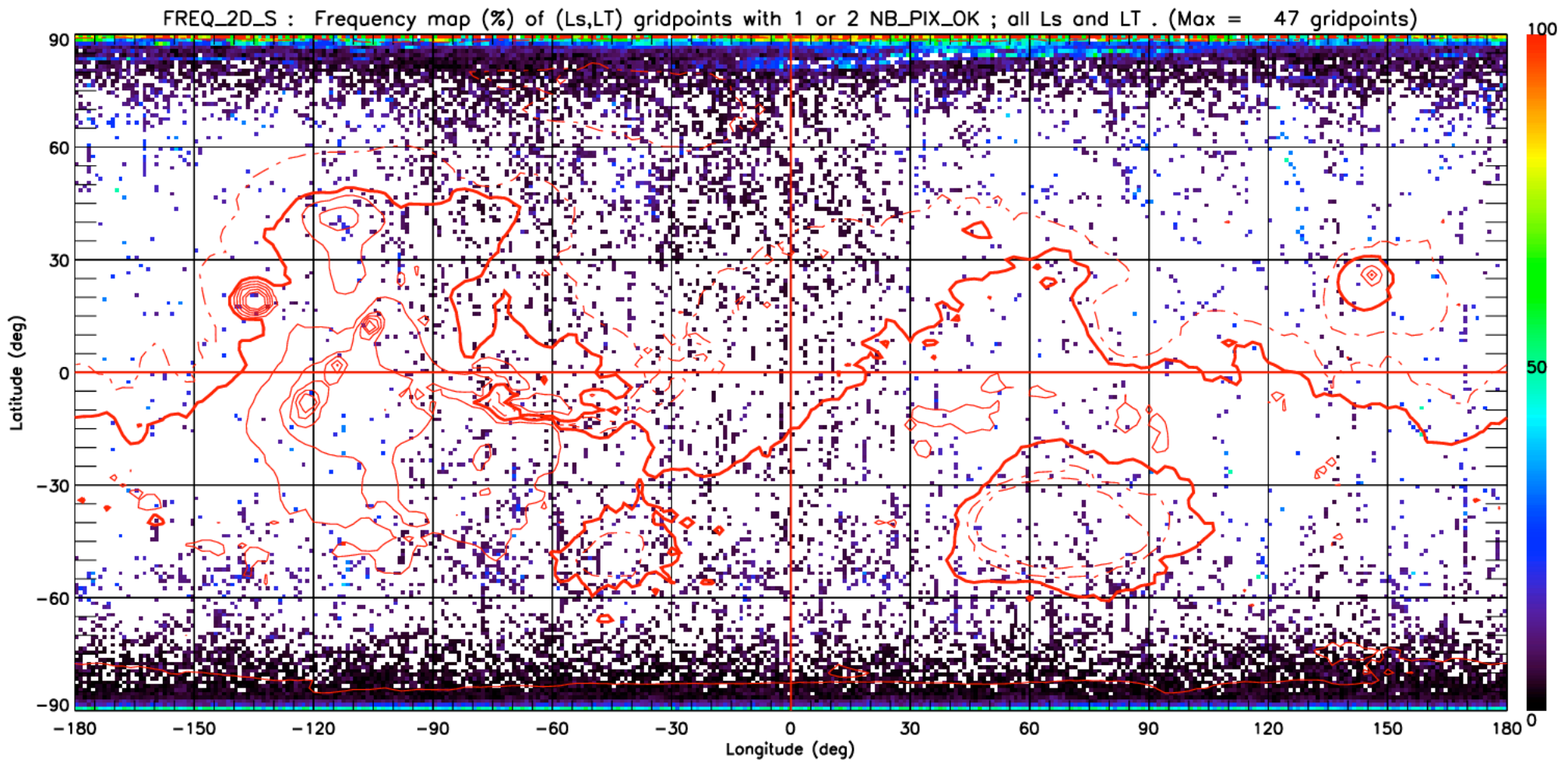

**Figure A1:** frequency of occurrence map (%) of ICIR gridded values for all available (Ls,LT) instants, derived from one or two original ICIR pixels. The equator, and contour lines (in red) corresponding to surface elevation are superimposed on this figure and on following maps: thick line: datum (0 m) ; thin continuous line: elevation > 0 ; thin dashed line: elevation < 0. Consecutive contour lines are spaced every 2500 m.

Figure A2 confirms the impact of latitude on the frequency of occurrence. Gridpoints at polar latitudes (above 84°N, and 87°S) are always derived from less than 100 pixels. The proportion of gridpoints derived from less than 100 pixels decreases with latitude, but is still important at relatively high latitudes (above 55°N and 45°S). A small proportion of gridpoints (< 10 %) is nevertheless present in midlatitude and tropical regions, with a more frequent coverage in longitude between 100°W and 30°E, corresponding to the area with a denser coverage by the OMEGA instrument (see fig. 2a).

Figure A3, shows the frequency of occurrence for gridpoints derived from a large number of pixels (more than 100). This figure is complementary to Fig. A2 and indicates that at midlatitudes and in the tropics, the ICIR is derived from a large number of pixels in all, or at least in a large majority of (Ls,LT) cases.

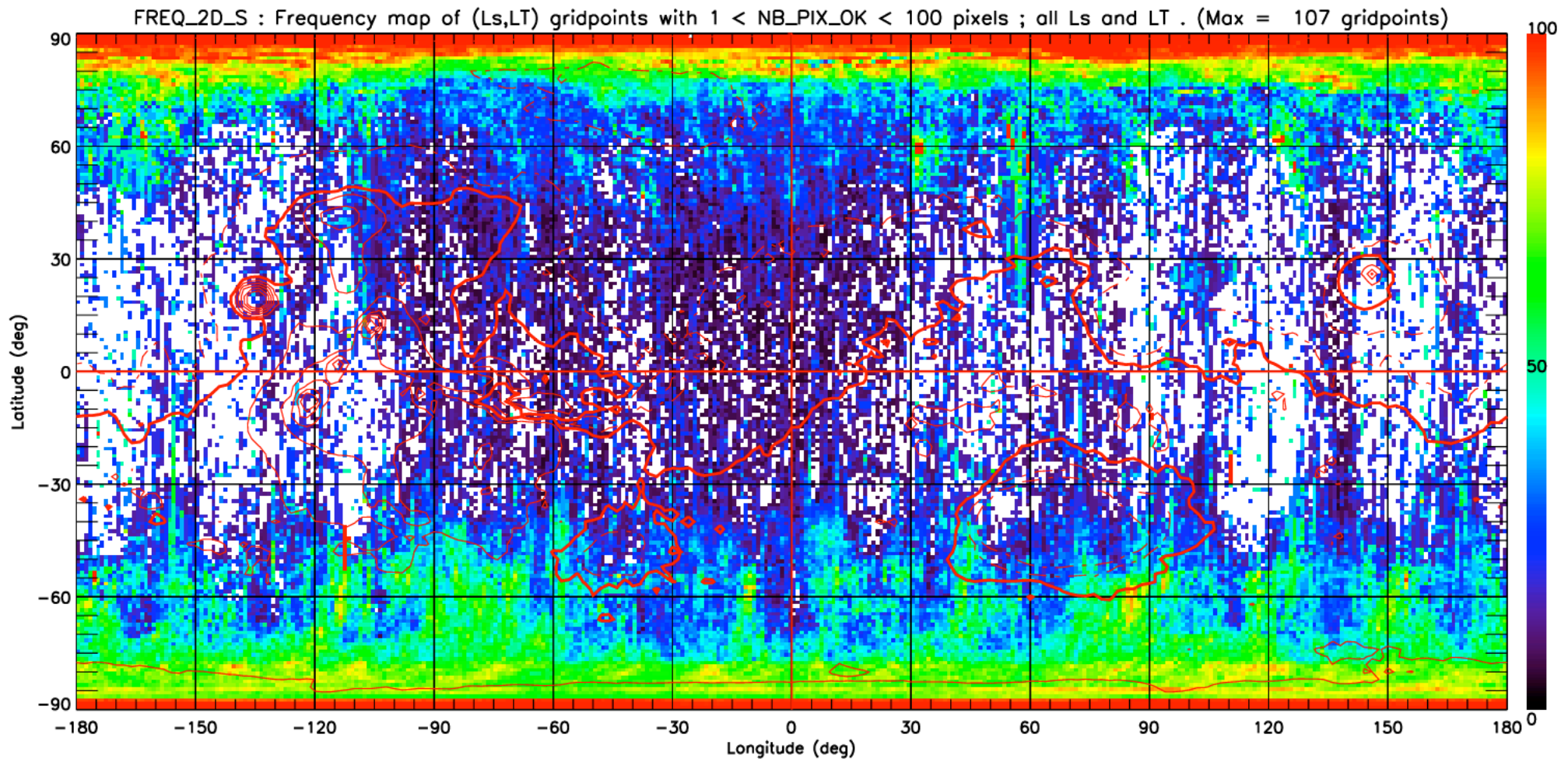

**Figure A2 :** frequency of occurrence map (%) of ICIR gridded values for all available (Ls,LT) instants, derived from less than 100 original ICIR pixels. The equator, and contour lines (in red) corresponding to surface elevation are superimposed on this figure and on following maps: thick line: datum (0 m) ; thin continuous line: elevation > 0 ; thin dashed line: elevation < 0. Consecutive contour lines are spaced every 2500 m.

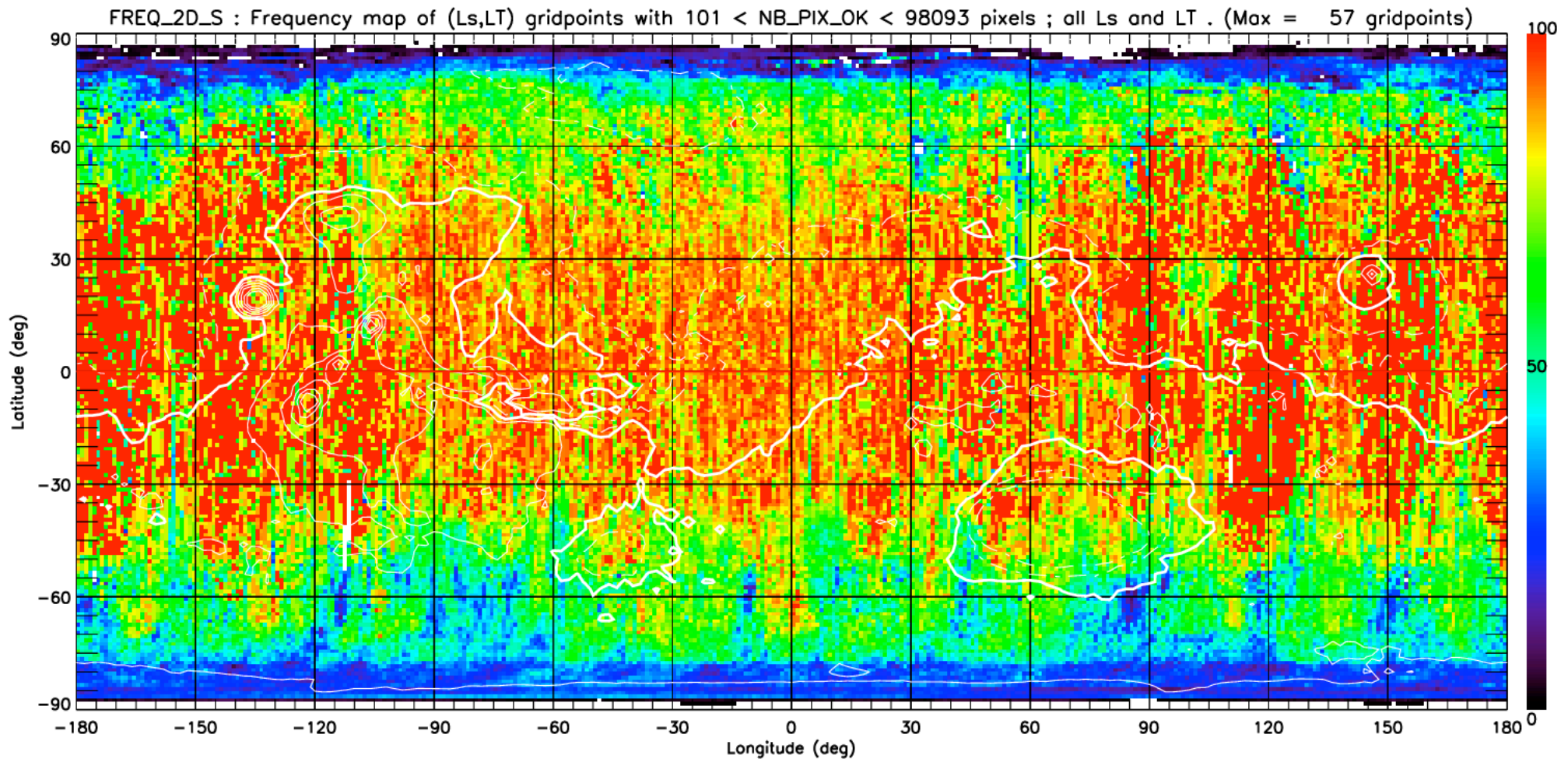

**Figure A3 :** frequency of occurrence map (%) of ICIR gridded values for all available (Ls,LT) instants, derived from 100 or more original ICIR pixels. The equator, and contour lines (in white) corresponding to surface elevation are superimposed on this figure and on following maps: thick line: datum (0 m) ; thin continuous line: elevation > 0 ; thin dashed line: elevation < 0. Consecutive contour lines are spaced every 2500 m.

The main conclusion from these results is that the number of observations (i.e. pixels) used to calculate gridded ICIR values is always small (below 100 pixels) at high latitudes, in particular close to the poles, and that this number is large at midlatitudes and in the tropics for a large majority of gridpoints. A study going more into detail would be necessary to investigate the impact of other factors, i.e. the quality filters applied on the data, the altitude of the satellite, the surface of the square degree, and also the instants (Ls,LT) when the number of pixels is small or large.

## Appendix B : Access to OMEGA ICIR and PCP data

An updated version of OMEGA ICIR, PCP and related data is currently being installed on the Planetary Surface Portal (PSUP) website at the Institut d'Astrophysique Spatiale, Orsay, France : http://psup.ias.u-psud.fr , where the data may be accessed and viewed in a near future. A description of the PSUP site is available in an article by Poulet et al. (2018).

## References

Amiri, S., McGrath, M., Al Awadhi, M., Almatroushi, H., Sharaf, O., AlDhafri, S., AlRais, A., Wali, M., AlShamsi, Z., AlQasim, I., AlHarmoodi, K., Ferrington, N., Withnell, P., Reed, H., AlTeneiji, N., Landin, B. and AlShamsi, M. (2018). Emirates Mars Mission (EMM) 2020 Overview. 42nd COSPAR Scientific Assembly, 14-22 July 2018, Pasadena, California. COSPAR Meeting, 42, B4.2-2-18.

Audouard, J., Poulet, F., Vincendon, M., Bibring, J.-P., Forget, F., Langevin, Y., and Gondet, B. (2014a). Mars surface thermal inertia and heterogeneities from OMEGA/MEX. Icarus, 233:194–213.

Audouard, J., Poulet, F., Vincendon, M., Milliken, R., Jouglet, D., Bibring, J.-P., Gondet, B., and Langevin, Y. (2014b). Water in the Martian regolith from OMEGA/Mars Express. J. Geophys. Res., 119(E16):1969–1989.

Benson, J. L., Bonev, B. P., James, P. B., Shan, K. J., Cantor, B. A., and Caplinger, M. A. (2003). The seasonal behavior of water ice clouds in the Tharsis and Valles Marineris regions of Mars: Mars Orbiter Camera observa tions. Icarus, 165:34–52.

Benson, J. L., James, P. B., Cantor, B. A., and Remigio, R. (2006). Interannual variability of water ice clouds over major martian volcanoes observed by MOC. Icarus, 184:365–371.

Benson, J. L., Kass, D. M., and Kleinböhl, A. (2011). Mars' north polar hood as observed by the Mars Climate Sounder. J. Geophys. Res., 116:E03008.

Benson, J. L., Kass, D. M., Kleinböhl, A., McCleese, D. J., Schofield, J. T., and Taylor, F. W. (2010). Mars' south polar hood as observed by the Mars Climate Sounder. J. Geophys. Res., 115:E12015.

Bibring, J.-P., Soufflot, A., Berthé, M., Langevin, Y., Gondet, B., Drossart, P., Bouyé, M., Combes, M., Puget, P., Semery, A., Bellucci, G., Formisano, V., Moroz, V., Kottsov, V., Bonello, G., Erard, S., Forni, O., Gendrin, A., Manaud, N., Poulet, F., Poulleau, G., Encrenaz, T., Fouchet, T., Melchiori, R., Altieri, F., Ignatiev, N., Titov, D., Zasova, L., Coradini, A., Capacionni, F., Cerroni, P., Fonti, S., Mangold, N., Pinet, P., Schmitt, B., Sotin, C., Hauber, E., Hoffmann, H., Jaumann, R., Keller, U., Arvidson, R., Mustard, J., and Forget, F. (2004). OMEGA: Observatoire pour la Minéralogie, l'Eau, les Glaces et l'Activité, pages 37–49. ESA SP-1240: Mars Express: the Scientific Payload.

Bonello, G., Bibring, J.-P., Soufflot, A., Langevin, Y., Gondet, B., Berthé, M., and Carabetian, C. (2005). The ground calibration setup of OMEGA and VIRTIS experiments: description and performances. Planet. Space Sci., 53:711–728.

Carrozzo, F. G., Bellucci, G., Altieri, F., D'Aversa, E., and Bibring, J.-P. (2009). Mapping of water frost and ice at low latitudes on Mars. Icarus, 203:406–420.

Chicarro, A., Martin, P., and Trautner, R. (2004). The Mars Express mission: an overview, pages 3–13. ESA Special Publication. ESA SP-1240: Mars Express: the Scientific Payload.

Curran, R. J., Conrath, B. J., Hanel, R. A., Kunde, V. G., and Pearl, J. C. (1973). Mars: Mariner 9 spectroscopic evidence for H2O ice clouds. Science, 182:381–383.

Forget, F., Hourdin, F., Fournier, R., Hourdin, C., Talagrand, O., Collins, M., Lewis, S. R., Read, P. L., and Huot, J.-P. (1999). Improved general circulation models of the Martian atmosphere from the surface to above 80 km. J. Geophys. Res., 104:24,155–24,176.

French, R. G., Gierasch, P. J., Popp, B. D., and Yerdon, R. J. (1981). Global patterns in cloud forms on Mars. Icarus, 45:468–493.

Hale, A. S., Tamppari, L. K., Bass, D. S., and Smith, M. D. (2011). Martian water ice clouds: A view from Mars Global Surveyor Thermal Emission Spectrometer. Journal of Geophysical Research (Planets), 116:E04004.

Kahn, R. (1984). The spatial and seasonal distribution of Martian clouds and some meteorological implications. J. Geophys. Res., 89:6671–6688.

Korablev, O., Montmessin, F., Trokhimovskiy, A., Fedorova, A. A., Shakun, A. V., Grigoriev, A. V., Moshkin, B. E., Ignatiev, N. I., Forget, F., Lefèvre, F., Anufreychik, K., Dzuban, I., Ivanov, Y. S., Kalinnikov, Y. K., Kozlova, T. O., Kungurov, A., Makarov, V., Martynovich, F., Maslov, I., Merzlyakov, D., Moiseev, P. P., Nikolskiy, Y., Patrakeev, A., Patsaev, D., Santos-Skripko, A., Sazonov, O., Semena, N., Semenov, A., Shashkin, V., Sidorov, A., Stepanov, A. V., Stupin, I., Timonin, D., Titov, A. Y., Viktorov, A., Zharkov, A., Altieri, F., Arnold, G., Belyaev, D. A., Bertaux, J. L., Betsis, D. S., Duxbury, N., Encrenaz, T., Fouchet, T., Gérard, J.-C., Grassi, D., Guerlet, S., Hartogh, P., Kasaba, Y., Khatuntsev, I., Krasnopolsky, V. A., Kuzmin, R. O., Lellouch, E., Lopez-Valverde, M. A., Luginin, M., Määttänen, A., Marcq, E., Martin Torres, J., Medvedev, A. S., Millour, E., Olsen, K. S., Patel, M. R., Quantin-Nataf, C., Rodin, A. V., Shematovich, V. I., Thomas, I., Thomas, N., Vazquez, L., Vincendon, M., Wilquet, V., Wilson, C. F., Zasova, L. V., Zelenyi, L. M., and Zorzano, M. P. (2018). The Atmospheric Chemistry Suite (ACS) of three spectrometers for the ExoMars 2016 Trace Gas Orbiter. Space Science Review, 214:7, 1–62.

Langevin, Y., Bibring, J.-P., Montmessin, F., Forget, F., Vincendon, M., Douté, S., Poulet, F., and Gondet, B. (2007). Observations of the south seasonal cap of Mars during recession in 2004-2006 by the OMEGA visible/near-infrared imaging spectrometer on board Mars Express. J. Geophys. Res., 112:E08S12.

Liu, J., Richardson, M. I., and Wilson, R. J. (2003). An assessment of the global, seasonal, and interannual spacecraft record of Martian climate in the thermal infrared. Journal of Geophysical Research (Planets), 108:8–1.

Madeleine, J.-B., Forget, F., Spiga, A., Wolff, M. J., Montmessin, F., Vincendon, M., Jouglet, D., Gondet, B., Bibring, J.-P., Langevin, Y., and Schmitt, B. (2012). Aphelion water-ice cloud mapping and property retrieval using the OMEGA imaging spectrometer onboard Mars Express. J. Geophys. Res., 117(E16):E00J07.

Mateshvili, N., Fussen, D., Vanhellemont, F., Bingen, C., Dodion, J., Montmessin, F., Perrier, S., Dimarellis, E., and Bertaux, J.-L. (2007). Martian ice cloud distribution obtained from SPICAM nadir UV measurements. J. Geophys. Res. (Planets), 112:E07004.

Millour, E., Forget, F., Spiga, A., Vals, M., Zakharov, V., Montabone, L., Lefevre, F., Montmessin, F., Chaufray, J.-Y., Lopez-Valverde, M., Gonzalez-Galindo, F., Lewis, S., Read, P., Desjean, M.-C., Cipriani, F., and MCD/GCM Development Team (2018). The Mars Climate Database (version 5.3). In From Mars Express to ExoMars Scientific Workshop, 27 - 28 February, ESA-ESAC Madrid, Spain, 2 pp.

Montabone, L., Forget, F., Millour, E., Wilson, R. J., Lewis, S. R., Cantor, B., Kass, D., Kleinböhl, A., Lemmon, M. T., Smith, M. D., and Wolff, M. J. (2015). Eight-year climatology of dust optical depth on Mars. Icarus, 251:65– 95.

Montmessin, F., Forget, F., Rannou, P., Cabane, M., and Haberle, R. M. (2004). Origin and role of water ice clouds in the Martian water cycle as inferred from a general circulation model. Journal of Geophysical Research (Planets), 109:E10004.

Navarro, T., Madeleine, J.-B., Forget, F., Spiga, A., Millour, E., Montmessin, F., and Määttanen, A. (2014). Global Climate Modeling of the Martian water cycle with improved microphysics and radiatively active water ice clouds. Journal of Geophysical Research (Planets), 119:1479–1495.

Olsen, K. S., Forget, F., Madeleine, J.-B., Szantai, A., Audouard, J., Geminale, A., Altieri, F., Bellucci, G., Oliva, F., Montabone, L., and Wolf, M. J. (2019). The distributions of retrieved properties from water-ice clouds in the Martian atmosphere using the OMEGA imaging spectrometer. Icarus (accepted; this issue).

Peale, S. J. (1973). Water and the Martian W cloud. Icarus, 18:497–501.

Pottier, A., Montmessin, F., Forget, F., Wolff, M., Navarro, T., Millour, E., Madeleine, J.-B., Spiga, A., Bertrand, T. ( 2015). Water ice clouds on Mars: a study of partial cloudiness with a global climate model and MARCI data. EGU General Assembly Conference Abstracts 17, 1751.

Pottier, A., Forget, F., Montmessin, F., Navarro, T., Spiga, A., Millour, E., Szantai, A., and Madeleine, J.-B. (2017). Unraveling the martian water cycle with high-resolution global climate simulations. Icarus, 291:82–106.

Poulet, F., Quantin-Nataf, C., Ballans, H., Dassas, K., Audouard, J., Carter, J., Gondet, B., Lozac'h, L., Malapert, J. C., Marmo, C., Riu, L., and S´ejourn´e, A. (2018). PSUP: A Planetary SUrface Portal. Planet.

Space Sci., 150:2–8.

Smith, M. D. (2009). THEMIS observations of Mars aerosol optical depth from 2002-2008. Icarus, 202: 444-452. Doi: 10.1016/j.icarus.2009.03.027

Smith, M. D. (2004). Interannual variability in TES atmospheric observations of Mars during 1999-2003. Icarus, 167:148–165.

Smith, S. A. and Smith, B. A. (1972). Diurnal and seasonal behavior of discrete white clouds on Mars. Icarus, 16:509–521.

Tamppari, L. K., Smith, M. D., Bass, D. S., and Hale, A. S. (2008). Water-ice clouds and dust in the north polar region of Mars using MGS TES data. Planet. Space Sci., 56:227–245.

Tamppari, L. K., Zurek, R. W., and Paige, D. A. (2003). Viking-era diurnal water-ice clouds. J. Geophys. Res., 108:9–1.

Vandaele, A. C., Lopez-Moreno, J.-J., Patel, M. R., Bellucci, G., Daerden, F., Ristic, B., Robert, S., Thomas, I. R., Wilquet, V., Allen, M., Alonso- Rodrigo, G., Altieri, F., Aoki, S., Bolsée, D., Clancy, T., Cloutis, E., Depiesse, C., Drummond, R., Fedorova, A., Formisano, V., Funke, B., Gonz´alez- Galindo, F., Geminale, A., Gérard, J.-C., Giuranna, M., Hetey, L., Ignatiev, N., Kaminski, J., Karatekin, O., Kasaba, Y., Leese, M., Lefèvre, F., Lewis, S. R., L´opez-Puertas, M., L´opez-Valverde, M., Mahieux, A., Mason, J., Mc- Connell, J., Mumma, M., Neary, L., Neefs, E., Renotte, E., Rodriguez-Gomez, J., Sindoni, G., Smith, M., Stiepen, A., Trokhimovsky, A., Vander Auwera, J., Villanueva, G., Viscardy, S., Whiteway, J., Willame, Y., and Wolff, M. (2018). NOMAD, an integrated Suite of three spectrometers for the ExoMars Trace Gas Mission: technical description, science objectives and expected performance. Space Science Review, 214:80, 1–47.

Vincendon, M., Audouard, J., Altieri, F., Ody, A. (2015) Mars Express measurements from albedo changes over 2004-2010. Icarus, 251: 145-163.

Vincendon, M., Pilorget, C., Gondet, B., Murchie, S., and Bibring, J.-P. (2011). New near-IR observations of mesospheric $CO_2$ and $H_2O$ clouds on Mars. Journal of Geophysical Research (Planets), 116:E00J02.

Wang, H. and Ingersoll, A. P. (2002). Martian clouds observed by Mars Global Surveyor Mars Orbiter Camera.

J. Geophys. Res., 107:8–1.

Wang, H. and Richardson, M. I. (2013). The origin, evolution and trajectory of large dust storms on Mars during Mars years 24-30 (1999-2011). Icarus, 251:112–127.

Willame, Y., Vandaele, A.-C., Depiesse, C., Lefevre, F., Letocart, V., Gillotay, D. and Montmessin, F. (2017). Retrieving cloud, dust and ozone abundances in the Martian atmosphere using SPICAM/UV nadir spectra. Planet. Space Sci., 142, 9-25.